\documentclass[twocolumn,fleqn,usenatbib]{mnras}

\usepackage{amssymb}
\usepackage{amsmath}

\usepackage{graphicx}
\usepackage{graphics}
\usepackage{color}
\usepackage{fancyhdr} 
\usepackage{graphicx}
\usepackage{float}
\usepackage[dvipsnames]{xcolor}
\usepackage{comment}

\usepackage{subfig}

\usepackage[utf8]{inputenc}

\usepackage[justification=centerlast, format=plain, labelfont=bf]{caption}

\def\VEV#1{\left\langle #1 \right\rangle}

\newcommand{\be}{\begin{equation}}
\newcommand{\ee}{\end{equation}}
\newcommand{\ba}{\begin{align}}
\newcommand{\ea}{\end{align}}

\newcommand{\Tcmb}{T_{\rm CMB}}

\newcommand{\Msun}{M_{\odot}}
\newcommand{\Mpcinv}{ {\rm Mpc}^{-1} }
\newcommand{\fesc}{f_{\rm esc}}
\newcommand{\xHI}{x_{\rm HI}}

\newcommand{\sfrd}{{\rm SFRD}}
\newcommand{\xa}{{x_\alpha}}
\newcommand{\codename}{{\tt Zeus21}}
\newcommand{\cmfast}{{\tt 21CMFAST}}
\newcommand{\class}{{\tt CLASS}}

\title{An Effective Model for the Cosmic-Dawn 21-cm Signal}

\author[J.B.~Mu\~noz]{
	Julian B.~Mu\~noz$^{1,2}$\thanks{E-mail: julianmunoz@austin.utexas.edu} \\
$^{1}$Department of Astronomy, The University of Texas at Austin, 2515 Speedway, Stop C1400, Austin, TX 78712, USA\\
$^{2}$Harvard-Smithsonian Center for Astrophysics, 60 Garden St., Cambridge, MA 02138, USA
}

\date{Accepted XXX. Received YYY; in original form ZZZ}

\pubyear{}

\begin{document}
	\label{firstpage}
	\pagerange{\pageref{firstpage}--\pageref{lastpage}}
	\maketitle

\begin{abstract}
	The 21-cm signal holds the key to understanding the first structure formation during cosmic dawn.
	Theoretical progress over the last decade has focused on simulations of this signal, given the nonlinear and nonlocal relation between initial conditions  and observables (21-cm or reionization maps).
	Here, instead, we propose an {\it effective} and fully analytic model for the 21-cm signal during cosmic dawn.
	We take advantage of the exponential-like behavior of the local star-formation rate density (SFRD) against densities at early times to analytically find its correlation functions including nonlinearities.
	The SFRD acts as the building block to obtain the statistics of radiative fields (X-ray and Lyman-$\alpha$ fluxes), and therefore the 21-cm signal.
	We implement this model as the public Python package {\tt Zeus21}.
	This code can fully predict the 21-cm global signal and power spectrum in $\sim 1$ s, with negligible memory requirements. 
	When comparing against state-of-the-art semi-numerical simulations from {\tt 21CMFAST} we find agreement to  $\sim10\%$ precision in both the 21-cm global signal and power spectra, after accounting for a (previously missed) underestimation of adiabatic fluctuations in {\tt 21CMFAST}.
	{\tt Zeus21} is modular, allowing the user to vary the astrophysical model for the first galaxies, and interfaces with the cosmological code {\tt CLASS}, which enables searches for beyond standard-model cosmology in 21-cm data.
	This represents a step towards bringing 21-cm to the era of precision cosmology.
\end{abstract}


\section{Introduction}

The cosmic-dawn era, which saw the formation of the first galaxies, is quickly becoming the next frontier of cosmology.
In addition to direct observations from telescopes such as {\it Hubble} and {\it James Webb}~(e.g., \citealt{Bouwens2021,Finkelstein_CEERS,Treu:2022iti,Robertson:2022gdk}), different 21-cm experiments are targeting neutral hydrogen through its spin-flip transition~\citep{Dewdney:SKA,DeBoer:2016tnn,vanHaarlem:2013dsa,Beardsley:2016njr,Voytek:2013nua}. These observatories are expected to provide us with tomographic information on the evolution of cosmic hydrogen from the beginning of cosmic dawn until the end of reionization ($z\sim 5-25$). 
As such, they have the potential to change our understanding of early-universe galaxy formation and cosmology.

Extracting physical insights from the upcoming 21-cm data is, however, challenging. 
Mapping initial conditions (matter densities and velocities) into observable 21-cm fields is a nonlinear and nonlocal process, one that is most often computed through simulations. 
These ought to account for the (nonlinear) physics of structure formation as well as the (nonlocal) propagation of the radiative fields.
Efforts in hydrodinamical simulations (e.g.,~\citealt{Gnedin:2014uta,Mutch2016,Ocvirk:2018pqh,Kannan21}) have improved the modeling of the latter epoch of reionization ($z\lesssim 10$).
Nevertheless, simulating the earlier cosmic-dawn era is more expensive, given the broad dynamical range spanned between the small first galaxies and the large mean-free path of photons.
Moreover, the parameter space to explore is vast, further increasing the cost of comparing a plethora of detailed simulations against data.
As an alternative, analytic models for the cosmic dawn were proposed in \citet{BarkanaLoeb2005} and \citet{Pritchard:2006sq} based on linear perturbation theory.
These were, however, eventually abandoned in favor of semi-analytic simulations such as~\cite{Santos:2009zk,Visbal:2012aw,Fialkov:2012su,Ghara:2014yfa,Battaglia:2012id,Mesinger:2007pd,Thomas:2008uq}, and the well-known \cmfast~(\citealt{Mesinger:2010ne,Murray:2020trn}, see however work based on the halo model~\citealt{Holzbauer:2011mv,Schneider:2020xmf,Schneider:2023ciq}). 
Despite the semi-analytic nature of these codes, computing the 21-cm signal still requires a simulation box with gas cells whose properties are evolved individually. 
For instance, a typical \cmfast\ box still takes $\sim 1$ hr to evolve through cosmic dawn, and sets stringent memory constraints\footnote{Alternative routes to speed through emulation have been explored~\citep{Kern:2017ccn,Saxena:2023tue}, though they still require generating expensive training sets; whereas linearized methods such as Fisher matrices are powerful for forecasts but cannot perform data inference~\citep{Mason:2022obt}.}. This problem is compounded by cosmic-variance considerations, as current 21-cm telescopes can only observe wavenumbers near the line of sight (with cosines $|\mu|\sim 1$,~\citealt{Pober:2014lva}), a subset of all modes that are simulated, which is thus plagued by larger sample noise.
Leaving aside computational cost, a fully analytic approach to cosmic dawn can provide new insights into early-universe astrophysics by laying bare the impact of each process, and is thus complementary to numerical simulations.

Here we present a new approach to analytically compute the 21-cm signal during cosmic dawn, building upon the  linear approach of \citet{BarkanaLoeb2005} and \citep[][jointly referred to as BLPF hereafter]{Pritchard:2006sq} but including nonlinearities as well as nonlocalities.
Since the 21-cm signal depends on the radiative (X-ray and Lyman-$\alpha$) fields sourced by the first galaxies, our building block will be the local star-formation-rate density (SFRD).
Historically, the main hurdle to obtain the 21-cm signal has been the complex behavior of the SFRD against density.
Our key insight is that the SFRD on a region of density $\delta_R$ (averaged over a radius $R$) scales as SFRD $\propto {\rm exp} (\gamma_R \delta_R)$ at high redshifts $z$, given the {\it effective bias} $\gamma_R$. 
This behavior is caused by the paucity of early-universe galaxies, which makes their abundance exponentially sensitive to over- and under-densities.
Under this assumption, and Gaussian initial conditions, the SFRD is a lognormal variable, for which correlation functions are analytically known~\citep{Coles:1991if}.
The 21-cm signal will depend upon a sum of SFRDs averaged over different radii $R$.
As such, we can {\bf analytically} compute the power spectrum of the 21-cm signal nonlinearly and nonlocally.

In this paper we present our effective theory for cosmic dawn, as well as the public package \codename\footnote{\url{https://github.com/JulianBMunoz/Zeus21}} where it is fully implemented.
\codename\ encodes the astrophysics desribed through this paper and it is built modularly in Python.
Moreover, \codename\ interfaces with the cosmic Boltzmann solver {\tt CLASS}\footnote{\url{http://class-code.net/}}~\citep{Blas:2011rf}, so it allows the user to vary the underlying cosmology as easily as the astrophysics.
We compare \codename\ against \cmfast\footnote{\url{https://github.com/21cmfast/21cmFAST}}, which has become the standard of semi-numerical 21-cm simulations.
When using the exact same inputs, we recover the same 21-cm global signal and power spectrum to within $20\%$ (which is the goal of current semi-numerical codes~\citealt{Zahn:2010yw,Majumdar:2014cza,Hutter:2018qxa,Ghara:2017vby}).
Yet, \codename\ takes 3 s to run in a laptop (down to $z=10$, or 5 s down to $z=5$), as opposed to $\sim 1$ hr for \cmfast, and has negligible memory requirements.
Fundamentally, semi-numerical simulations like \cmfast\ compute the same non-linear correlation function of weighted SFRDs, though by numerically placing it in a grid rather than analytically. As such, it is not surprising to find remarkable agreement between both approaches.

The rest of this paper is structured as follows.
In Sec.~\ref{sec:basics} we cover the basics of the 21-cm signal, and in Sec.~\ref{sec:effective_model} our effective model for cosmic dawn.
We use this model in Secs.~\ref{sec:cosmicdawn} and~\ref{sec:full21cmsignal} to compute the evolution of the 21-cm signal and related quantities such as the IGM neutral fraction and kinetic temperature.
We compare \codename\ against \cmfast\ in Sec.~\ref{sec:compare21cmfast}, and conclude with further discussion in Sec.~\ref{sec:conclusions}.
All units here will be comoving unless specified.

\section{Basics of the 21-cm signal}
\label{sec:basics}

We begin with a basic introduction to the 21-cm signal during cosmic dawn and reionization.
The interested reader is encouraged to visit \citet{Furlanetto:2006jb}, \citet{Pritchard:2011xb}, or~\citet{Liu:2019awk} for in-depth reviews.

Observationally, the 21-cm brightness temperature $T_{21}$ measures the deviation (i.e., absorption or emission) from the cosmic-microwave background (CMB) backlight.
Physically, this deviation is sourced by intervening neutral hydrogen.
As such, $T_{21}$ at a redshift $z$ and position $\mathbf x$ is determined by the thermal and ionization state of the intergalactic medium (IGM).
In particular, given the temperature $\Tcmb$ of the CMB, 
\be
T_{21} = \dfrac{T_S-\Tcmb}{1+z} \left(1-e^{-\tau_{21}}\right),
\label{eq:T21def_nonlinear}
\ee
where $T_S$ is the spin temperature, determined by the population ratio of the triplet and singlet states of neutral hydrogen.
The 21-cm optical depth is calculated through~\citep{Barkana:2000fd}
\be
\tau_{21} = (1+\delta) x_{\rm HI} \dfrac{T_0}{T_S} \dfrac{H(z)}{\partial_r v_r} (1+z),
\ee 
where $ x_{\rm HI}$ is the neutral hydrogen fraction, $\delta$ its density\footnote{In principle hydrogen may not trace matter fluctuations perfectly, even at large scales, so one should distinguish $\delta_H$ from $\delta$.
We will ignore this subtlety here for ease of comparison with past literature, and will revisit it in future work.}, $H(z)$ is the Hubble expansion rate, and $\partial_r v_r$ is the line-of-sight gradient of the velocity.
We have additionally defined a normalization factor
\be
T_0 = 34\,{\rm mK} \times \left(\dfrac{1+z}{16}\right)^{1/2}
\left(\dfrac{\Omega_b \,h^2}{0.022}\right)
\left(\dfrac{\Omega_m \,h^2}{0.14}\right)^{-1/2}.
\ee
Through this paper we will work with a {\it Planck} 2018 cosmology~\citep{Aghanim:2018eyx} unless otherwise specified, which fixes the (reduced) Hubble parameter $h$, as well as the baryon and matter densities $\Omega_b$ and $\Omega_m$.
Furthermore, we will drop  $\mathbf x$ and $z$ dependencies unless non-obvious.

Detecting a 21-cm signal requires hydrogen to be out of equilibrium with the photon background, as otherwise $T_S=\Tcmb$ in Eq.~\eqref{eq:T21def_nonlinear}, which yields $T_{21}=0$.
At early times, during the dark ages, collisions between hydrogen atoms couple the spin $T_S$ and kinetic $T_k$ temperatures~\citep{Loeb:2003ya}\footnote{In this $z>30$ pre-galaxy era there are fast and precise analytic predictions, e.g.,~\citet{Lewis:2007kz}.}.
Since the gas is colder than the CMB after the two fluids thermally decouple at $z\sim 200$, this coupling produces 21-cm absorption from then until $z\sim 30$, when collisions become too rare to keep the kinetic and spin temperatures coupled (so $T_S$ rises to $\Tcmb$).
That would be the end of the story, were it not for astrophysical sources.
When the first galaxies form they generate a UV background that permeates the universe, exciting hydrogen through the Wouthuysen-Field (WF,~\citealt{Wout,Field,Hirata:2005mz}) effect.
This again couples $T_S$ and $T_k$ and produces 21-cm absorption.
Later on, X-rays from the first galaxies will heat up the hydrogen in the IGM, driving $T_k$ (and thus $T_S$) above the CMB temperature $\Tcmb$, and thus bringing 21-cm into emission.
Eventually, reionization will drive $x_{\rm HI}$ towards zero, and with it the 21-cm signal.

We will focus on the cosmic-dawn epoch, where collisional excitations can be ignored.
In that case, the spin temperature can be obtained through
\be
T_s^{-1} = \dfrac{\Tcmb^{-1} + \xa T_c^{-1}}{ 1 + \xa},
\ee
where $\xa$ is the dimensionless WF coupling parameter (given by the local Lyman-$\alpha$ flux, as we will detail in Sec.~\ref{sec:cosmicdawn}),
and $T_c$ is the color temperature, closely associated with the gas kinetic temperature $T_k$ and numerically approximated by~\citep{Hirata:2005mz}
\be
T_c^{-1} \approx T_k^{-1} + g_{\rm col} T_k^{-1} \left( T_S^{-1} - T_k^{-1} \right),
\ee
given $g_{\rm col} \approx 0.4055$ K.
In practice, the difference between $T_c$ and $T_k$ during cosmic dawn is always below 5\% for the models we study, so one can keep $T_k$ in mind for intuition, though we do use the correct $T_c$ through this work.

By taking the approximation $\tau_{21} \ll 1$, and linear-order redshift-space distortions ($\delta_v$, ~\citealt{Barkana:2005jr,Mao:2011xp}), we can write the 21-cm temperature as
\be
T_{21} = T_0(z) (1+\delta - \delta_v)  \left (\dfrac{x_\alpha}{1 + x_\alpha} \right) \left( 1 - \dfrac{\Tcmb}{T_c} \right) x_{\rm HI}.
\label{eq:T21def}
\ee
This equation neatly separates the four different quantities that determine the 21-cm signal, namely: 
(i) the large-scale structure, through the local density and velocity;
(ii) the WF effect, which measures the Lyman-$\alpha$ flux emitted by the first galaxies; 
(iii) the gas kinetic/color temperature, determined by the competition between adiabatic cooling and X-ray heating due to the first stars; 
and (iv) reionization.
Through these four terms the 21-cm signal is a powerful tracer of the first stellar formation and the growing cosmic web at high redshifts.

\begin{figure}
	\centering
	\includegraphics[width=0.46\textwidth]{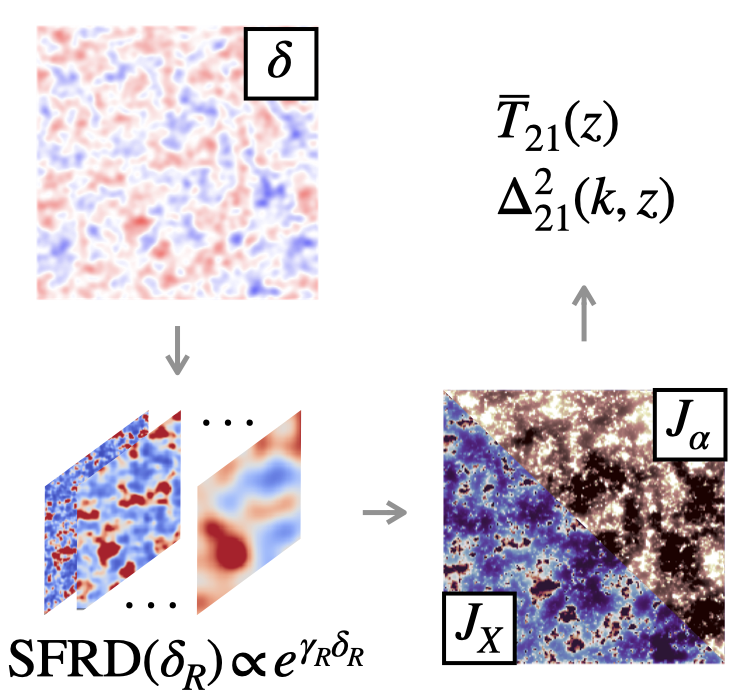}
	\caption{Schematic diagram of how \codename\ computes the 21-cm signal (global and fluctuations). 
	Starting from cosmology (over/underdensities $\delta$, {\bf top left}), \codename\ calculates the SFRD as a function of the density $\delta_R$ (smoothed over different radii $R$, {\bf bottom left}) with our lognormal model, given the calculated effective biases $\gamma_R$ (which depend on the halo-galaxy connection). The Lyman-$\alpha$ and X-ray fluxes $J_{\alpha/X}$ ({\bf bottom right}) are computed as a weighted sum of those SFRDs (with coefficients that depend on stellar parameters like the SEDs).
	These fluxes finally determine the 21-cm signal (global $\overline T_{21}$ and power spectrum $\Delta^2_{21}$).
	The slices shown here are for visualization purposes only, as \codename\ does not require simulation boxes. 
	The $\delta$ realization is generated with the aid of {\tt powerbox}~\citep{powerbox_Murray}, the SFRD$(\delta_R)$ slices follow our exponential effective model, and the $J_{X/\alpha}$ fluxes are sums of those SFRDs, given the weights defined in Eqs.~(\ref{eq:xaflucts},\ref{eq:Txflucts}).
	}
	\label{fig:diagram_code}
\end{figure}

The evolution of the 21-cm line is intimately linked to the intensity of the Lyman-$\alpha$ (through $\xa$) and X-ray (through $T_c$) backgrounds sourced by the first galaxies.
Both radiative fields can travel significant distances before being absorbed, so their fluxes at a point $\mathbf x$ depend on the emission over the past lightcone.
Schematically, they are given by integrals of the type~\citep{Pritchard:2006sq}
\be
J_{\alpha/X} \propto \int d R\,c_{\alpha/X}(R) \,{\rm SFRD}(R),
\label{eq:schematic_JaX}
\ee
where $R$ is the lightcone (comoving) radius, over which the star-formation-rate density (SFRD) is averaged, and $c_{\alpha/X}$ are coefficients that account for photon propagation, and depend on the specific astrophysics and cosmological parameters.
As such, one can think of $J_{\alpha/X}$ (and thus $T_{21}$) as nonlocal tracers of the SFRD. The SFRD is, in turn, a nonlinear function of the matter density field $\delta$.
Translating from initial conditions in $\delta$ to $T_{21}$ is, thus, a nonlinear and nonlocal process, as expected.

Rather than constructing simulation boxes, here we will account for these nonlinearities and nonlocalities with a fully analytic model.
We show a diagram of our model in Fig.~\ref{fig:diagram_code}.
This model uses the SFRD as a building block, which we find with a lognormal approximation to the nonlinear process of structure formation~\citep{Coles:1991if}.
The X-ray and Lyman-$\alpha$ fluxes are then obtained as weighted sums of the SFRD averaged over different radii $R$.
The 21-cm signal is straightforwardly computed from these fluxes, and thus accounts for nonlocalities as well as nonlinearities.
Note that we have shown a (simulation-like) realization of this algorithm in Fig.~\ref{fig:diagram_code}, but our effective model in \codename\ does not need to draw from a realization.
We will find the 21-cm signal, both global ($\overline T_{21}$) and fluctuations (through the power spectrum $\Delta^2_{21}$) fully analytically, i.e., without simulation boxes.
This will provide a sizeable computational advantage, both in terms of speed and memory usage, and a new way to understand the different processes at play during cosmic dawn.
Yet, our model can reproduce the results of more complicated semi-numerical simulations.
This approach relies on our effective lognormal model for the SFRD, which we now describe.

\section{An effective model for the SFRD}
\label{sec:effective_model}

The building block that determines the 21-cm signal is the SFRD.
Before delving into its density dependence, let us begin by computing its spatially averaged value~\citep{Madau:1996yh},
\be
\overline \sfrd(z) = \int dM_h \dfrac{dn}{dM_h} \dot M_*(M_h), 
\label{eq:sfrd}
\ee
where $ {dn/dM_h}$ is the halo mass function (HMF, which depends only on cosmology), and $\dot M_* (M_h,z)$ is the star-formation rate (SFR) of a galaxy hosted in a halo of mass $M_h$, which thus also holds information about the halo-galaxy connection. 
The main assumption taken so far is that stars form in galaxies, which are hosted in (dark-matter) haloes.
As such, this formula is very generic.

To compute the SFRD we will take a \citet*{Sheth:1999su} fit for the HMF, 
\be
\dfrac{dn}{dM_h} = -A_{\rm ST}  \dfrac{\rho_M}{M_h \sigma} \dfrac{d\sigma}{dM}  \nu (1 + \nu^{-2 p_{\rm ST}}) e^{-\nu^2/2} 
\label{eq:HMF_ST}
\ee
where $\sigma^2$ is the variance of matter fluctuations on the scale of $M_h$, $\nu = \sqrt{q_{\rm ST} } \delta_{\rm crit}/\sigma$ is a dimensionless variable, with $\delta_{\rm crit}=1.686$ as the usual critical barrier for collapse, and $q_{\rm ST}=0.85$ found to be a good fit in high-$z$ simulations~\citep{Schneider:2018xba}.
Moreover, $p_{\rm ST}=0.3$ corrects the abundance of small-mass objects~\citep{Sheth:1999su}, and $A_{\rm ST} = 0.3222\, \sqrt{2/\pi}$ is a normalization factor\footnote{We reproduce the numerical factors here for completeness, but encourage the reader to visit the \codename\ site for the implementation.}.

\begin{figure}
	\centering
	\includegraphics[width=0.46\textwidth]{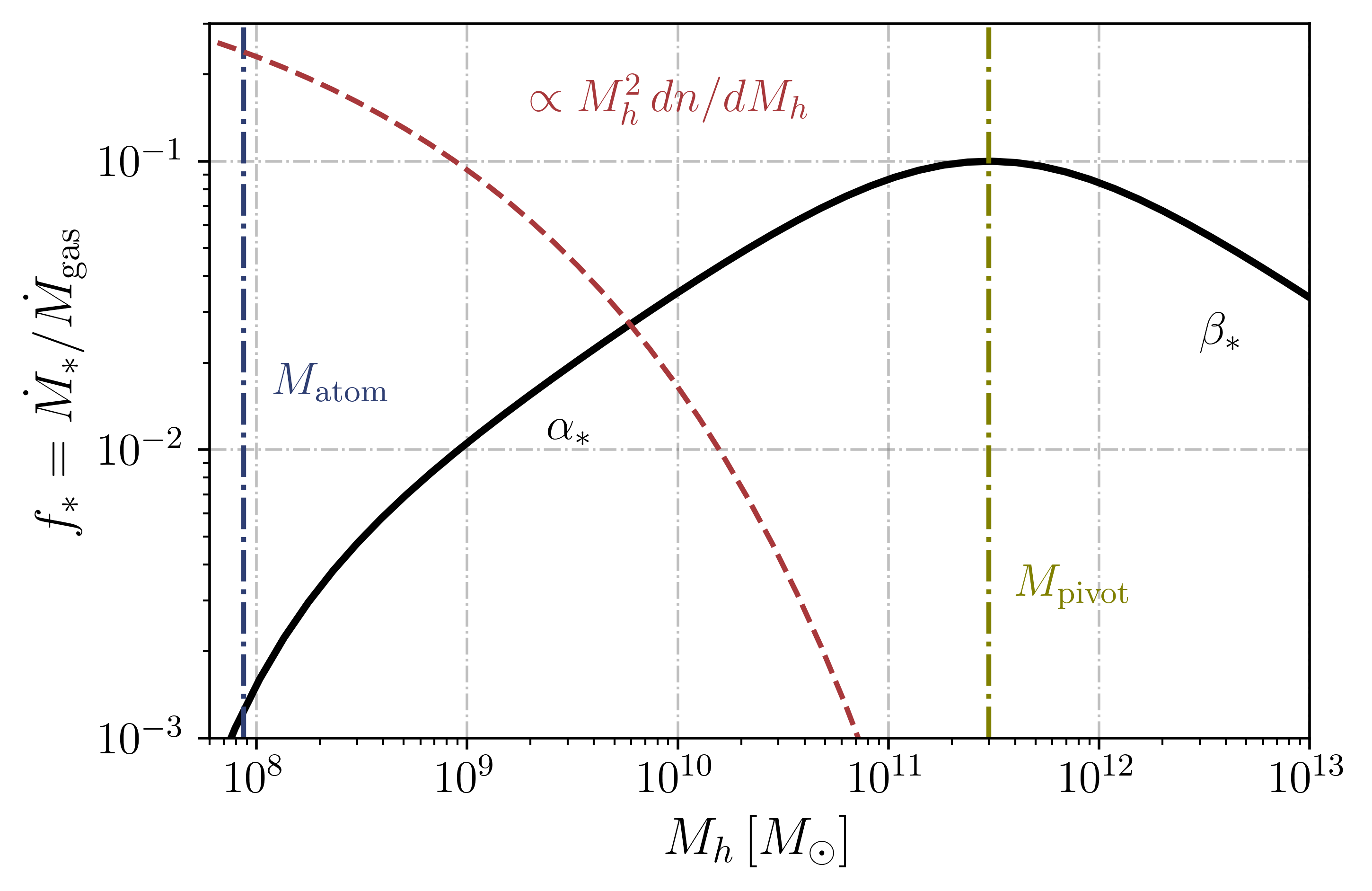}
	\caption{Star-formation efficiency $f_*$ in our parametric model from Eq.~\eqref{eq:fstar} at $z=10$.
	The power-law indices $\alpha_*$ and $\beta_*$ control the steepness of the faint and bright ends, respectively; $M_{\rm pivot}$ sets the pivot point; and the amplitude $\epsilon_*$ rescales the entire curve.
	Values for these parameters have been chosen to broadly reproduce the HST UVLFs. 
	Also shown is the minimum mass $M_{\rm atom}$ that a halo needs to cool gas through atomic transitions (blue vertical line), as well as a curve proportional to the HMF at $z=10$ (red dashed), which shows that most haloes are in the faint end during cosmic dawn.
	}
	\label{fig:fstar_vs_Mhalo}
\end{figure}

The SFR of high-$z$ galaxies is highly uncertain.
As such, we will make progress through flexible models that can capture different behaviors.
In particular, we implement two approaches to link $\dot M_*$ to $M_h$ based on recent analyses of galaxy data from the {\it Hubble} (HST) and {\it James Webb} (JWST) Space Telescopes.
Our main model, based on~\citet*[see also \citealt{Mason:2015cna,Furlanetto2016:feedback,Tacchella:2018qny,Mirocha2016_UVLF_GS,Schneider:2020xmf}]{Sabti:2021unj}, assumes that some fraction of the gas that is accreted by a galaxy is converted into stars, i.e., 
\be
\dot M_* = f_* f_b \dot M_h,
\label{eq:SFR_fiducial}
\ee
where $f_b=\Omega_b/\Omega_m$ is the baryon fraction (which we take to be mass independent) and the mass accretion rate $\dot M_h$ is found from the extended Press-Schechter formalism fitted in~\citet{Neistein:2006ak} (we also include an exponential model, where $M_h(z) \propto e^{\alpha z}$ with $\alpha=0.5$~\citealt{Schneider:2020xmf} as an alternative).
In both cases we assume a functional form for the efficiency
\be
f_*(M_h) = \dfrac{2\,\epsilon_*}{(M_h/M_{\rm pivot})^{-\alpha_*} + (M_h/M_{\rm pivot})^{-\beta_*} } f_{\rm duty} ,
\label{eq:fstar}
\ee
where
\be
f_{\rm duty} = \exp(-M_{\rm turn}/M_h)
\ee 
is a duty fraction, with $M_{\rm turn}$ the turn-over mass below which gas does not cool into stars efficiently.
We will assume that galaxies can form down to the atomic-cooling threshold, so $M_{\rm turn} = M_{\rm atom}(z)$~(corresponding to a virial temperature $T_{\rm vir}=10^4$ K,~\citealt{Oh:2001ex}).
We will study the effect of molecular-cooling haloes and their feedback~\citep{Johnson:2006mt} in future work.
Our star-formation efficiency $f_*$ has four free parameters: two power-law indices $\alpha_*$ and $\beta_*$ for the small- and large-mass regimes, a normalization $\epsilon_*$, and the pivot mass $M_{\rm pivot}$.
For our fiducial case we will take $\epsilon_* = 0.1$, $\alpha_* = -\beta_*=0.5$, and $M_{\rm pivot} = 3 \times 10^{11}\,\Msun$, which broadly fits the UV luminosity functions (UVLFs) from HST in the range $z=4-10$~\citep{Sabti:2020ser,Sabti:2021unj}.
We show our functional form for $\dot M_*$ in Fig.~\ref{fig:fstar_vs_Mhalo}, where we also illustrate how for the cosmic-dawn epoch ($z \geq 10$) the HMF is exponentially suppressed at high masses, so faint galaxies will dominate the emission.

For ease of comparison with \cmfast\ in Sec.~\ref{sec:compare21cmfast} we also implement their model (from ~\citealt{Park:2018ljd}), which takes
\be
\dot M_* = \tilde f_*(M_h)  M_h H(z)/t_*,
\label{eq:SFR_21cmfast}
\ee
with $t_* = 0.5$ as a dimensionless constant.
Here $\tilde f_*(M_h)$ is a single power-law in mass (with a ceiling at unity).
Both of these models can be made $z$ dependent, and enhanced by adding scattering~\citep{Zahn:2006sg,Whitler:2019nul}, which we will explore in future work.

We note that these are two example models inspired by seminumerical simulations. The effective approach we will present here is agnostic about the SFR parametrization, and can be extended to other models.

\subsection{An effective lognormal model for the SFRD}

The ``effective" nature of our model consists of approximating the dependence of the SFRD on density in such a way that arbitrary two-point functions---i.e., power spectra---can be computed analytically.
Let us describe how.
For notational clarity, we define the fluctuation on a quantity $q$ to be $\delta q(\mathbf x) = q(\mathbf x) - \overline q = \overline q \delta_q (\mathbf x)$, given its global average $\overline q$.
Through this work will use the reduced power spectrum
\be
\Delta^2_q = P_q \dfrac{k^3}{2\pi^2}
\ee
of different quantities $q$, as customary, and refer to it as power spectrum unless confusion can arise.
Moreover, we will define $\delta_R$ to be the linear matter overdensity averaged over a radius $R$ with a spherical tophat window (unless otherwise specified), and $\delta$ without a subscript is the usual (unwindowed) density.

We build upon our model for the average value of the SFRD, from Eq.~\eqref{eq:sfrd}, to find the SFRD of a region of comoving radius $R$ overdense by $\delta_R$ as~\citep{BarkanaLoeb2005}
\be
\sfrd(z | \delta_R) = (1 + \delta_R)\,\int dM_h \dfrac{dn}{dM_h}(\delta_R) \dot M_*(M_h), 
\label{eq:SFRD_EPS}
\ee
where the extra factor of $(1+\delta_R)$ in front with respect to~\citet{BarkanaLoeb2005} accounts for the conversion from Lagrangian to Eulerian space (see e.g.,~\citealt{Mesinger:2010ne}), and we have implictly assumed that the local density only modulates the cosmology (HMF) but not the astrophysics ($\dot M_*$), i.e., we neglect assembly bias~\citep{Wechsler:2001cs}.
One can find the density-modulated HMF by using the extended PS (EPS) formalism~\citep{Press:1973iz,Bond:1990iw}. 
Here we follow~\citet{BarkanaLoeb2005} and take
\be
\dfrac{dn}{dM_h}(\delta_R) = \dfrac{dn}{dM_h} \dfrac{dn^{\rm PS}/dM_h(\delta_R)}{\VEV{dn^{\rm PS}/dM_h(\delta_R)} },
\label{eq:HMF_EPS}
\ee
which has been shown to match well numerical simulations~\citep{Schneider:2020xmf}, and is constructed to return the correct average HMF from Eq.~\eqref{eq:HMF_ST}.
The density behavior follows~\citep{Lacey:1993iv}
\be
\dfrac{dn^{\rm PS}/dM_h(\delta_R)}{\VEV{dn^{\rm PS}/dM_h(\delta_R)} } = \mathcal C \dfrac{\tilde \nu}{\nu_0} \dfrac{\sigma^2}{\tilde \sigma^2} e^{a_{\rm EPS} (\tilde \nu^2 - \nu_0^2)/2}
\label{eq:HMF_PS}
\ee
where $\tilde \nu = \tilde \delta_{\rm crit}/\tilde \sigma$, $\nu_0=\delta_{\rm crit}/\sigma$, $a_{\rm EPS}$ is a numerical constant (well-fit by $a_{\rm EPS} = q_{\rm ST}$,~\citealt{Schneider:2020xmf}), and
where we have defined
\ba
\tilde \delta_{\rm crit} &= \delta_{\rm crit} - \delta_R,\ \rm and \nonumber \\
\tilde \sigma^2 &=  \sigma^2 - \sigma_R^2,
\label{eq:deltasigmatilde}
\end{align}
for a region of overdensity $\delta_R$ and variance $\sigma_R^2$.
We will find the value of the amplitude $\mathcal C$ numerically.
This EPS formalism is used in public 21-cm seminumerical codes, such as \cmfast, though with a slight variation as we will explain in Sec.~\ref{sec:compare21cmfast}.

\begin{figure}
	\centering
	\includegraphics[width=0.5\textwidth]{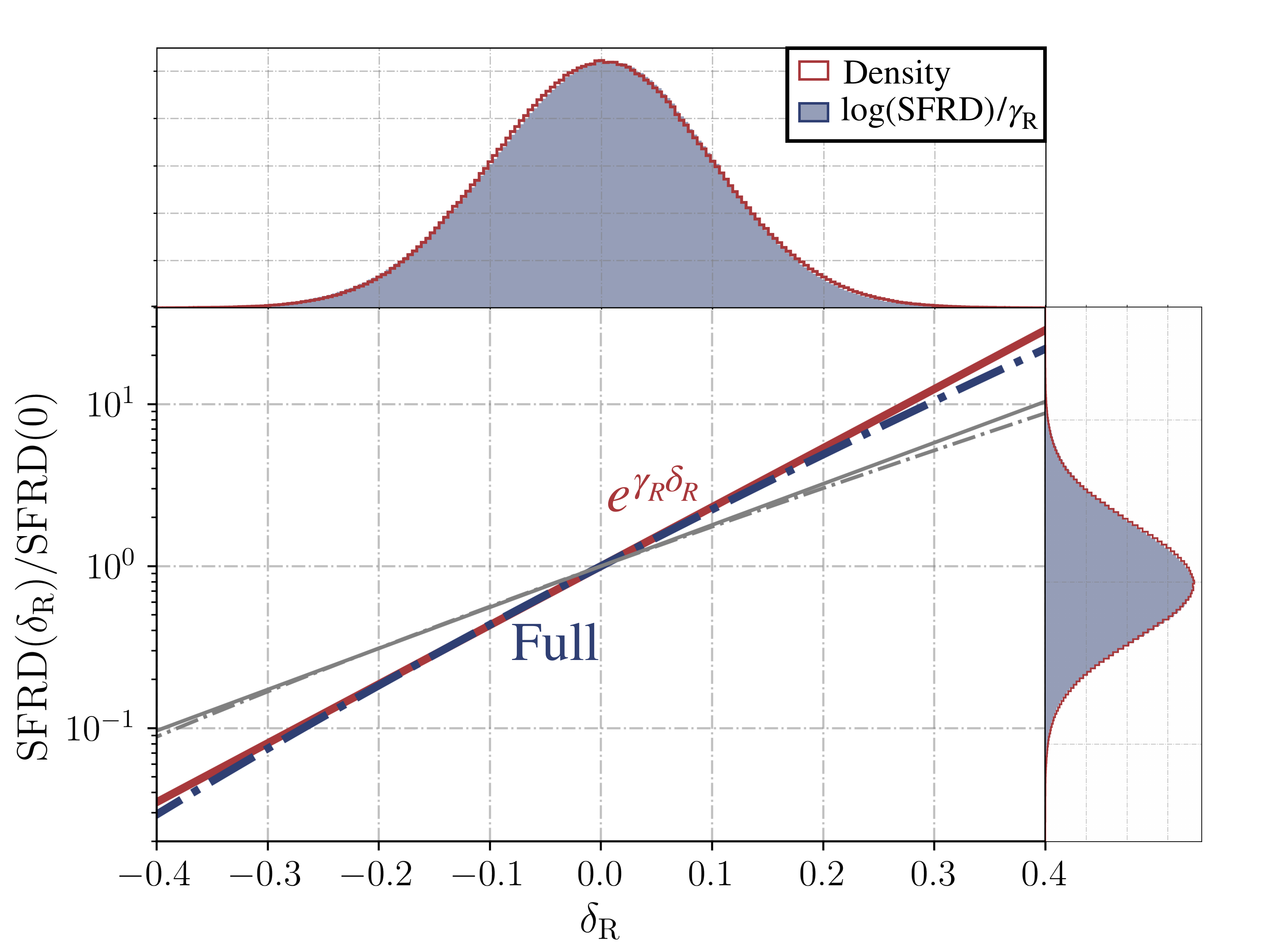}
	\caption{Behavior of the SFRD against density $\delta_R$ (smoothed with a Gaussian kernel of radius $R_G=3$ Mpc) at $z=15$ (thick colored lines) and $z=10$ (thin gray lines).
	Blue dot-dashed line represents the full result, compared to the red lognormal approximation we follow in this work.
	The right-hand side shows the histograms of both SFRDs from a simulation box (as detailed below in Fig.~\ref{fig:slice_sfrd}).
	Likewise, the top shows histograms for the density (red line), and the logarithm of the SFRD (divided by SFRD(0) and the exponent $\gamma_R$) at $z=15$, which also are in remarkable agreement. 
	}
	\label{fig:sfr_v_delta}
\end{figure}

\begin{figure*}
	\centering
	\includegraphics[width=0.8\textwidth]{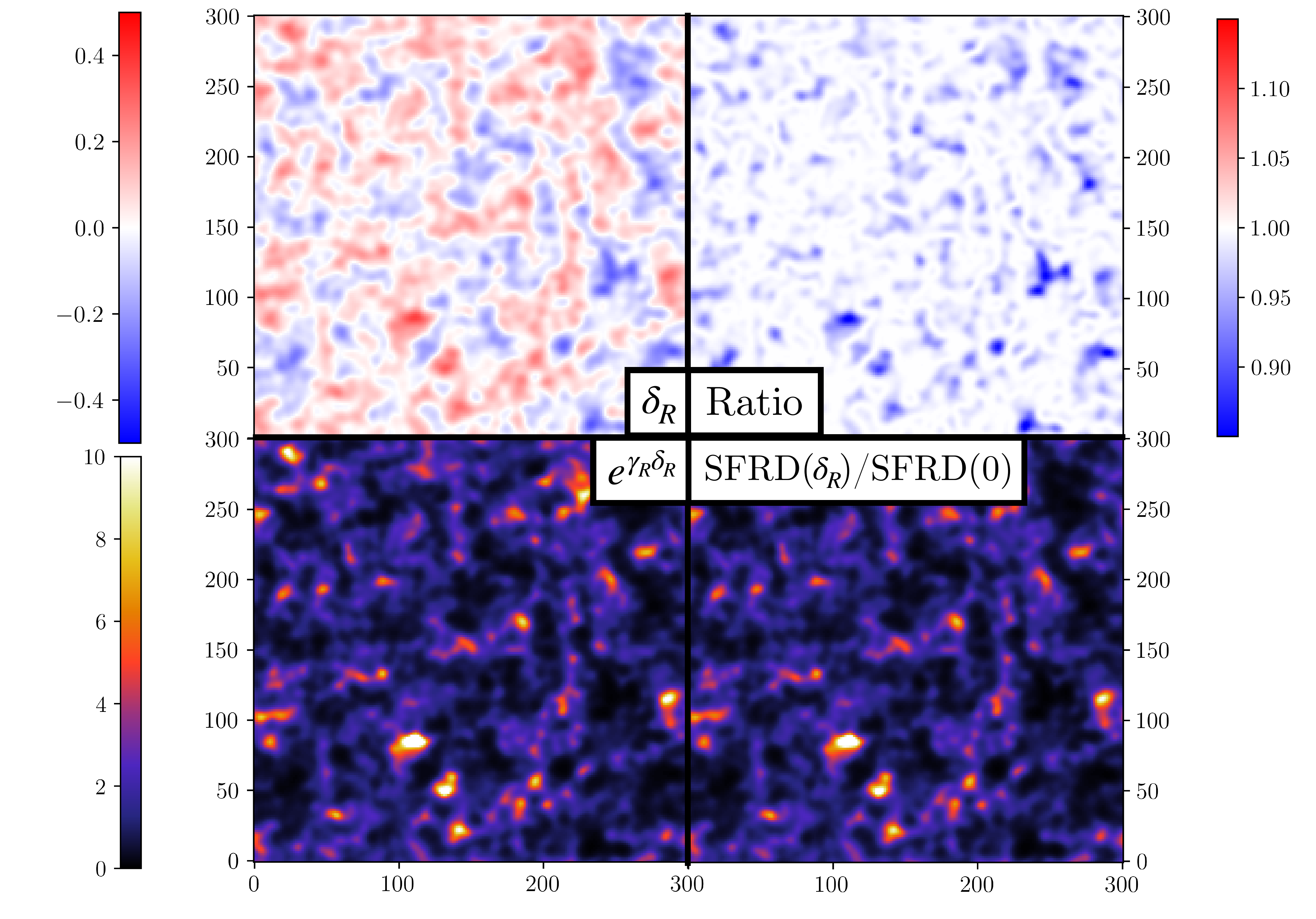}
	\caption{Slice of different quantities  at $z=15$ across a simulated box 300 Mpc in side. Top left is the density $\delta_R$, averaged with a 3 Mpc Gaussian kernel.
		The bottom shows both our approximation to the SFRD (left, a lognormal), and the full SFRD (right, from Eq.~\ref{eq:SFRD_EPS}), which are remarkably similar. The top right showcases their ratio, which deviates from unity by $\mathcal O(10\%)$ only at extreme $\delta_R$ values, in line with the expectation from Fig.~\ref{fig:sfr_v_delta}.
	}
	\label{fig:slice_sfrd}
\end{figure*}

We show the behavior of the SFRD as a function of density in a relevant cosmic-dawn scenario (smoothed with a Gaussian kernel of $R_G=3$ Mpc at $z=15$ and 10, chosen for illustration purposes) in Fig.~\ref{fig:sfr_v_delta}.
The strong dependence of this quantity on the overdensity owes to the extreme rarity of the first galaxies during early structure formation.
They reside in haloes with exponentially small abundances, which are therefore exponentially sensitive to the local matter density (through the threshold in Eq.~\ref{eq:deltasigmatilde}).
This poses an obstacle to the usual perturbative methods\footnote{One could in principle Taylor expand the SFRD in powers of $\delta_R$~\citep{Assassi:2014fva}. This would, however, not guarantee a positive definite solution, or one that grows monotonically with $\delta_R$, as shown in~\citet{Wu:2021rzh} in the context of the low-$z$ large-scale structure.}.

There is, however, a better alternative.
Encouraged by the SFRD shown in Fig.~\ref{fig:sfr_v_delta}, we will take a simplifying approximation, which will allow us to make progress analytically. 
We will posit that
\be
\sfrd(z | \delta_R) \approx \overline{\sfrd}(z) e^{\gamma_R \tilde \delta_R}
\label{eq:sfrd_exponential}
\ee
for $\delta_R \ll \delta_{\rm crit}$, which occupies the entirety of the cosmic-dawn epoch.
For notational convenience we have defined 
\be
\tilde \delta_R = \delta_R - \gamma_R \sigma_R^2/2,
\label{eq:tilde_deltaR}
\ee
which is renormalized to recover $\VEV{\exp (\gamma_R \tilde \delta_R)} = 1$ for any value of $\gamma_R$ (\citealt{Xavier:2016elr}, see also our Appendix~\ref{app:normalization} for details).
This approximation works very well\footnote{This is not surprising, since for highly biased objects the exponential term of the EPS correction dominates, in which case the contribution from each mass is multiplied by an exponent $\propto \tilde \delta_c^2$ - $\delta_{\rm crit}^2$, which is linear in $\delta_R$ to first order.} in Fig.~\ref{fig:sfr_v_delta} for $R_G=3$ Mpc, and becomes increasingly accurate for larger $R$, as there we will have $\delta_R \sim \sigma_R \ll \delta_{\rm crit}$. At very small $\delta_R$ (or large $R$) it recovers the usual linear bias, as originally used in BLPF.
As such, our lognormal effective model allows us to analytically compute the SFRD fluctuations accurately including nonlinearities.

To sharpen our intuition, and to test that the fit in Fig.~\ref{fig:sfr_v_delta} is a representative test case for cosmic dawn, we show in Fig~\ref{fig:slice_sfrd} a simulation slice of the SFRD at $z=15$ given a linear density field, along with the result from our lognormal approximation.
The two slices are indistinguishable by eye.
Both the full SFRD and our approximation from Eq.~\eqref{eq:sfrd_exponential} predict that the densest regions will dominate the SFRD.
We also plot the ratio of the two predictions, which even at the small scales shown ($R_G=3$ Mpc)  deviates by less than 10\%. Note that the ratio skews smaller than one, as we have not normalized those SFRD boxes.
As we will see in Sec.~\ref{sec:compare21cmfast} when comparing with \cmfast, these differences shrink to the percent level for $R \gtrsim 10$ Mpc, which are the scales most readily observable by 21-cm interferometers.

\begin{figure}
	\centering
	\includegraphics[width=0.46\textwidth]{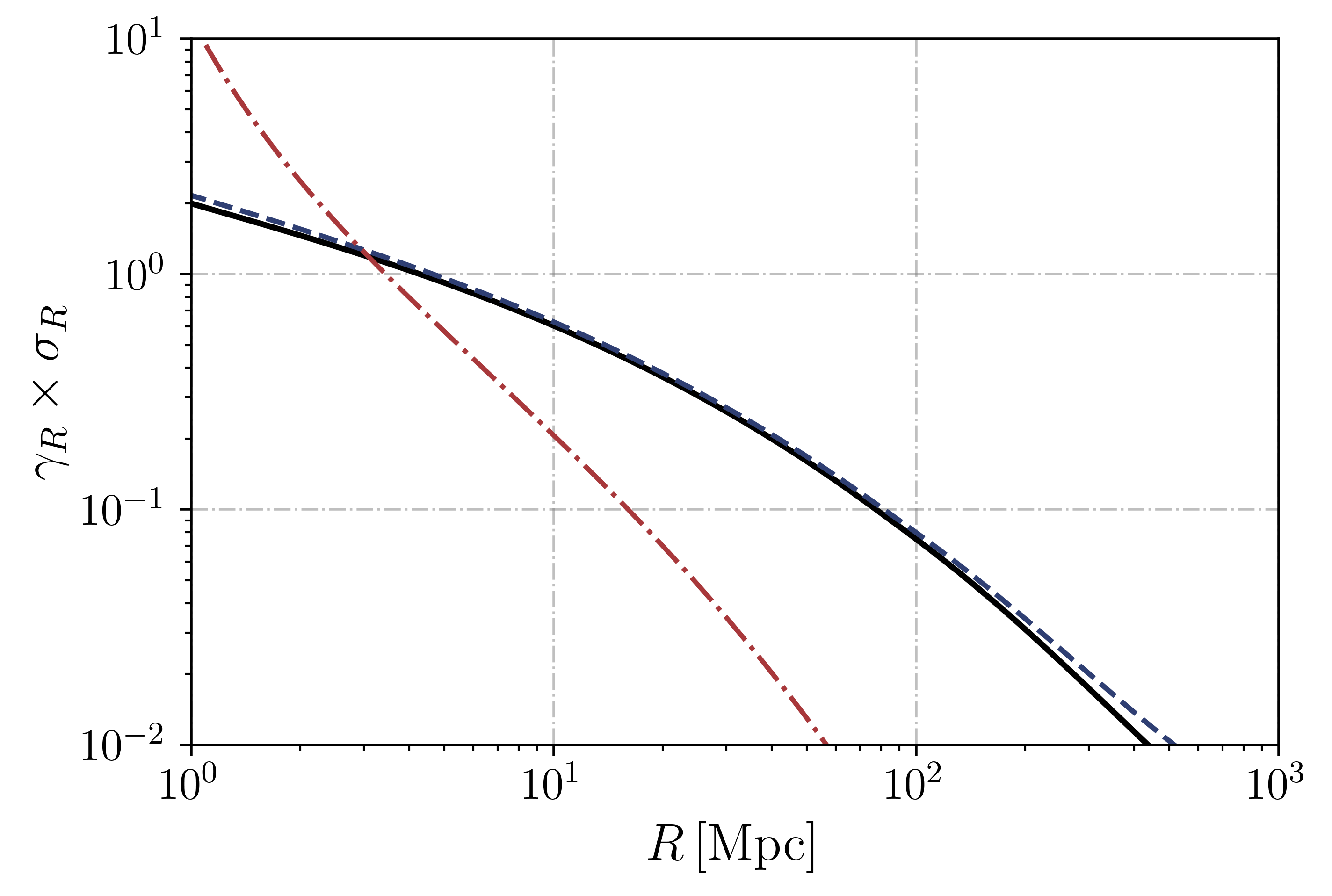}
	\caption{A prediction of our effective bias parameter $\gamma_R$ as a function of the smoothing scale $R$. We multiply it by $\sigma_R$ as a proxy for the 
	$\delta_R$ factor that it always accompanies.
	We show results at $z=10$ (solid) and 20 (dashed), though they are largely overlapping for our chosen parameters.
	We also show as a red dash-dotted curve an estimate of the correction due to non-linearities, defined as $(e^{x_R^2}-1)/x_R^2$ for $x_R=\gamma_R \sigma_R$.
	We thus expect our lognormal model to predict $\sim 10-100\%$ larger power than the linear prediction for $R\lesssim 20$ Mpc, which will increase the $k\gtrsim 0.1\Mpcinv$ power spectrum of different quantities as we will show in Sec.~\ref{sec:cosmicdawn}.
	}
	\label{fig:gamma_vs_R}
\end{figure}

This exponential approximation will allow us to analytically calculate correlation functions of the SFRD given those of $\delta_R$, which are a well known cosmological output.
As such, all the nonlinearities are encoded into the {\it effective biases} $\gamma_R$.
In practice, \codename\ numerically calculates the SFRD$(z|\delta_R)$ at each $z$ and $R$ using the formalism outlined in this Section, and simply fit for the $\gamma_R$ parameters on the fly.
These $\gamma_R$ coefficients form the base parameters of our effective model of the cosmic dawn.
We show a prediction for $\gamma_R$ as a function of $R$ in Fig.~\ref{fig:gamma_vs_R}, both at $z=10$ and 20.
We plot $x_R = \gamma_R \times \sigma_R$ to include the relevant scale of the $\delta_R$ term that multiplies the effective bias $\gamma_R$.
Small values of this product indicate linear-like behavior (as $e^{x_R}\sim 1 + x_R$ for $x_R\ll 1$).
This will only be the case for large $R$, whereas the lower $R$ will have nonlinear corrections, also estimated in Fig.~\ref{fig:gamma_vs_R}.
These corrections reach  $\mathcal O(1)$ for $R \sim 10$ Mpc.
As we will explore in Sec.~\ref{sec:cosmicdawn}, this will give rise to a factor of two larger power spectra of relevant quantities at scales $k=0.1-1\,\Mpcinv$.

While here the effective biases $\gamma_R$ are calculated under a specific model for the SFR (Eq.~\ref{eq:SFR_fiducial}) and assuming the HMF is modulated by the EPS formalism (Eq.~\ref{eq:HMF_EPS}), these biases could be used as free parameters and fit to high-$z$ clustering data, simulations, or directly to the 21-cm signal, bypassing the SFR modeling.

\subsection{Correlation functions}

The benefit of a lognormal approximation for the SFRD is that we can analytically find its correlation function.
In real space, the 2-pt function of two lognormal variables is~\citep{Coles:1991if,Xavier:2016elr}
\be
\VEV {e^{\gamma_{R_1} \tilde \delta_{R_1}} e^{\gamma_{R_2} \tilde \delta_{R_2}} } = e^{\gamma_{R_1}  \gamma_{R_2} \xi^{R_1,R_2}}  - 1,
\ee
where $\xi^{R_1,R_2} $is the real-space correlation function of the density field smoothed over radii $R_1$ and $R_2$, i.e., 
\be
\xi^{R_1,R_2} (r) = {\rm FT} \left[ P_m (k) W_{R_1}(k) W_{R_2}(k) \right],
\label{eq:corrfunc_with_windows}
\ee
where FT denotes Fourier transform, $P_m$ is the matter power spectrum, and $W_{R}(k)$ is a window function of radius $R$. 
The EPS formalism is usually calibrated with a (3D) tophat:
\be
W_R(k) = 3\,\left[\cos(k R) - (k R) \sin (k R) \right]/(k R)^3.
\ee
For computational convenience we will assume linear growth in the correlation functions, so  $\xi^{R_1,R_2}$ always refers to the $z=0$ result (and the biases $\gamma_R$ will be multiplied by the corresponding growth factor).

\begin{figure}
	\centering
	\includegraphics[width=0.46\textwidth]{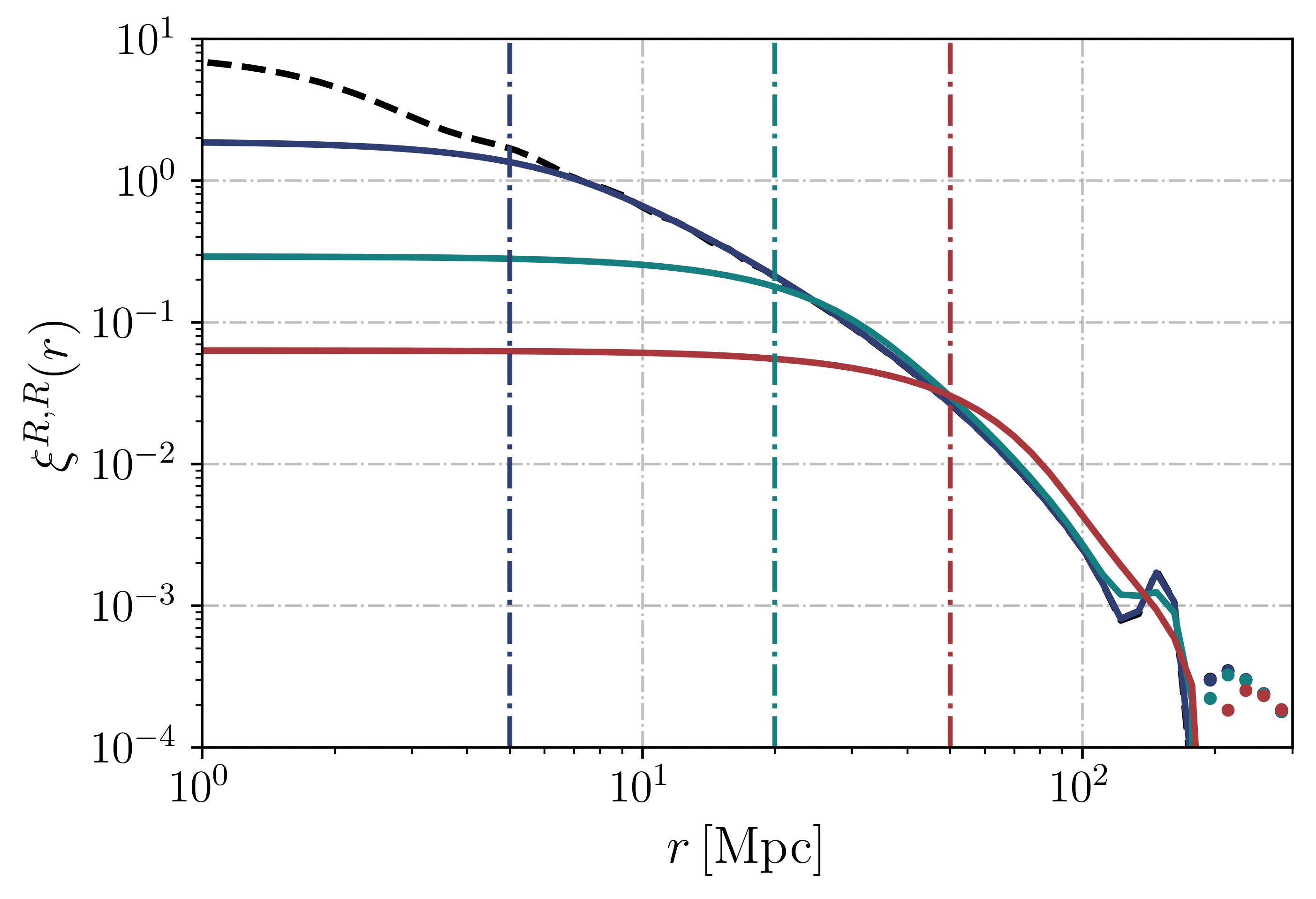}
	\caption{Correlation function of (matter) density fluctuations as a function of separation $r$, linearly extrapolated to $z=0$. In black dashed we show the standard (unwindowed) result and the three colors represent different smoothing scales relevant to cosmic-dawn observations ($R=5, 20, 50$ Mpc), also shown as vertical lines. The correlation function becomes negative for large $r$, beyond the sound horizon, shown as dotted points.
	}
	\label{fig:corrFunc_dens_R}
\end{figure}

In order to build intuition, we show the diagonal elements (i.e., $R_1=R_2=R$ for four values of $R$) of the $\xi^{R_1,R_2}$ matrix in Fig.~\ref{fig:corrFunc_dens_R} as a function of separation $r$, all linearly extrapolated to $z=0$ using {\tt CLASS} and {\tt mcfit}\footnote{\url{https://github.com/eelregit/mcfit}. Note that \codename\ does this transformation on the fly for each cosmology.}.
Larger smothing scales $R$ suppress the peak of the correlation function at low separations $r$. For reference, $\xi^{R,R}(r=0) \equiv \sigma_R^2$, showing that scales with $R\lesssim 10$ Mpc are expected to be nonlinear at $z=0$  (though less so during cosmic dawn owing to the smaller growth factor).
In practice, we expect our exponential model for the SFRD to not hold for $\delta_R \sim \delta_{\rm crit}$, so we will not consider correlation functions with $\sigma_R\gtrsim 1$, or $R\lesssim 2$ Mpc.

While EPS is calibrated with a 3D window function, in the model of BLPF the fluxes ($J_{\alpha/X}$) are obtained with a 1D window function, $W_R^{(\rm 1D)}(k) = \sin(k R)/(k R)$ (see also~\citealt{Dalal:2010yt}).
We will follow their model and compute the linear part of the power spectra with a 1D window, and simply add the nonlinear corrections on top (computed with nonlinear EPS, and thus a 3D tophat).
This allows us to keep the success of the linear BLPF model, to which we add the nonlinear corrections calibrated from the EPS formalism.
The semi-numeric model of \cmfast\ assumes a 3D window throughout, which we have also implemented in \codename, as we will show in Sec.~\ref{sec:compare21cmfast}.

\section{The IGM during Cosmic Dawn}
\label{sec:cosmicdawn}

So far we have discussed the SFRD and our effective model for it.
Physically, the 21-cm signal is sensitive to the thermal and excitation state of the IGM, and thus to the radiative fields.
The X-ray and Lyman-$\alpha$ fluxes at a point $\mathbf x$ are built from integrating the SFRD over $R$, the past lightcone, as outlined schematically in Eq.~\eqref{eq:schematic_JaX}.
What we have to determine are the coefficients $c_{\alpha/X}$ that act as weights of the contribution from each $R$.
While the SFRD depends on cosmology (through the HMF) and the halo-galaxy connection (through $\dot M_*$), these coefficients will also depend on the stellar properties of the first sources of light.
In particular, how much each radius contributes to the Lyman-$\alpha$ and X-ray flux is determined by photon propagation, and thus by the spectral energy distribution (SED) of the first galaxies.
We now compute these weights, and show in more technical detail how we find the 21-cm signal.
In both cases we will start calculating the average (or global) fluxes before finding their fluctuations.

\subsection{WF coupling}

We begin with the process that likely started first: the Wouthuysen-Field (WF) coupling of the gas and spin temperatures.
This coupling depends on the local flux of Lyman-$\alpha$ photons, which we can compute as~\cite{BarkanaLoeb2005}
\be
J_\alpha(\mathbf x, z) = \dfrac{(1+z)^2}{4\pi} \int_0^\infty dR \, {\rm SFRD}(\mathbf x, R) \epsilon_\alpha^{\rm tot}(\nu')
\label{eq:JaofR}
\ee
given the SFRD, where $\nu' =  \nu\, [1+z'(R)]/(1+z)$ is the redshifted frequency~\footnote{We have chosen to use the same symbol $\nu$ for spectral frequency and normalized density in Eq.~\eqref{eq:HMF_ST}, as both are deeply ingrained symbols in cosmology, and confusion is unlikely to arise.},
and compared to previous literature we have phrased the integral in terms of the comoving radius $R$ over which photons contribute, rather than their corresponding redshift\footnote{We define $z'(R)$ at the edge of a shell of radius $R$ by default in \codename. When comparing to \cmfast, however, we will follow their prescription and take it halfway to the previous shell.} $z'(R)$. The reason will become apparent when we compute fluctuations below.
Comparing Eqs.~\eqref{eq:JaofR} and~\eqref{eq:schematic_JaX} we see that the weights $c_{\alpha}(R)$ that determine the contribution of the SFRD at each $R$ depend on how far the photons, and thus on the SED $\epsilon_\alpha^{\rm tot}$ of the first galaxies (in the Lyman-$\alpha$ to continuum regime, as lower-energy photons cannot excite hydrogen, and higher-energy photons get absorbed locally through ionizations).

We have defined the ``total" SED as a sum over possible transitions that eventually redshift into the Lyman-$\alpha$ line,
\be
\epsilon_\alpha^{\rm tot} (\nu') \equiv \epsilon_\alpha(\nu') \sum_{n=2}^{n_{\rm max}} f_{\rm rec}(n)   w_\alpha(n),
\ee
where $f_{\rm rec}(n)$ are the recycled fractions from~\citet{Pritchard:2005an}, and $w_\alpha(n)$ are weights equal to unity for $z<z_{\rm max}(n)$, and zero above\footnote{In order to reduce the impact of ``kinks" whenever one of these shells turns on we account for a last partial shell with a weight $w_n\propto [z - z_{\rm max}(n)]$.}, with $[1 + z_{\rm max}(n)]/(1+z) = [1 - (1+n)^{-2}]/(1 - n^{-2})$.
As an intrisic spectrum we simply take two power laws as
\be
\epsilon_\alpha (\nu) = \dfrac{N_\alpha}{\mu_b} \mathcal A_i \left(\dfrac{\nu}{\nu_\beta}\right)^{\alpha_i},
\ee
for frequencies between Lyman-$\alpha$ and the Lyman limit, with a break at Lyman-$\beta$ and power-law indices $\alpha_i = \{0.14,-8.0\}$ below and above that break.
We set the amplitudes $\mathcal A_i$ so that 68\% of the flux is between Lyman-$\alpha$ and $\beta$, and 32\% above it, and take a total number $N_\alpha=9690$~\citep{BarkanaLoeb2005} of photons emitted in this band per star-forming baryon (as $\mu_b$ is the mean baryonic mass).
This can be easily modified by reading more complicated stellar spectra both for PopII and III stars (e.g.~\citealt{Bromm:2003vv}), which we defer to future work.

\begin{figure}
	\centering
	\includegraphics[width=0.46\textwidth]{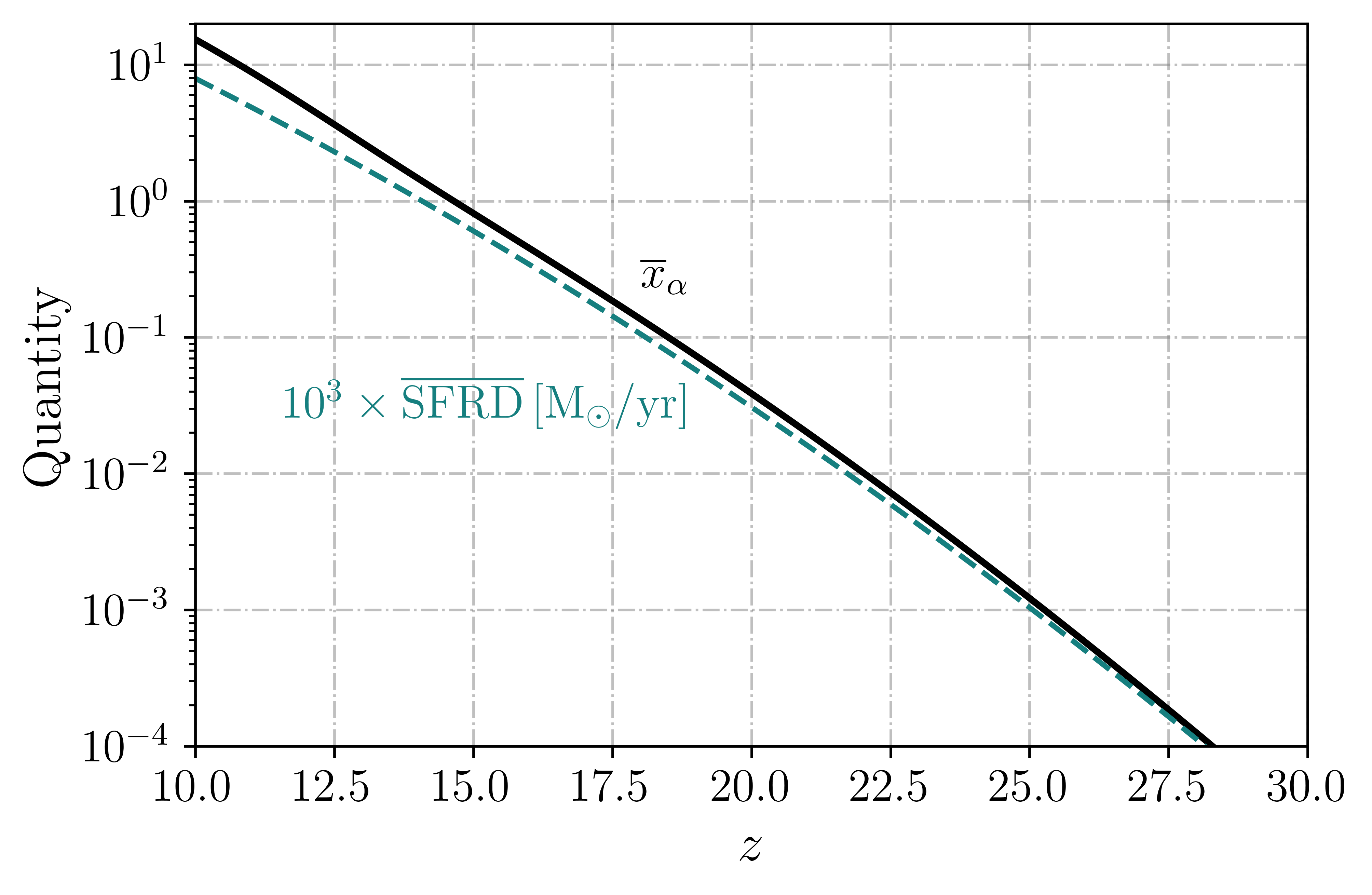}
	\caption{Evolution of the average SFRD (dashed, multiplied by $10^3$) and Lyman-$\alpha$ coupling parameter $\xa$ (solid, defined in Eq.~\ref{eq:xadef}) as a function of $z$.
	In order to have 21-cm absorption due to efficient WF coupling one needs $\xa\gtrsim 0.1$, which for our chosen parameters occurs for $z\lesssim 20$.
	}
	\label{fig:xalpha_and_SFRD}
\end{figure}

As advanced in Sec.~\ref{sec:basics}, we will compute the spin temperature $T_S$ in the IGM in terms of the (dimensionless) coupling coefficient~\citep{Furlanetto:2006jb}
\be
x_\alpha = S_\alpha \dfrac{J_\alpha}{J_\alpha^c},
\label{eq:xadef}
\ee
with $1/J_\alpha^c = 1.811 \times 10^{11}/(1+z)\, (2.725\,{\rm K}/T_{\rm CMB}^{(0)})$ cm$^2$ s Hz sr is a numerical constant (for which we have set the CMB temperature today $T_{\rm CMB}^{(0)}$ to the {\it Planck} 2018 value).
We find the correction factor $S_\alpha$ iteratively as detailed in~\citet{Hirata:2005mz}.
This factor, which reduces the expected WF coupling by $\sim 10-20\%$ during cosmic dawn, requires finding the kinetic temperature $T_k$ and free-electron fraction $x_e$ as well, which we will detail in the next subsection. 
Note that for now we only use the average correction for $S_\alpha$ in \codename, though in principle this quantity can be perturbed in our formalism too.

We show the evolution of $x_\alpha$ in Fig.~\ref{fig:xalpha_and_SFRD}, along with our average SFRD (rescaled).
Clearly these two variables trace each other, so in principle a clean measurement of $\xa$ can be invaluable to find the SFRD at high redshifts (cf.~\citealt{Madau:1996yh}).
The 21-cm line allows us to infer $\xa$, as the gas kinetic and spin temperatures will be coupled when $J_\alpha \sim J_\alpha^c$, i.e., when $x_\alpha \sim 1$.
This requires modeling the rest of ingredients in the 21-cm signal, which we will do in turn.

\subsubsection*{Fluctuations}

The integral form of Eq.~\eqref{eq:JaofR} immediately makes it clear how to connect the fluctuations of the SFRD, which we computed above, to those of $\xa$ in the IGM.
By converting the integral over $R$ into a sum we can write
\be
\xa (z) =  c_{1,\alpha} (z) \sum_R \,c_{2,\alpha} (z, R) e^{\gamma_R \tilde \delta_R},
\label{eq:xaflucts}
\ee
where 
\be
c_{1,\alpha} = \dfrac{(1+z)^2}{4\pi} \dfrac{S_\alpha}{J_\alpha^c}
\ee
is an $R$-independent coefficient, and 
\be
c_{2,\alpha} = \overline{\rm SFRD}\, \epsilon_\alpha^{\rm tot}\, \Delta R
\ee
includes all the $R$-dependent factors in Eq.~\eqref{eq:JaofR}, as well as the step\footnote{We keep the $\Delta R$ explicit to use sums rather than integrals, which keeps the notation and computations tidy. We also remind the reader that $\overline{\rm SFRD}$ is evaluated at the redshift $z'(R)$ relevant for each $z$ and $R$.} $\Delta R$.
Finally, $\gamma_R$ are the exponents of the SFRD against $\delta_R$ that we found in Sec.~\ref{sec:effective_model}.

With this simple formula we can compute the (auto)correlation function of $\xa$ as
\ba
\xi_{\alpha}(z) =& \, c_{1,\alpha}^2 (z) \sum_{R_1,R_2} [c_{2,\alpha} (z, R_1) c_{2,\alpha} (z, R_2)] \nonumber \\ &\times \left ( \exp \left [ \gamma_{R_1} \gamma_{R_2} \xi^{R_1,R_2} \right] - 1 \right),
\label{eq:alpha_corrf}
\end{align}
by taking advantage of our lognormal building blocks.
This expression may seem daunting to evaluate, given the double sum.
However, all the coefficients have been stored when computing the global evolution.
As such, for standard precision in \codename\ it takes $\approx 1$ s to do all the correlation-function sums down to $z=5$ (compared to $\approx 3$ s for running the entire 21-cm and SFRD evolution, or $\approx 5$ s for running the {\tt CLASS} cosmology to the necessary high $k\approx 500\,\Mpcinv$.).

\begin{figure}
	\centering
	\includegraphics[width=0.46\textwidth]{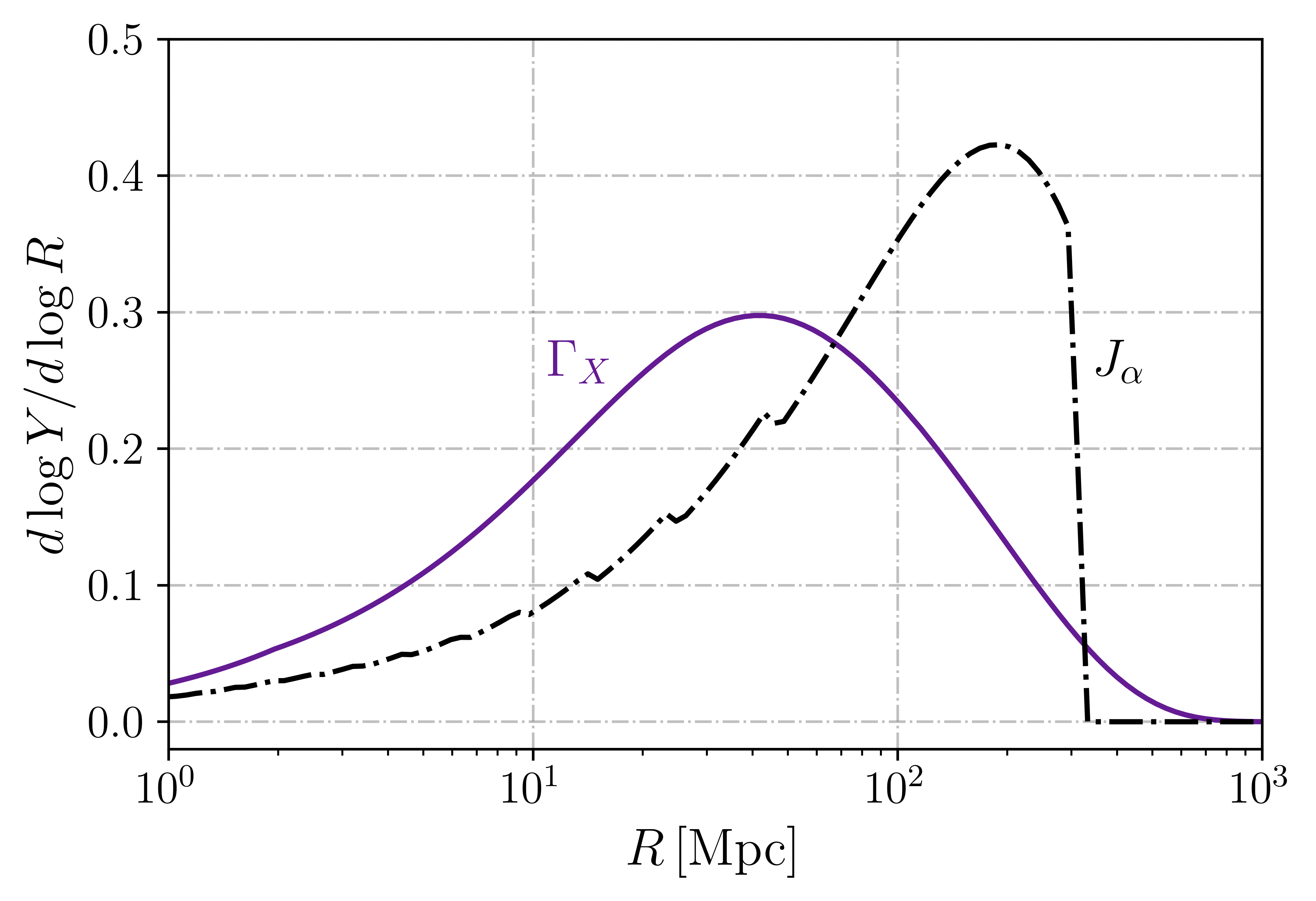}
	\caption{Logarithmic derivative of the integrand in the Lyman-$\alpha$ flux ($Y=J_\alpha$ in black dashed, Eq.~\ref{eq:JaofR}) and X-ray heating term ($Y=\Gamma_X$ in purple, obtained integrating the X-ray flux $J_X$ over energies as in Eq.~\ref{eq:GammaX}) versus comoving radius $R$.
	The Lyman-$\alpha$ flux receives its largest contributions from large radii $R$, up to the maximum distance for photons to redshift from Lyman-$\beta$ into Lyman-$\alpha$.
	The X-rays tend to come from closer distances, though they have a tail at large $R$.
	Both curves depend sensitively on our choice of UV and  X-ray SEDs.
	The kinks in the $J_\alpha$ curve appear whenever an atomic transition crosses one of the radii we sum over (and could be eliminated with interpolation as in \cmfast).
	}
	\label{fig:coeffs2_R}
\end{figure}

The coefficients $c_{2,\alpha}$ capture the nonlocality of the Lyman-$\alpha$ flux.
To illustrate their behavior, we show the logarithmic derivative of $J_\alpha$ with respect to radius $R$ in Fig.~\ref{fig:coeffs2_R} (which is proportional to $c_{2,\alpha}$), at $z=15$.
One can think of this quantity as the Fourier transform of a window function, as each point carries the weight of modes at that radius $R$.
We see that the weight is spread broadly, growing towards larger radii until $R \sim 350$ Mpc, where it drops.
This corresponds to the comoving distance from $z=15$ to $z_{\rm max}=17.9$, over which photons from Lyman-$\beta$ can redshift into Lyman-$\alpha$.
We also show in Fig.~\ref{fig:coeffs2_R} the same quantity for X-rays, which we will describe in the next subsection.

\begin{figure}
	\centering
	\includegraphics[width=0.46\textwidth]{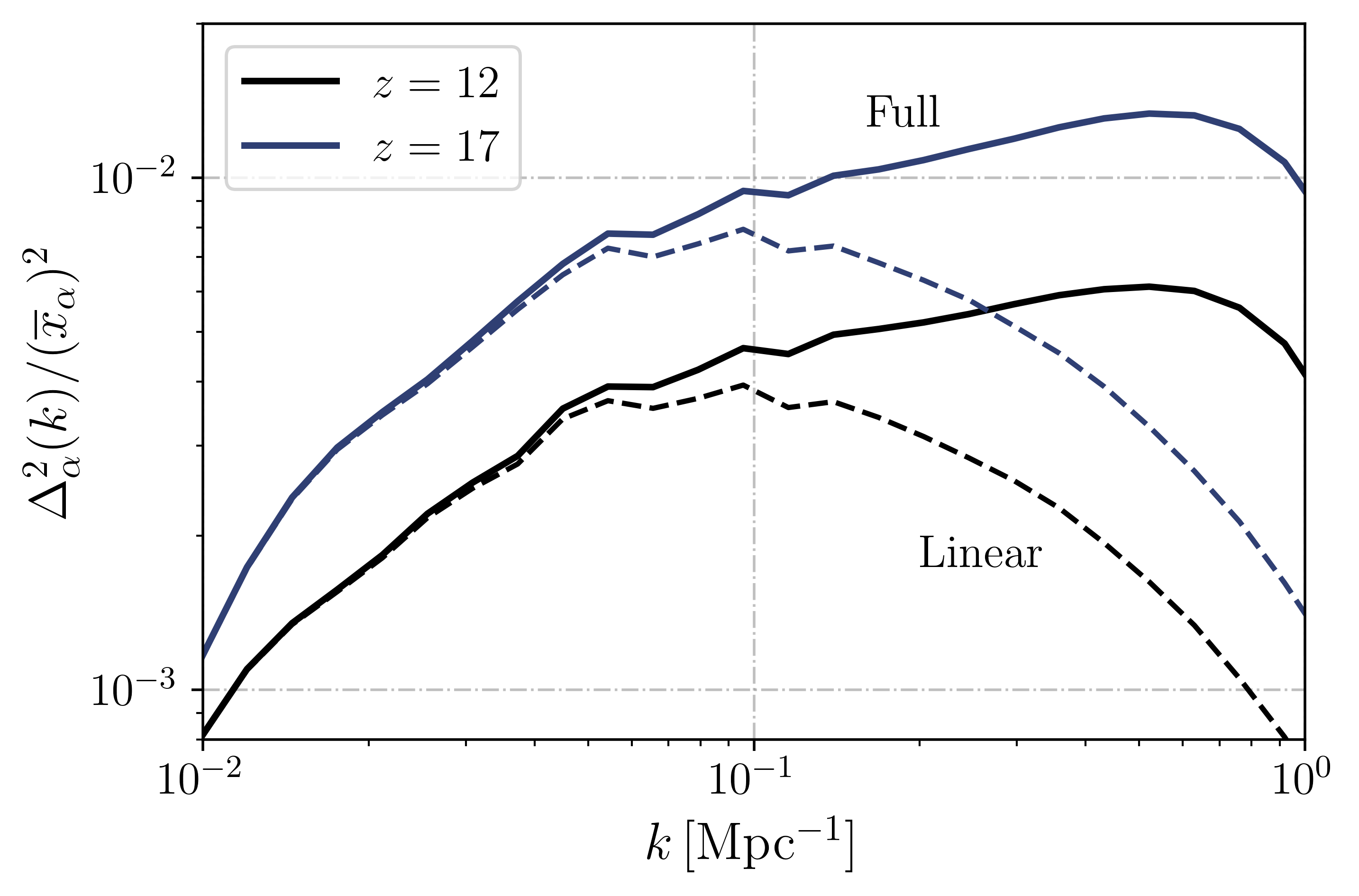}
	\caption{Power spectrum of the WF coupling parameter $\xa$ (as a tracer of the Lyman-$\alpha$ flux $J_\alpha$).
	Solid lines show our calculation with \codename, which includes nonlinearities; versus the linear estimate in dashed, which severely underpredicts the power at $k\gtrsim 0.1$ Mpc$^{-1}$, the scales most relevant for 21-cm experiments.
	}
	\label{fig:Deltasq_xa}
\end{figure}

We now Fourier transform the $\xi_\alpha$ correlation function to obtain the power spectrum of $\xa$ fluctuations, which we show for our fiducial parameters in Fig.~\ref{fig:Deltasq_xa}.
The fluctuations are rather large, reaching $\Delta_{\alpha}^2\sim 10^{-2} \, \overline{x}_\alpha$ at small scales for $z=12$.
We compare our result with a linear calculation as originally proposed in BLPF, though adapted to our astrophysical model (which has a non-constant $f_*(M_h)$).
For low wavenumbers $k\lesssim 0.1$ Mpc$^{-1}$ the linear and full calculations agree fairly well.
However, they deviate by $\mathcal O(1)$ in the $k = 0.1-1$ Mpc$^{-1}$ range, where the nonlinear calculation predicts significantly more power.
This is precisely the range of scales that is relevant for 21-cm observations, highlighting the need for a nonlinear approach for cosmic dawn.

One may wonder why the nonlinear calculation provides additional power (rather than decrease it) in Fig.~\ref{fig:Deltasq_xa}.
While we will compare against \cmfast\ simulations in Sec.~\ref{sec:compare21cmfast}, which show the same trend (see also~\citealt{Santos:2007dn}), let us also provide an intuitive argument here.
The SFRD grows faster than linear with $\delta$ (e.g., Fig.~\ref{fig:sfr_v_delta}).
As such, its correlation function will receive beyond-linear corrections, which tend to grow towards high $k$ where the $\delta$ fluctuations are larger.
Another way to see this is through the SFRD slices in Fig.~\ref{fig:slice_sfrd}.
These show significant small-scale fluctuations in the SFRD, which are less pronounced in the densities themselves, corresponding to more high-$k$ power in the former than the latter. 
Our lognormal model successfully predicts this behavior.

Finally, we note that the impact of nonlinearities depends on the stellar parameters (through the $c_{i}$ coefficients) and cosmology+halo-galaxy connection (through the exponents $\gamma_R$).
Yet, the SFRD will remain as the building block, and our lognormal approximation will allow us to calculate its power spectrum quickly and accurately.

\subsection{X-rays and the Temperature of the IGM }

We now turn our attention to the second component of the 21-cm signal during cosmic dawn: the gas kinetic temperature $T_k$.

During cosmic dawn $T_k$ is determined by the competition between adiabatic cooling (due to cosmic expansion), and heating.
The latter is due to both coupling to the CMB (which is however fairly inefficient after $z\sim 150$) and X-rays from the first galaxies.
We evolve the temperature through
\be
\dfrac{3}{2} T_k' = T_k \dfrac{d \log \rho}{dz} - \dfrac{\Gamma_C}{H(z) (1+z)}  - \dfrac{\Gamma_X}{H(z) (1+z)},
\label{eq:Tk_total_ODE}
\ee
where $\Gamma_C$ is the standard Compton heating rate to the CMB~\citep{Ma:1995ey}, and $\Gamma_X$ is the X-ray term, which will be given by the X-ray flux $J_X$ integrated over frequencies as we will detail below.
We can divide the temperature evolution in two parts (see a similar discussion in~\citealt{Schneider:2020xmf})
\be
T_k =  T_{\rm cosmo} + T_X,
\ee
where the ``cosmology" term $T_{\rm cosmo}$ takes into account the Compton coupling
\be
\dfrac{3}{2} T_{\rm cosmo}' = T_{\rm cosmo} \dfrac{d \log \rho}{dz} - \dfrac{\Gamma_C}{H(z) (1+z)}, 
\label{eq:Tk_adiabatic}
\ee
and is obtained self-consistently using {\tt CLASS}, whereas the X-ray term follows
\ba
\dfrac{3}{2} T_X' &= T_X \left(\dfrac{3}{1+z}\right) - \dfrac{\Gamma_X}{H(z) (1+z)}, \label{eq:Txprime} 
\end{align}
where we have taken the approximation that $\rho \propto (1+z)^3$ regardless of $\delta$ (which we will however lift when accounting for adiabatic fluctuations below).
Given the simple $z$ dependence of Eq.~\eqref{eq:Txprime} we can rewrite
\be
T_X(z) = (1+z)^2 \int_z^\infty dz' \dfrac{1}{(1+z')^2} \left [- \dfrac{2}{3} \dfrac{\Gamma_X}{H(z') (1+z')} \right]. 
\label{eq:Txintegral}
\ee
The next step is, then, to define $\Gamma_X$. 

\subsubsection*{The heating term and X-ray flux}

The X-ray heating rate at a point $\mathbf x$ (and redshift $z'$) is given by an integral over all frequencies of the X-ray flux $J_X$ at that point~\citep{Pritchard:2006sq},
\be
\Gamma_X(\mathbf x) = (4\pi) f_{\rm heat} \int d\nu J_X(\mathbf x, \nu) \sum_{i=\rm HI, HeI} (\nu - \nu_{\rm ion}^{(i)}) \sigma_i(\nu) f_i,
\label{eq:GammaX}
\ee
weighted by the energy injected per photon and the ionization cross sections $\sigma_i(\nu)$ for $i=$HI and HeI (as we ignore the small correction from HeII), which we take from~\citet{Verner:1996th}.
Here $f_i$ are the (number) fractions of both species, and  $\nu_{\rm ion}^{(i)}$ their ionization energies/frequencies.
We approximate the deposition efficiency by $f_{\rm heat} = \overline x_e^{0.225}$~\citep{Schneider:2020xmf}, which forces us to compute the average free-electron fraction $\overline x_e$.
For this we simply take the ansatz
\be
\overline {x}_e = \overline {x}_e^{\rm cosmo} + \overline {x}_e^X,
\ee
where $\overline x_e^{\rm cosmo}$ is the result from \class\ (using {\tt Hyrec}~\citealt{AliHaimoud:2010dx}), and the X-ray term we estimate by $\overline x_e^X = \int dz \Gamma_{X,\rm ion}$~\citep{Mirocha:2014faa}, for $\Gamma_{X,\rm ion} = (3/2) f_{\rm ion} \Gamma_X/\VEV{\nu_{\rm ion}}$, with $\nu_{\rm ion}$ the averaged ionization frequency of HI and HeI, and where we take the approximation $f_{\rm ion}=0.4 \exp(-\overline x_e/0.2)$ for the fraction of energy that goes into ionization, from \citet{Furlanetto:2009uf}.
This is a simplified treatment, but it will suffice for this first work.

The X-ray flux is given by
\be
J_X (\mathbf x, \nu) = \dfrac{(1+z)^2}{4\pi} \int_0^\infty dR \, {\rm SFRD}(\mathbf x, R) \epsilon_X(\nu') e^{-\tau_X},
\label{eq:JXofR}
\ee
similarly to Eq.~\eqref{eq:JaofR} for $J_\alpha$, but with an opacity term determined by the optical depth at each energy $\nu$~\citep{Pritchard:2006sq}
\be
\tau_X(\nu, z,R) = \int_z^{z(R)} dR \ n_b \sum_{i=\rm HI, HeI} f_i \sigma_i(\nu')
\ee
where as before $\nu'$ is the redshifted energy, and we will take the average $n_b$ here for computational simplicity (as done for instance in \cmfast).
Moreover, we will keep track of the $\tau_X$ numerically computed, though \codename\ has a flag to make the opacity $e^{-\tau_X}$ either 0 or 1 as done in \cmfast, for ease of comparison to their results in Sec.~\ref{sec:compare21cmfast}.

The heating rate in Eq.~\eqref{eq:GammaX} depends sensitively on the X-ray SED $\epsilon_X(\nu)$, as lower-energy photons travel shorter distances $R$ and are more likely to heat up the IGM.
We will assume an SED
\be
\epsilon_X(\nu) = L_{40} \times  \dfrac{10^{40}{\rm erg/s}}{(\Msun/\rm yr)} \dfrac{I_X(\nu)}{\nu},
\label{eq:LX_def}
\ee
where $I_X(\nu)$ is the X-ray spectrum, normalized to integrate to unity over the band considered, which we take to cover from $\nu_0=0.5$ keV up to $\nu_{\rm max}=2$ keV, as for energies above the mean free path is too large so they barely contribute to heating~\citep{Greig:2015qca}.
This allows us to define the luminosity $L_{40}$ (in units of $10^{40}$ erg/s/SFR) as a free parameter.
In practice we will use a power-law SED, $I_X(\nu) \propto \nu^\alpha_X$ with $\alpha_X=-1$, for $\nu> \nu_0$, which reasonably fits the spectra of high-mass X-ray binaries~\citep{Fragos:2013bfa}.
Both the power-law parameter $\alpha_X$ and the cutoff frequency $\nu_0$ are free parameters in \codename, and the spectrum $I_X$ can be enhanced for arbitrary SEDs.

Given all this, we can calculate the X-ray heating rate $\Gamma_X$.
We show the contributions to $\Gamma_X$ coming from different radii $R$ in Fig.~\ref{fig:coeffs2_R}, as we did for the Lyman-$\alpha$ flux.
The X-ray term has broader support over $R$, and it peaks at slightly lower $R$ than its Lyman-$\alpha$ counterpart ($R\sim 30$ Mpc, rather than $\sim 300$ Mpc). 
As such, the X-ray power will remain unsuppressed until smaller scales (larger $k$).
This is, however, very dependent on the X-ray SED, which determines how locally the IGM heating proceeds during cosmic dawn~\citep{Pritchard:2006sq,Pacucci:2014wwa,Fialkov:2014kta}.
We take a closer look at the effect of the X-ray SED in \codename, and compare it against that of Lyman-$\alpha$ photons in Appendix~\ref{app:Xrays}.

In this first work we will neglect several small corrections in the name of simplicity.
We will ignore fluctuations on the free-electron fraction $x_e$, which could affect the distribution of energy deposited.
We note, though, that $f_{\rm heat}$ is a shallow function of $x_e$, and even by the end of our simulations (at $z=10$ where heating is saturated) the free-electron fraction is still fairly small, $x_e\approx 10^{-3}$.
We also ignore the excitations from X-rays, which would produce a small amount of WF coupling, as well as energy transfer with Lyman-$\alpha$ photons~\citep{Venumadhav:2018uwn} which would heat the gas modestly.
These can be straightforwardly introduced into \codename.

\begin{figure}
	\centering
	\includegraphics[width=0.46\textwidth]{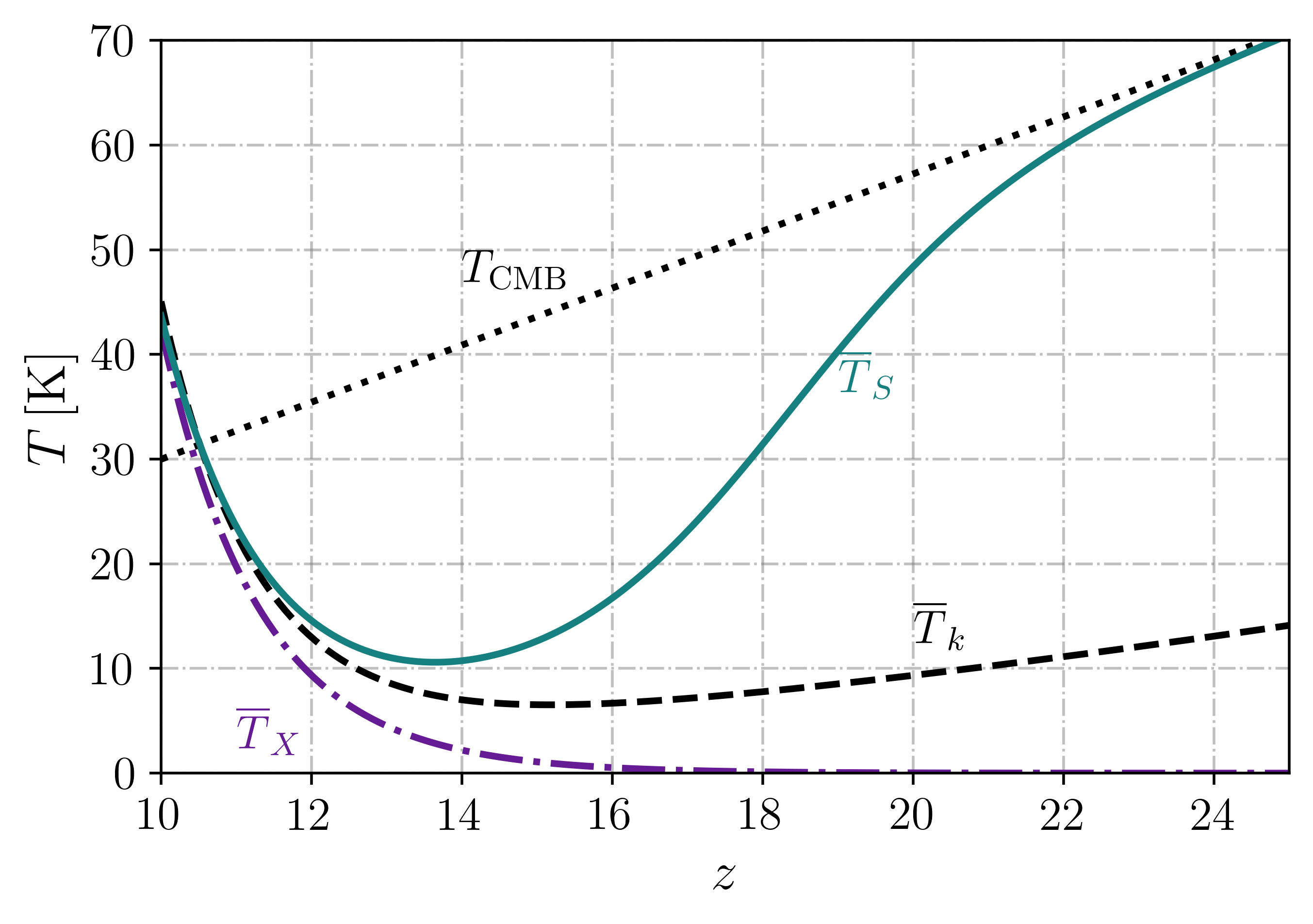}
	\caption{Evolution of the different relevant temperatures during cosmic dawn, all spatially averaged.
	The kinetic temperature $\overline T_k$ originally follows the adiabatic prediction at high $z$ (which includes Compton coupling to photons $\Tcmb$, through {\tt CLASS}), and \codename\ computes the X-ray heating $\overline T_X$ component, see Eq.~\eqref{eq:Tk_total_ODE}.
	The spin temperature $\overline T_S$ additionally depends on the WF coupling $\xa$, turning on when the first galaxies form at $z\sim 20$. There will be 21-cm signal whenever $T_S \neq \Tcmb$.
	}
	\label{fig:Temperatures}
\end{figure}

With all these caveats, we can finally compute the evolution of the gas temperature, and we show its spatial average in Fig.~\ref{fig:Temperatures}.
For our fiducial parameters the X-ray heating begins in earnest at $z\sim15$, and fully saturates ($\overline T_k\gg \Tcmb$) by the end of our calculation at $z\sim 10$).
We also show the resulting spin temperature $\overline T_S$ in Fig.~\ref{fig:Temperatures}, which accounts for the kinetic temperature as well as the WF coupling.
It is this quantity that determines the 21-cm temperature: if $\overline T_S < \Tcmb$ we will have absorption (which will occur at $z \sim 10-20$ for our parameters), and for $\overline T_S > \Tcmb$ emission.
Note that we are ignoring reionization for now, so we stop our calculations at $z\sim 10$, since below that $z$ the 21-cm signal will be saturated.

\subsubsection*{Fluctuations}

Let us now compute the fluctuations on the gas kinetic temperature.

We begin with the X-rays, for which we will mirror the formalism that we followed for $\xa$.
We re-write the X-ray temperature (i.e., the heating due to X-rays) as
\be
T_X (z) =  \sum_{z'\geq z} c_{1,X} (z') \sum_R c_{2,X} (z', R) e^{\gamma_R, \tilde \delta_R}
\label{eq:Txflucts}
\ee
where the $c_{i,X}$ coefficients are determined from the equations above, and $\gamma_R$ are the same SFRD effective biases as we had in Eq.~\eqref{eq:xaflucts}.
This equation has, however, an additional nonlocality in time than its $\xa$ counterpart. 
The kinetic temperature of a gas parcel at $\mathbf x$, $z$ depends on the heating rate $\Gamma_X$ at all previous times, so Eq.~\eqref{eq:Txflucts} is integerated over $z'$ (as a consequence $c_{1,X}$ will include a $\Delta z'$ step, much like the $\Delta R$ in $c_{2,\alpha/X}$).

The building block for the X-ray-heating term is still the SFRD, and thus we can compute its correlation functions in terms of the same $\gamma_R$ that we calculated in Sec.~\ref{sec:effective_model} to be
\ba
\xi_{X} =& \sum_{z_1',z_2'\geq z} c_{1,X} (z_1')  c_{1,X}(z_2') \sum_{R_1,R_2} \, c_{2,X}(z_1',R_1)  \nonumber \\ &\times \, c_{2,X}(z_2',R_2)  \left ( \exp \left [ \gamma_{R_1} \gamma_{R_2} \xi^{R_1,R_2} \right] - 1 \right),  \\
\xi_{\alpha X} =& c_{1,\alpha}(z) \sum_{z'\geq z} c_{1,X}(z') \sum_{R_1,R_2} \, c_{2,\alpha} (z, R_1) \,   \nonumber \\ &\times \, c_{2,X}(z_2',R_2) \left ( \exp \left [ \gamma_{R_1} \gamma_{R_2} \xi^{R_1,R_2} \right] - 1 \right).  
\end{align}
As was the case for $\xa$, these expressions appear computationally expensive, even more so given the $z'$ integrals involved (due to the non-locality in time of X-ray heating).
Nevertheless, only one $z'$ sum has to be carried over, as we cumulatively track $\xi_X(z)$ from high to low $z$, which highly speeds up the calculation.

\begin{figure}
	\centering
	\includegraphics[width=0.46\textwidth]{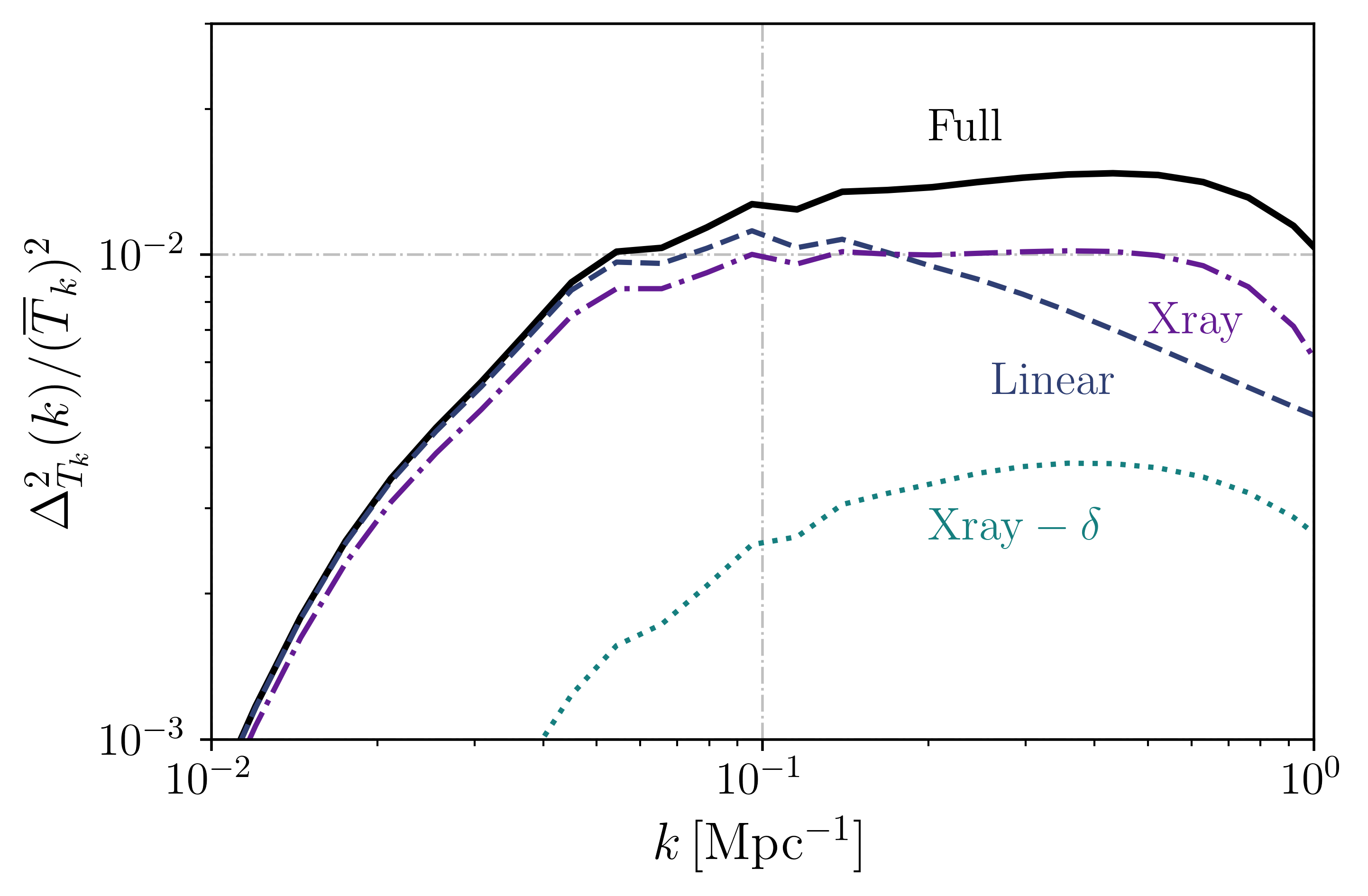}
	\caption{Same as Fig.~\ref{fig:Deltasq_xa} but for the kinetic temperature $T_k$.
	We only show the result at $z=12$, both with full nonlinear (black solid) and linear (blue dashed) fluctuations,  including X-rays and adiabatic.
	We also separate the X-ray term (purple dot-dashed, nonlinear) and its cross term with density (the $T_X-\delta$  adiabatic part, teal dotted) to showcase each contribution.
	}
	\label{fig:Deltasq_Tk}
\end{figure}

We show the power spectrum of X-ray fluctuations at $z=12$ in Fig.~\ref{fig:Deltasq_Tk}.
Notice that it is divided by the average kinetic temperature $\overline T_k$, rather than the X-ray only component $\overline T_X$.
As in Fig.~\ref{fig:Deltasq_xa}, the nonlinearities in the SFRD change the power spectrum significantly for $k\gtrsim 0.1\,\Mpcinv$.
Fig.~\ref{fig:Deltasq_Tk} also shows the effect of anisotropic adiabatic cooling, which is important at early times. Let us now describe it.

\subsubsection*{Adiabatic Fluctuations}

The adiabatic cooling of the IGM determines its kinetic temperature prior to the X-ray epoch.
The adiabatic cooling rate depends on the local matter density $\delta$, and as such $T_k$ has fluctuations sourced by it.
Looking at Eq.~\eqref{eq:Tk_adiabatic}, one can see that if we ignored the coupling to the CMB (i.e., if we set $\Gamma_C=0$), the adiabatic temperature would follow $T_k \propto (1+\delta)^{c_T}$, with $c_T=2/3$ the adiabatic index.
This well-known result is, however, complicated by the non-trivial thermal evolution of gas.
Electrons keep scattering off of the photon bath after recombination, so their thermal evolution (and thus that of the IGM) retains some memory of the photon temperature down to the cosmic-dawn epoch.
Rather than assume a value of $c_T$, we follow \citet{Munoz:2015eqa} and calculate the adiabatic index by solving for the evolution of $T_k$ with $z$ including the Compton coupling term $\Gamma_C$.
This calculation will turn out to uncover a missing element in the standard setup of \cmfast, so we now take a brief detour to describe it.

\begin{figure}
	\centering
	\includegraphics[width=0.46\textwidth]{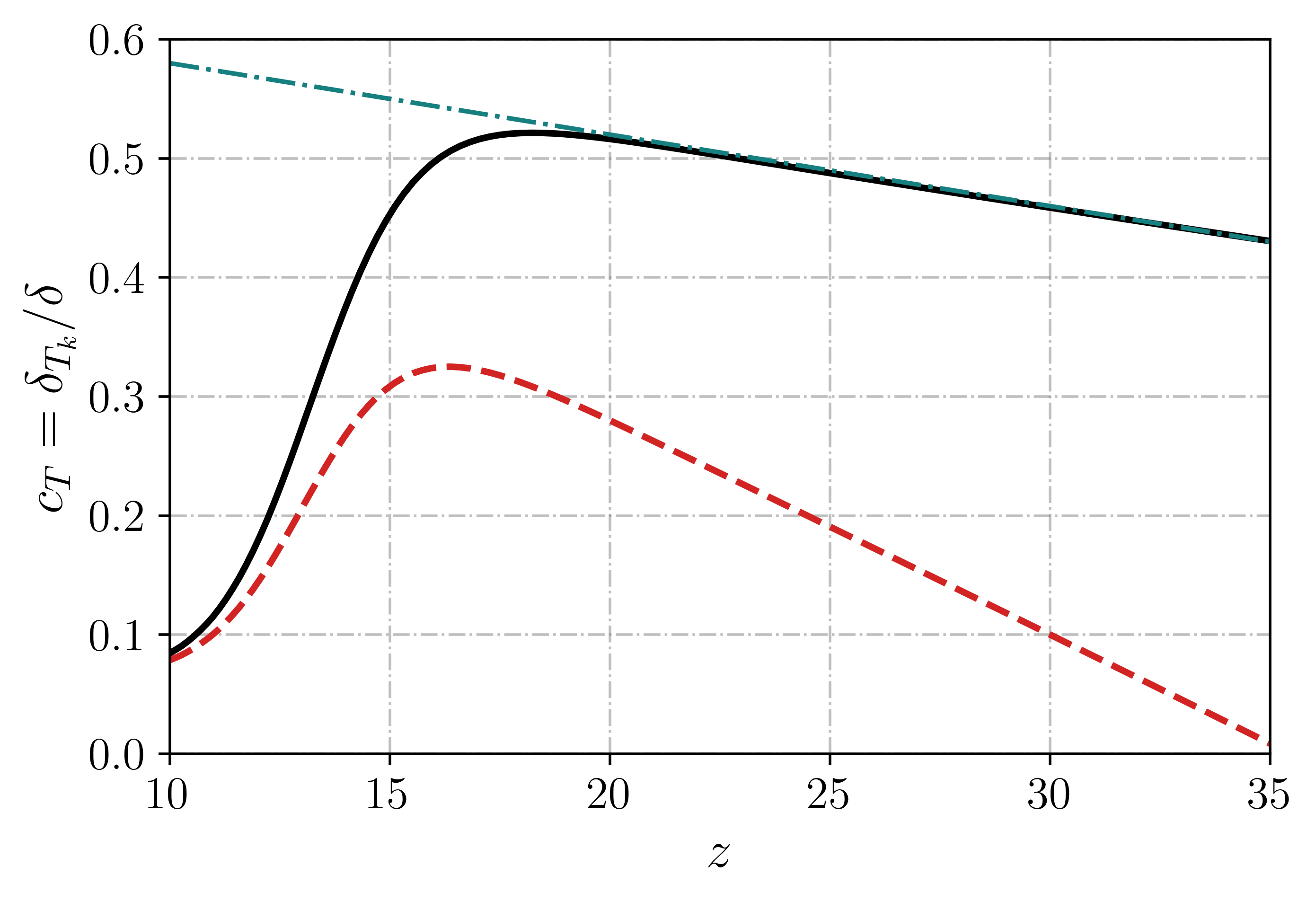}
	\caption{Evolution of the adiabatic index $c_T$, as defined in Eq.~\eqref{eq:cT_adiabatic}, through cosmic time.
	Our prediction in black accounts for the Compton coupling of the IGM to the CMB, which lowers $c_T$ from its idealized value of 2/3. We also show the fit from \citet{Munoz:2015eqa} in teal dot-dashed (reproduced here in Eq.~\ref{eq:cT_fit}), which ignores X-ray heating; and the default \cmfast\, assumption in red dashed, which fixes $T_k$ to be homogeneous at $z=35$ and underpredicts adiabatic fluctuations during cosmic dawn.
	}
	\label{fig:cT_adiabatic}
\end{figure}

Assuming a global kinetic temperature $\overline{T}_k$, we can find its adiabatic-cooling fluctuations to linear order in $\delta$ by expanding\footnote{Note that here we do not separate $T_k$ into ``cosmology" and X-ray terms.} Eq.~\eqref{eq:Tk_total_ODE},
\be
\delta T_k^{\rm (ad)}(z) = \dfrac{2}{3} (1+z)^2 \int_z^\infty \dfrac{dz'}{(1+z')^2} \overline{T}_k(z') \delta'(z') ,
\ee
mirroring Eq.~\eqref{eq:Txintegral}, where as before primes denote derivatives with respect to redshift.
We define the adiabatic index through the linearized relation
\be
\delta T_k^{\rm (ad)}(k,z) = c_T(z) \delta(k,z)\,\overline{T}_k(z).
\label{eq:cT_adiabatic}
\ee
In that case, we can find the index to be
\be
c_T(z) = \dfrac{2}{3} \dfrac{(1+z)^2}{D(z)\,\overline{T}_k(z)} \int_z^\infty \dfrac{dz'}{(1+z')^2} \overline{T}_k(z') D'(z') ,
\label{eq:cT_def}
\ee
assuming that $D(z)$ is the (scale-independent) growth factor.
One can see that by setting $T_k = T_0 (1+z)^2$, and $D(z)=D_0/(1+z)$ at all $z$ we would recover $c_T=2/3$. 
Even in the most standard cosmology, $\overline T_k$ does not follow that exact scaling, which changes the value of $c_T$.
We use the background thermal evolution from {\tt CLASS/HyREC} (which accurately includes the electron scattering in $\Gamma_C$ after recombination),  to find the adiabatic index $c_T$, which we show in Fig.~\ref{fig:cT_adiabatic} as a function of redshift.
At early times the gas and CMB temperatures are coupled, which erases adiabatic fluctuations and drives $c_T$ towards zero.
However, as we approach cosmic dawn the index tends to its value of 2/3, though never reaching it.
For reference, in \citet{Munoz:2015eqa} we suggested the approximation 
\be
c_T = c_T^{(0)} - c_T^{(1)} (z-10)
\label{eq:cT_fit}
\ee
with $c^{(0)}=0.58$ and $c^{(1)} = 6\times 10^{-3}$, which is accurate to better than 3\% for our standard cosmology (in the absence of heating) in the range $z=6-50$, as we also show in Fig.~\ref{fig:cT_adiabatic}.
At lower $z$ these curves diverge as the gas is heated by X-rays, which drives $\overline T_k$ up but does not generate further adiabatic fluctuations, lowering $c_T$.

We note that it is critical to integrate up to very early times (as high as $z\sim 200$) in Eq.~\eqref{eq:cT_def} to find $c_T$, as the integral retains memory of the high-$z$ temperature.
Cutting off the high-$z$ part of the integral is equivalent to neglecting fluctuations (i.e., setting $T_k=\overline T_k$).
This is an issue for current \cmfast\ simulations, for which $T_k$ is assumed to be homogeneous at their initial $z=35$.
This assumption leads to an underprediction of the adiabatic fluctuations by a $\mathcal O(1)$ factor even at $z=15$, as illustrated in Fig.~\ref{fig:cT_adiabatic} by setting the $z>35$ part of the integral to zero.
This will affect the predictions of the 21-cm power spectrum by a similar amount, as we will explore in Sec.~\ref{sec:compare21cmfast}.

\begin{figure}
	\centering
	\includegraphics[width=0.46\textwidth]{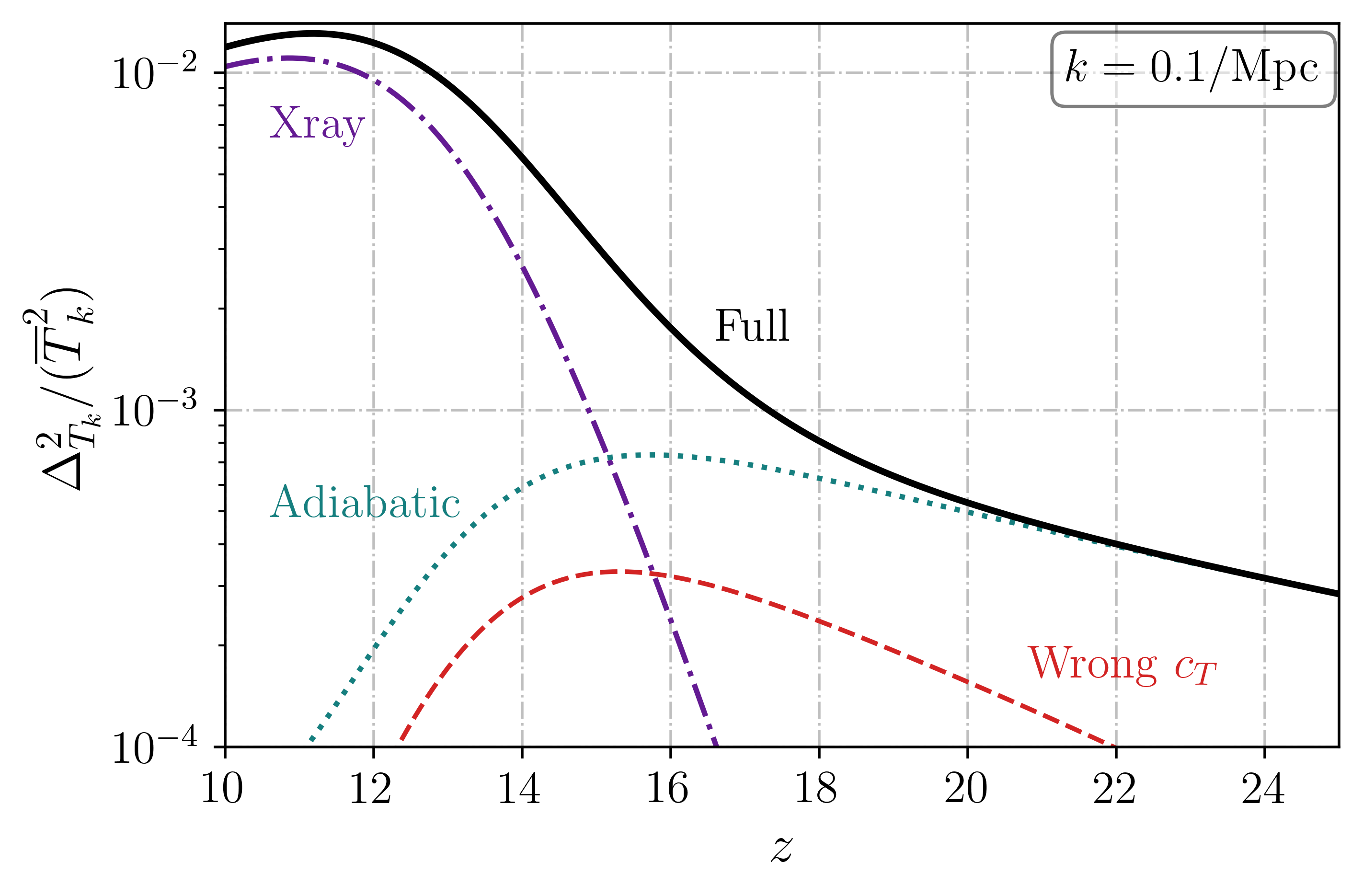}
	\caption{Dimensionless power spectrum of $T_k$ as a function of redshift, at $k=0.1\,\Mpcinv$.
	We show the contribution from X-rays (nonlinear, in purple dot-dashed line), from adiabatic fluctuations (teal dotted), as well as the total (which includes their cross term, in black).
	At early times ($z\gtrsim15$) the temperature fluctuations are dominated by adiabatic expansion and contraction, whereas later on X-rays are more important.
	We also show, for illustration purposes, the adiabatic prediction if one set the initial $T_k$ to be homogeneous at $z=35$ (in red dashed, as in \cmfast, see Fig.~\ref{fig:cT_adiabatic}). This would underpredict temperature fluctuations by nearly an order of magnitude during cosmic dawn.
	}
	\label{fig:Pk_contributions_vsz}
\end{figure}

In order to evaluate the impact of adiabatic fluctuations during cosmic dawn we show the $T_k$ power spectrum (again divided by its average) as a function of redshift in Fig.~\ref{fig:Pk_contributions_vsz}.
We have separated the X-ray and adiabatic terms, and also show the total, which includes their cross term.
Before X-ray heating is in full swing ($z\gtrsim 15$ for our parameters) adiabatic fluctuations reign, giving rise to a sizable power spectrum ($\Delta^2_{T_k}/T_k^2\sim10^{-3}$, or $\sim $ few\% rms fluctuations).
After X-rays turn on, the temperature largely will follow the SFRD; adiabatic fluctuations will slowly fade away as the total power grows.
For reference, the power spectrum that we showed in Fig.~\ref{fig:Deltasq_Tk} was at $z=12$, where the adiabatic fluctuations are small (though they still contribute to the $T_k$ power spectrum by $\sim 10\%$ through their cross term).
We also show in Fig.~\ref{fig:Pk_contributions_vsz} the prediction if one incorrectly set the temperature to be homogeneous at $z=35$ (following the red line in Fig.~\ref{fig:cT_adiabatic}).
This ``wrong" $c_T$ would underpredict the $T_k$ power spectrum by a factor of $3-10$ before X-ray heating.

We conclude that adiabatic fluctuations ought to be properly included up to high $z$ (see Fig.~\ref{fig:cT_adiabatic}) in order to recover the full $T_k$ fluctuations during cosmic dawn, and thus to predict the correct 21-cm signal.
In practice we will group the adiabatic term with the ``large-scale structure" term $(1+\delta)$ in \codename, as it depends on the local density $\delta$ rather than the SFRD.

\subsection{Large-scale Structure}

We now move to study the contribution from the density and RSD terms in Eq.~\eqref{eq:T21def}, which we group under the ``large-scale structure" (LSS) label. 

There is a fundamental difference between these and the terms that give rise to $\xa$ and $T_X$.
One can think of the latter two terms to live ``in Lagrangian space", as the SFRD depends on the initial overdensities linearly extrapolated; whereas the LSS terms are ``Eulerian", as they are given by local densities and velocities.
In principle these LSS terms also suffer from nonlinearities.
At the redshifts and scales of interest ($z\gtrsim 10$ and $k\lesssim 1\,\Mpcinv$) we expect the density field to be rather linear, so we will assume a simple model for nonlinearities, which we aim to refine in future work.

Let us define $\Delta = (1+\delta_{\rm NL})$ as the (non-linear) density term. 
For notational consistency we will assume a lognormal model as we did for the SFRD, based on the work of~\citet{Coles:1991if}.
In that case the auto- and cross-correlations can be found trivially to be
\ba
\xi_{\Delta} =& \exp( \xi) - 1 \nonumber \\
\xi_{\Delta,\xa} =& \,c_{1,\alpha} \sum_R \, c_{2,\alpha} \left( \exp [\gamma_R \xi^{R,0}] - 1\right)  \nonumber\\
\xi_{\Delta,T_X} =& \sum_{z' \geq z} c_{1,X} \sum_R \, c_{2,X}  \left( \exp  [\gamma_R \xi^{R,0}] - 1\right). 
\label{eq:crosscorrLSS} 
\end{align}
For the redshifts and scales of interest to cosmic dawn we have tested that this correction to the auto and cross-spectrum of $\Delta$ is a few \%.
We have included a flag to allow the user to toggle this correction on and off.
We will improve this model in future work, for instance through perturbation theory in Eulerian~\citep{Shaw:2008aa} or Lagrangian~\citep{Crocce:2006ve} space.

Our model for redshift-space distortions will be likewise simple, as these are not the focus of our paper (for more thorough studies see~\citealt{Mao:2011xp,Ross:2020tnz,Shaw:2023zrs} for instance).
We will take the linear relation
$\delta_v = -\mu^2 \delta,$
where $\mu = k_{||}/k$ is the line-of-sight cosine.
We have implemented three different options for RSDs in \codename.
These are (i) no RSDs (or real-space, $\mu=0$), (ii) spherical RSDs (as standard for simulations, which we calculate by setting $\mu=0.6$, so that $(1+\mu)^2\approx1.87$), and (iii) foreground-avoided RSD (as observed by interferometers outside the wedge, which we obtain by setting $\mu = 1$).
Each of these will be revisited and compared to simulations in future work.
Unless otherwise specified, we will show results for the spherical RSD mode for the 21-cm signal in order to better compare to other results in the literature, but real space for the rest of quantities like SFRD, $T_X$, and $\xa$.

\subsection{Reionization}

The final ingredient for computing $T_{21}$ in Eq.~\eqref{eq:T21def} is the neutral hydrogen fraction $x_{\rm HI}$.
The epoch of cosmic reionization will see $\xHI$ evolve from near unity during cosmic dawn to zero by $z=5$~\citep{Becker:2015lua}.
This is a complicated process, as originally small and isolated bubbles of HII will grow and merge, eventually percolating to reionize the entire universe~\citep{Furlanetto:2015hqp}.
A successful analytic approach to model this era is that pioneered by \citet{Furlanetto:2004nh}, where the universe is populated with reionization bubbles built on top of galaxies.
Rather than reproduce their model, or recent work based on perturbative reionization~\citealt{McQuinn:2018zwa,Qin:2022xho,Sailer:2022vqx}, we will focus on the cosmic-dawn epoch at $z\gtrsim 10$, and simply compute the average evolution of the neutral fraction $\xHI$ for reference.
We will tackle models of  the reionization fluctuations (or bubbles) in future work\footnote{In the meantime, the interested user could plug \codename\ into the reionization-only code from \citet{Mirocha:2022pys}, taking the $\overline T_S$ and $\overline x_{\rm HI}$ output from \codename\ as inputs and modeling the cross terms.}.

We will calculate the evolution of reionization through the ionizing emissivity $\dot N_{\rm ion}$, which is computed in terms of the SFRD as~\citep{Mason:2019oeg}
\be
\dot N_{\rm ion} = \int dM_h \left( \dfrac{d\rm SFRD}{d M_h}\right) f_{\rm esc}(M_h),
\ee
by accounting for the fraction $f_{\rm esc}$ of ionizing photons that can escape the galaxy where they were produced.
We model this fraction as a simple power-law in mass~\citep[e.g.,][]{Park:2018ljd},
\be
f_{\rm esc} (M_h) = f_{\rm esc}^{(0)} \, (M_h/10^{10} \Msun)^{\alpha_{\rm esc}},
\label{eq:fesc}
\ee
and for our fiducial we set $f_{\rm esc}^{(0)}=0.1$ and $\alpha_{\rm esc}=0$ for simplicity, which enforce full reionization by $z=5.3$, as suggested by Lyman-$\alpha$ data~\citep{Bosman:2021oom},
though both of these are free parameters in \codename.
For instance, one can set a positive index $\alpha_{\rm esc}>0$, which implies a faster epoch of reionization (EoR), in which heavier/brighter galaxies dominate reionization, as suggested by recent results from simulations~\citep{Yeh:2022nsl} and observations~\citep{Naidu:2019gvi}.

Rather than working with $x_{\rm HI}$, we define the filling factor $Q = 1 - x_{\rm HI}$ of ionized hydrogen, whose evolution is given by~\citep{Madau:1998cd}
\be
\dot Q = \dfrac{\dot N_{\rm ion}}{n_H} - \dfrac{Q}{t_{\rm rec}},
\label{eq:dotQ}
\ee
where $n_H$ is the number density of hydrogen, and $t_{\rm rec}$ is its recombination time.
We will follow~\citet{Mason:2019oeg} and find the recombination time through
\be
t_{\rm rec}^{-1}(z) = \alpha_B(T_k) C n_e^{(0)} (1+z)^3
\ee
with a constant $C=3$ and $\alpha_B$ evaluated at $T_k=10^4$ K for simplicity.
Eq.~\eqref{eq:dotQ} can be suggestively rewritten as
\be
(g Q)'/g = \dfrac{\dot N_{\rm ion}}{n_H} \dfrac{dt}{dz}
\ee
where as before primes indicate derivative with respect to $z$, and we have defined a function $g$ that satisfies
\be
g'/g = t_{\rm rec}(z) \dfrac{dt}{dz}. 
\ee
For matter domination (the epoch of interest) and a constant $C$ we can analytically solve this function to be 
\be
g = \exp\left (\dfrac{2}{3 \tau_0}\, (1+z)^{3/2}\right)
\ee
where $\tau_0 = t_{\rm rec}(0) H_0 \sqrt{\Omega_m}$.
Then, we can find the filling factor as
\be
Q(z) = \int_z dz' \dfrac{dt}{dz'} \dfrac{\dot N_{\rm ion}}{n_H} \exp\left (\dfrac{2}{3 \tau_0}\, [(1+z')^{3/2}-(1+z)^{3/2}]\right),
\label{eq:Qintz}
\ee
which accounts for recombinations (though only homogeneously, cf.~\citealt{Madau:1998cd} for a similar solution in terms of $t$ rather than $z$).
This equation can be generalized to clumping factors $C$ that evolve with $z$, though it is technically difficult to include imhonogeneities, so we defer those to future work.

\begin{figure}
	\centering
	\includegraphics[width=0.46\textwidth]{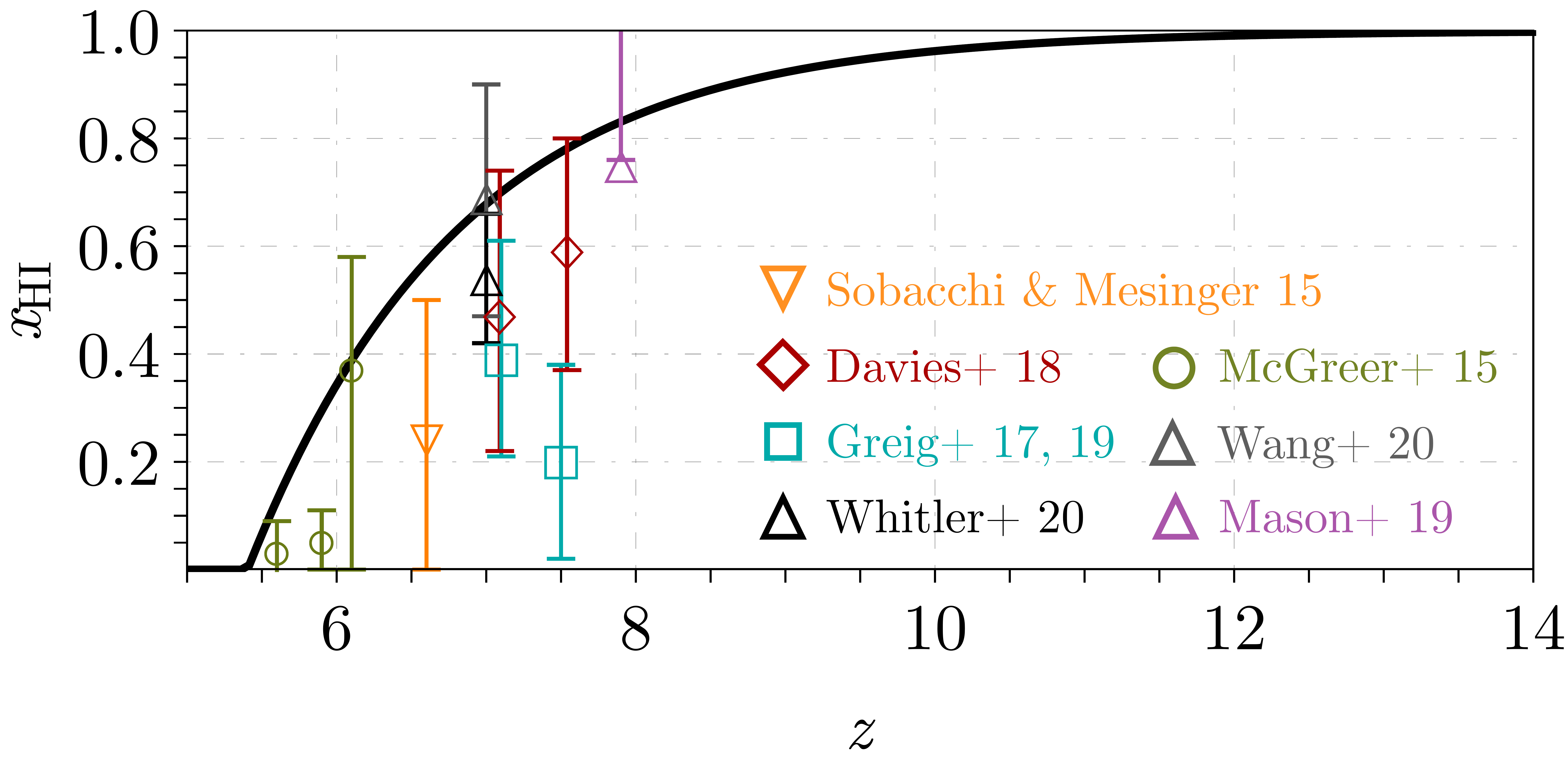}
	\caption{Average reionization history for our fiducial model, along with a subset of current constraints. We parametrize the escape fraction as in Eq.~\eqref{eq:fesc} and set a fixed clumping factor $C=3$ for simplicity. 
	}
	\label{fig:xHI_avg}
\end{figure}

We can now integrate Eq.~\eqref{eq:Qintz} to find the global evolution of the neutral fraction, which we show in Fig.~\ref{fig:xHI_avg}.
Indeed, reionization is over by $z\approx 5.3$ for our chosen parameters, and begins in earnest below $z\approx 10$. 
This is in broad agreement with current reionization data (also shown in the Figure, summarized in \citealt{Mason:2019oeg}), as well as the CMB (our model gives rise to $\tau_{\rm reio} = 0.05$, well within the 1-$\sigma$ {\it Planck} mesurement~\citealt{Aghanim:2018eyx}).
The evolution of $\xHI$ is fairly fast in this model, as the bright galaxies dominate the SFRD.
We focus on the cosmic-dawn era proper ($z>10$) in this work, so we will set $\xHI=1$ through the rest of this paper unless otherwise specified.

\section{A full analytical calculation of the 21-cm signal}
\label{sec:full21cmsignal}

In the previous sections we have explained how we compute each of the ingredients of the 21-cm temperature $T_{21}$ (Eq.~\ref{eq:T21def}).
We now show how we combine them to find the 21-cm global signal and its fluctuations in \codename.

So far we have not made any assumptions on the size of $\xa$, $T_X$, or their fluctuations. 
However, in order to keep an analytic closed form we will now expand their contributions to the 21-cm signal as~\citep{BarkanaLoeb2005,Pritchard:2006sq}
\ba
\dfrac{x_\alpha}{1 + x_\alpha} \approx& \dfrac{\overline x_\alpha}{1 + \overline  x_\alpha} + \dfrac{\delta \xa }{(1 + \overline  x_\alpha)^2} 
\label{eq:Taylor_expansions}
\\ \nonumber
\left ( 1 - \dfrac{\Tcmb}{T_c} \right)\approx& \left ( 1 - \dfrac{\Tcmb}{\overline T_c} \right) + \delta T_k \dfrac{\Tcmb }{\overline T_k \overline T_c}
\end{align}
where as before an overline means spatial average (and we assume a homogeneous $T_c-T_k$ connection in this work). We remind the reader that $T_{21}$ depends multiplicatively on each of these terms.
These relations are linear in $\delta {\xa}$ and $\delta {T_k}$, but not on the matter density $\delta$. The SFRD at each $R$ shell will behave as an exponential of $\delta_R$, which gets added over all $R$ to find the total power on $\xa$ and $T_k$ (and thus on $T_{21}$).
One can then infer that $T_{21}$ will be a sum of lognormal variables, with well-known statistical properties close to a lognormal itself~\citep{Mitchell:68}.
While we are working to linear order in the $\xa,\ T_k$ fluctuations (and as such the $T_X$ and $T_{\rm cosmo}$ terms can be computed independently as argued in Sec.~\ref{sec:cosmicdawn}), we will show the full expressions in Appendix~\ref{app:Taylor}, where we demonstrate that the corrections from higher-order terms are sub 10\% for the models and redshifts of interest, making our approximations sufficient.

\subsection{An Example Global signal with \codename}

We begin by computing the 21-cm global signal.
To first order, it can be found by simply using the average of each of the quantities at play (the temperature $\overline T_k$, Lyman-$\alpha$ coupling $\overline x_{\alpha}$, and density/RSD $\delta = \delta_v = 0$)~\footnote{This is the approach followed in e.g., {\tt ARES}~\citep{Mirocha:2014faa}.}.

\begin{figure}
	\centering
	\includegraphics[width=0.48\textwidth]{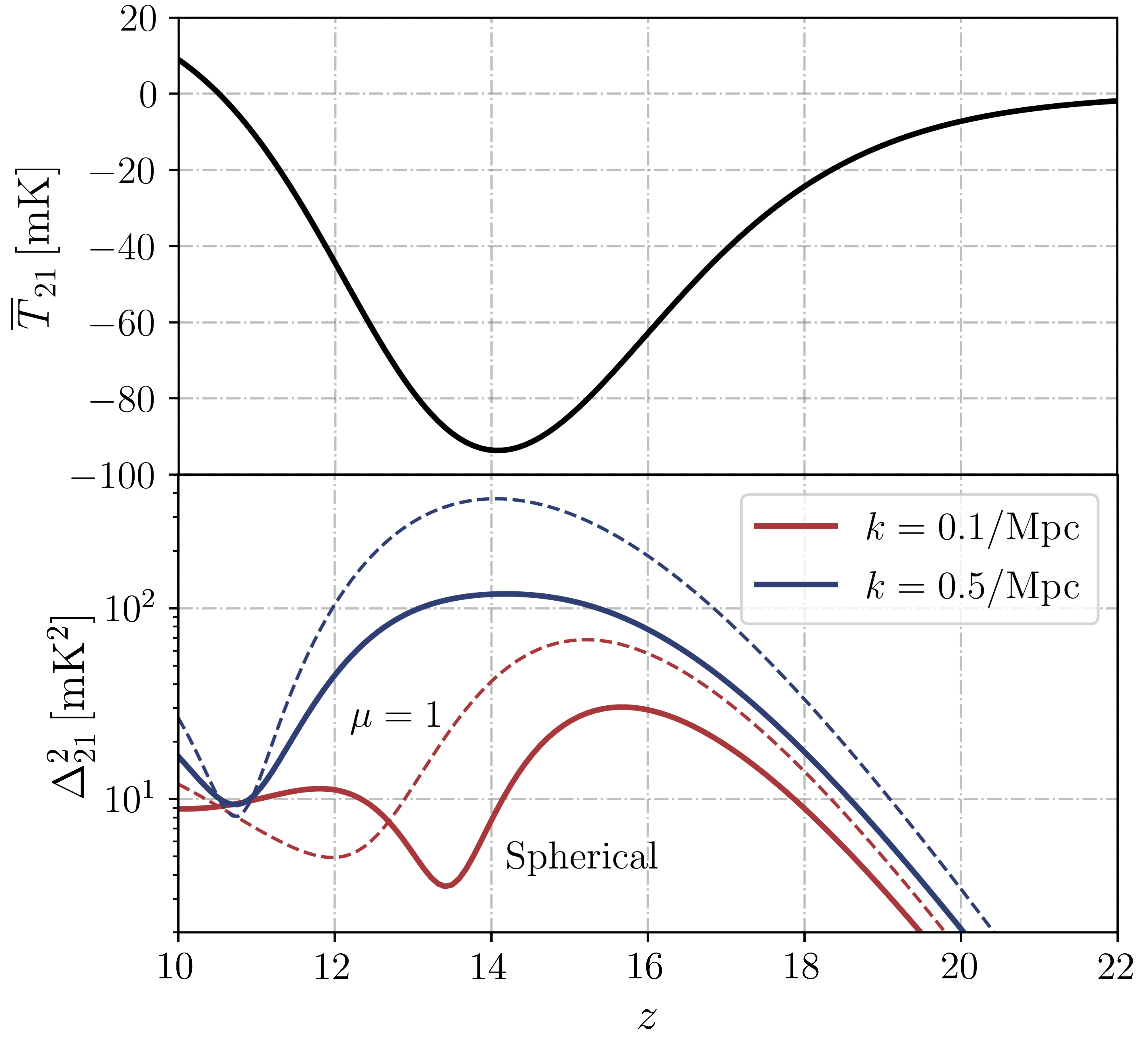}
	\caption{Cosmic evolution of the 21-cm global signal (top) and fluctuations (bottom) calculated with \codename.
	We have set a Planck 2018 $\Lambda$CDM cosmology, and ``minimal" astrophysical parameters as summarized in the main text (which do not include Pop III stars in minihaloes). 
	Solid lines show the spherically averaged prediction, whereas dashed lines have $|\mu|=1$, as expected of the modes that survive the foreground wedge, showing significantly larger power during cosmic dawn.
	}
	\label{fig:T21_P21_z}
\end{figure}

We show the \codename\ prediction for the 21-cm global signal in Fig.~\ref{fig:T21_P21_z}.
As expected of our fiducial (which contains only galaxies above the atomic-cooling threshold), the cosmic-dawn signal turns on at $z\sim 20$, peaks at $z\sim 15$, and turns into emission by $z\sim 10$ (cf.~\citealt{Mirocha2016_UVLF_GS,Park:2018ljd,Munoz:2019rhi}).
For our choice of X-ray efficiency $L_{40}$ the signal reaches a $\sim -90$ mK depth, far from the $\sim -250$ mK allowed by the adiabatic temperature of gas (assuming full WF coupling), and farther even from the $\sim -500$ mK required by the EDGES claimed detection~(\citealt{Bowman:2018yin}, though see of course \citealt{Hills:2018vyr,Singh:2021mxo} for criticisms).
In upcoming work we will implement dark-matter models (such as millicharged particles~\citealt{Munoz:2018jwq,Driskell:2022pax}) into \codename, which will allow the user to self-consistently account for the gas cooling from the CMB epoch to cosmic dawn and beyond (as \codename\ interfaces with {\tt CLASS}, allowing for joint CMB+21-cm analyses).

We now explore the 21-cm fluctuations in our full nonlinear and nonlocal model.

\subsection{An Example Power Spectrum with \codename}

We use the auto- and cross-power spectra for each of the 21-cm components (LSS, $\xa$, and $T_k$) derived in Sec.~\ref{sec:cosmicdawn}, along with the weights we show in Eq.~\eqref{eq:Taylor_expansions}, to find the 21-cm power spectrum analytically in \codename.
The full calculation at all $z$ takes a mere 3-5 s in a single-core laptop.
Further, the separation of components will allow us to test each of their contributions during cosmic dawn.

We show our prediction for the 21-cm power spectrum as a function of $z$ at two wavenumbers in Fig.~\ref{fig:T21_P21_z}.
Guided by observational constraints, we will focus on wavenumbers outside the ``foreground wedge"~\citep{Parsons:2012qh,Liu:2014bba,Liu:2014yxa}, as those within it are deemed unusable for cosmology.
As such, we choose wavenumbers roughly at the low-$k$ edge of the foreground wedge ($k=0.1\,\Mpcinv$), and before thermal noise spikes up ($k=0.5\,\Mpcinv$).
Moreover, we show results assuming either spherical RSDs (as commonly done in 21-cm simulations) or foreground-avoiding RSDs ($\mu=1$), which evade the wedge, and in which case the power is larger by a factor of a few. 
In both cases the 21-cm power shows the characteristic growth from high to low $z$ as the 21-cm signal grows in absorption ($\overline T_{21}$ becomes more negative), until the trough at $z\sim15$, after which the power begins to decrease.
The two power spectra shown reach different amplitudes at their peak, which occurs at slightly different $z$.
This showcases the power of the 21-cm fluctuations to unearth the astrophysics of cosmic dawn, holding more information than the global signal alone.
We do not run a full detectability study with interferometers such as HERA or the SKA here, as that is not our goal.
However, we note that a similar \cmfast\ model in \citet{Munoz:2021psm} had overall lower power, but still boasted a signal-to-noise ratio of $\approx 100-200$ for both HERA and the SKA, so we would expect a high-significance detection of our predicted 21-cm signal.

Our approach in \codename\ diverges from previous analytic approaches (eg BLPF or {\tt ARES}, see also~\citealt{Schneider:2020xmf}) in that it accounts for nonlinearities in the SFRD. This is key to obtaining accurate estimates of the 21-cm power spectrum.
We showed in Figs.~\ref{fig:Deltasq_xa} and~\ref{fig:Deltasq_Tk} that nonlinearities change the power spectrum of the $\xa$ and $T_k$ components by $\mathcal O(1)$ in the $k=0.1-1\,\Mpcinv$ range, where the 21-cm signal is most readily observable.
As such, we expect the 21-cm power spectrum to be affected by these nonlinearities that we are modeling.

\begin{figure}
	\centering
	\includegraphics[width=0.46\textwidth]{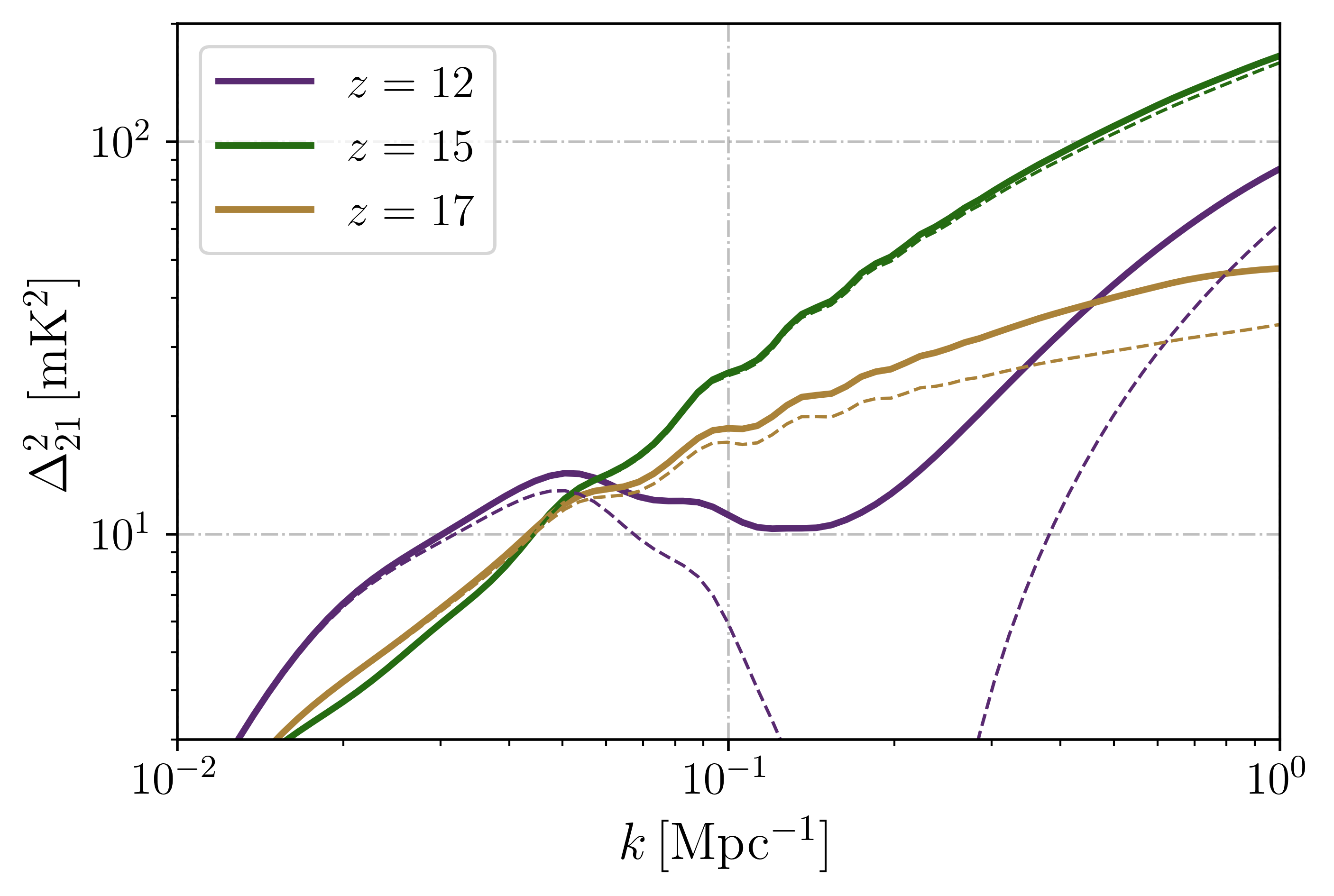}
	\caption{Power spectrum of the 21-cm signal during cosmic dawn as a function of wavenumber $k$, at three redshifts $z$.
		Solid lines show the full nonlinear result from $\codename$, whereas dashed are a linear approximation as in~\citet{BarkanaLoeb2005,Pritchard:2006sq}. 
	}
	\label{fig:P21_vs_k}
\end{figure}

We showcase this point in Fig.~\ref{fig:P21_vs_k}, where we plot the 21-cm power spectrum against wavenumbers $k$ at three redshifts, chosen roughly to be near the beginning and end of cosmic dawn ($z=12$ and 17), and midway (at $z=15$).
In all cases the linear calculation is accurate at very large scales ($k\lesssim 0.05\,\Mpcinv$), which are however inaccessible with current interferometers.
At the observationally relevant scales ($k=0.1-1\,\Mpcinv$) we see that the nonlinear calculation in \codename\ can change the 21-cm power spectrum significantly.
At the highest $z$ we show in Fig.~\ref{fig:P21_vs_k} the nonlinear corrections increase the power by roughly 50\%, as the Lyman-$\alpha$ term dominates the fluctuations.
This term follows the SFRD, and as we saw Fig.~\ref{fig:Deltasq_xa} the nonlinear behavior of this quantity increases the $\xa$ power significantly.
Near the trough ($z=15$) the Lyman-$\alpha$ and X-ray fluctuations are expected to roughly cancel at large scales~\citep{Pritchard:2006sq,Munoz:2020itp}, and the 21-cm power rises steeply towards higher $k$, roughly tracking the density. As such, the nonlinear SFRD corrections are small in Fig.~\ref{fig:Deltasq_xa}.
At the lowest $z=12$ that we show in that Figure we see that the linear prediction would cancel at $k\sim 0.2\,\Mpcinv$, whereas our full calculation is radically different.
At this low $z$ the Lyman-$\alpha$ fluctuations are small, so the 21-cm power is determined by a competition between the X-ray and LSS terms.
These two terms are anti-correlated here, as larger densities increase $T_{21}$ through the $(1+\delta)$ term, but also provide more heating, which would decrease the contrast with the CMB and thus $T_{21}$.
This correlation will flip to positive at $z\lesssim 11$, when our global signal turns into emission.
Our nonlinear modeling of the X-ray term is thus key to infer the correct 21-cm power at $k\sim 0.1\,\Mpcinv$ scales.

The three cases in Fig.~\ref{fig:P21_vs_k} are examples of the richness of information encoded in the 21-cm fluctuations.
They also showcase the level of detail that has to be modeled to extract said information from upcoming 21-cm data.
We note that, unlike simulated power spectra, our curves correspond to a specific $k$, rather than a broad bin of wavenumbers. One can average over $k$ in \codename\ to emulate the output of simulations, if desired.
We will compare our results with semi-numerical simulations from \cmfast\ in the following Section, but we note in passing that when comparing with the approach in \citet{Schneider:2020xmf} we also find good agreement (in the linear case specially, as their nonlinear model relies on the halo model, rather than the Eulerian picture we follow here).

\section{Comparison with \cmfast}
\label{sec:compare21cmfast}

We now move on to compare our analytic results from \codename\ to the well-known semi-numeric simulations from \cmfast. 
The excursion-set approach that we use for the heating and Lyman-$\alpha$ terms is very similar to \cmfast, as in both cases it is based on the work of~\citet{BarkanaLoeb2005,Pritchard:2006sq}.
The parametrization of the sources themselves is, nevertheless, fairly different.
In order to fairly compare the two, we will implement an astrophysical model within \codename\ that closely resembles \cmfast, as we now detail.
The busy reader may want to skip ahead to Figs.~\ref{fig:GS_vs_21cmFAST} and \ref{fig:PS_vs_21cmFAST} to see a comparison of the 21-cm global signal and fluctuations between the two codes, given the same inputs.

\subsection{The SFRD in \cmfast}

We begin by modeling the SFRD (and hence the halo-galaxy connection) as in \cmfast.
Throughout this paper we have followed the EPS formalism to perturb the SFRD on regions of density $\delta_R$. 
In this formalism one takes the halo abundance to depend on density as in Eq.~\eqref{eq:HMF_EPS}, which feeds into the SFRD.
\cmfast, instead, follows an approach where the SFRD is itself modulated by densities using a Press-Schechter HMF.
That is, in \cmfast~\citep{Mesinger:2010ne}
\be
\sfrd^{\rm PS}(z | \delta_R) = \int dM_h \dfrac{dn^{\rm PS}}{dM_h}(\delta_R) \dot M_*(M_h), 
\label{eq:sfrd_21cmfast}
\ee
which is then renormalized by a factor
\be
\sfrd(z | \delta_R) \to \overline \sfrd(z)  \dfrac{\sfrd^{\rm PS}(z | \delta_R)}{\VEV{\sfrd^{\rm PS}(z | \delta_R)}},
\ee
where $\overline \sfrd$ is the ``true" SFRD (calculated as in Eq.~\ref{eq:sfrd} with the spatially averaged Sheth-Tormen HMF), but $\sfrd^{\rm PS}(z | \delta_R)$, and its spatial average $\VEV{\sfrd^{\rm PS}(z | \delta_R)}$, are computed with the Press-Schechter HMF.
We will not dwell here on the validity of this formula against its counterpart in Eq.~\eqref{eq:SFRD_EPS}, and instead simply show that it can also be fit by an exponential of $\delta_R$, as it fundamentally relies on the same principle: in the early universe the abundance of haloes--and thus galaxies--is exponentially sensitive to the density field.

\begin{figure}
	\centering
	\includegraphics[width=0.34\textwidth]{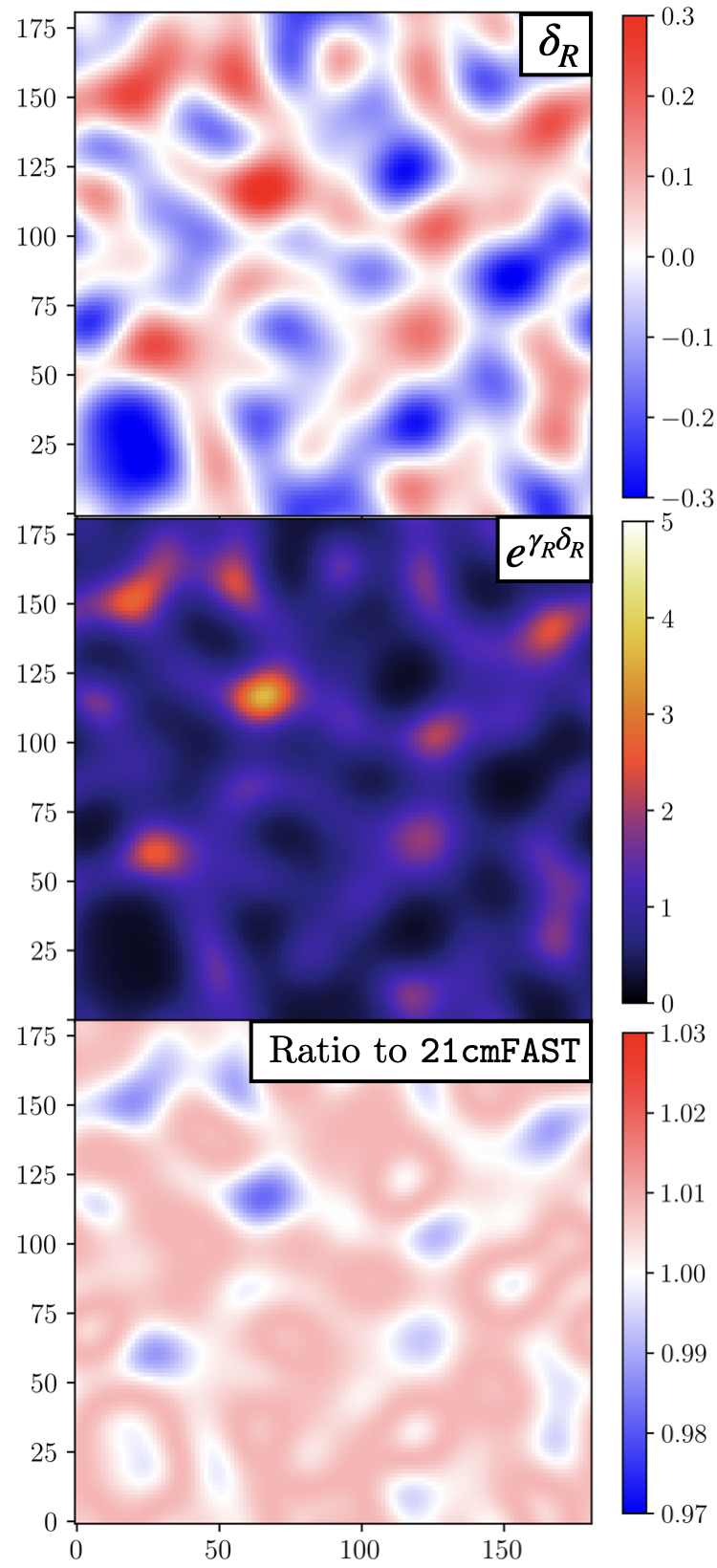}
	\caption{Slice with the density from a \cmfast\ simulation (on top), and our lognormal model for the SFRD (in the middle). We also show the ratio with respect to the SFRD output from \cmfast\ in the bottom, which only deviates from unity at the few percent level. The density field has been smoothed on $R=10$ Mpc scales, and is extracted from a simulation with $120^3$ cells and 180 Mpc side.
	}
	\label{fig:fcollslice_21cmfast}
\end{figure}

We showcase the agreement of the SFRD predicted by our lognormal model and \cmfast\ in Fig.~\ref{fig:fcollslice_21cmfast}.
We show a slice of densities from \cmfast, as well as our lognormal approximation to the SFRD and its ratio to the direct output of the \cmfast\ simulation (from their {\tt Fcoll} output box).
Our analytic model agrees with the \cmfast\ output to within $\lesssim3\%$.
This Figure shows a smoothing scale of $R = 10$ Mpc, comparable to the scales that determine the power spectrum at the observable wavenumbers $k\lesssim 1$ Mpc$^{-1}$, as we will confirm below when comparing power spectra of $T_k$ and $J_\alpha$.
We have implemented this alternative EPS algorithm in \codename, which can be turned on with the {\tt FLAG\_EMULATE\_21CMFAST}.

\subsection{Astrophysical and Simulation Parameters}

In addition to the SFRD, the \cmfast\ model for astrophysical sources differs from our baseline in a few key elements, which we now describe.

On the astrophysics side, the SFR ($\dot M_*$) for each halo is computed from Eq.~\eqref{eq:SFR_21cmfast}, which relies on a timescale $t_*$ rather than EPS or exponential accretion. 
We have also implemented this model in \codename, and through this section we will choose the same astrophysical parameters as the PopII-only model of~\citet{Munoz:2021psm}, which fit current reionization data ($\xHI$, HST UVLFs, and the Lyman-$\alpha$ forest, see~\citealt{Qin:2021gkn}).
We will take a number $N_\alpha = 11,000$ of Lyman-$\alpha$ photons per baryon, as is inferred from the stellar spectra on \cmfast\ (though we assume a simpler spectrum as described in Sec.~\ref{sec:cosmicdawn}).
As for X-ray propagation, we will force the \cmfast\ approximation that the X-ray opacity $(e^{-\tau})$ be a Heaviside theta function, either 0 or 1 (only for this section, the user can switch this feature on or off in \codename\ using the flag {\tt XRAY\_OPACITY\_MODEL}).
We additionally change the constant conversion factor from $x_\alpha$ to $J_\alpha$ to the \cmfast-coded value of $1.66\times 10^{11}$ (rather than our 1.81, both in the appropriate units, see Eq.~\ref{eq:xadef}), and we do not include the Hydrogen fraction $f_H\sim 0.9$ in the optical depth $\tau_{\rm GP}$  to imitate their code.
As introduced in Sec.~\ref{sec:cosmicdawn}, \cmfast\ ignores adiabatic fluctuations before its starting redshift $z=35$ by default. We will mimic this (erroneous) behavior by setting the high-$z$ part of the $c_T$ integral to zero (see Figs.~\ref{fig:cT_adiabatic} and~\ref{fig:Pk_contributions_vsz} for comparison).

On the cosmology side, we have modified the HMF parameters to match \cmfast\ (from \citealt{Jenkins:2000bv}), and use their approximate ``Dicke" growth factor rather than the {\tt CLASS} output.
We additionally take a 3D tophat window function in Eq.~\eqref{eq:corrfunc_with_windows} for both the linear and non-linear correlation functions, as that is what the \cmfast\ algorithm assumes.
We will cut scales smaller than the resolution of the \cmfast\ simulations, which will be $R_{\rm min}=0.93$ Mpc (corresponding to 1.5 Mpc resolution in a cubic cell as recommended in~\citealt{Kaur:2020qsa}); as well as larger than 500 Mpc, as \cmfast\ does not sum beyond that scale.

For the \cmfast\ simulations we have evolved the density field linearly, since that is supposed to be the input of the EPS algorithm.
We have also turned off photo-heating feedback~(\citealt{Sobacchi:2014rua}, and in fact we set a fixed $M_{\rm turn} = 10^{7.5} \Msun$ in both codes), as we have not implemented that feature in \codename\ yet. Moreover, we have set to zero the excitation contribution from X-rays (which is a small correction), and have fixed a homogeneous value of $x_e = 2\times10^{-4}$ for the free electron fraction (though it only makes a $<10\%$ difference). This serves a double purpose, as \cmfast\ uses {\tt RECFAST}\footnote{Because of this their average prediction for $T_k$ is off before X-rays, which we account for by lowering our $T_{\rm cosmo}$ by 5\%.} with a clumping factor $C=2$ for evolving $x_e$, which leads to discrepancies when compared to {\tt HyREC} and {\tt CLASS}, and we are currently not tracking spatial fluctuations in $x_e$ in \codename.
In all cases we calculate real-space power spectra and focus on the $f_{\rm esc} = 0$ (no reionization) limit, which will allow us to make crisp comparisons between the two codes.

\subsection{The IGM properties}

We begin by comparing the properties of the IGM (chiefly the X-ray heating and Lyman-$\alpha$ flux) between the \cmfast\ simulation boxes and our analytic predictions from \codename.
We choose two indicative redshifts during cosmic dawn, $z=17$ (when the signal will be turning down, as the first galaxies start to excite the Hydrogen and slowly heat it), and at $z=10$, when we finish our simulations and the gas is fully heated.

\begin{figure}
	\centering
	\includegraphics[width=0.46\textwidth]{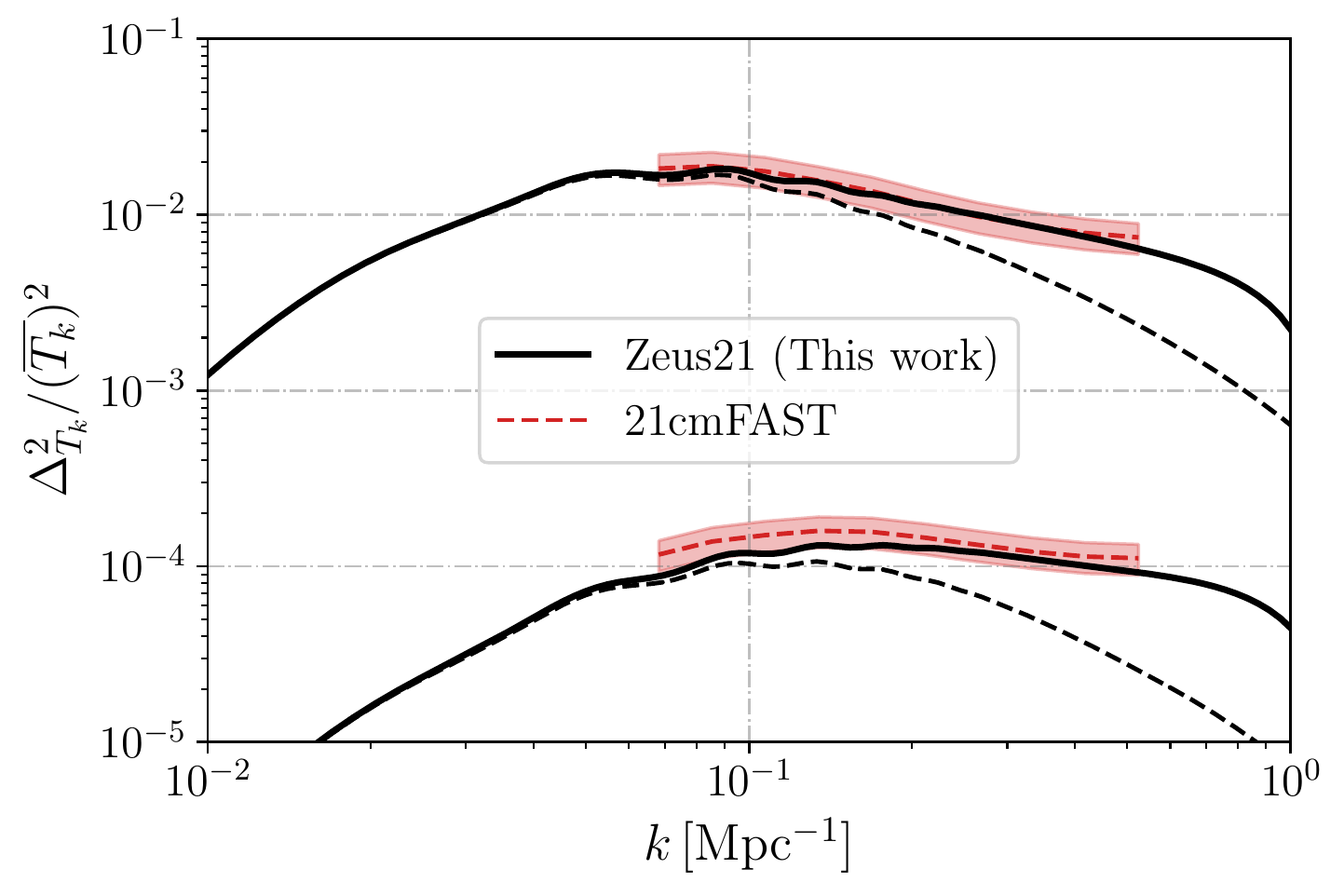}
	\caption{Power spectrum of the $T_k$ fluctuations (dimensionless) sourced by X-rays only (i.e., without adiabatic). Top curves correspond to $z=17$, and lower for $z=10$. Black lines are our \codename\ predictions, and the red region is from \cmfast\ simulations with different $L_{\rm box}$, from 300 to 1000 Mpc in side (all with $100^3$ cells).
		The black dashed line is a linear prediction, which falls short of the true power spectrum at the relevant scales for interferometers ($k\gtrsim 0.1\,\rm Mpc^{-1}$).
		}
	\label{fig:PTk_vs_21cmfast}
\end{figure}

\begin{figure}
	\centering
	\includegraphics[width=0.46\textwidth]{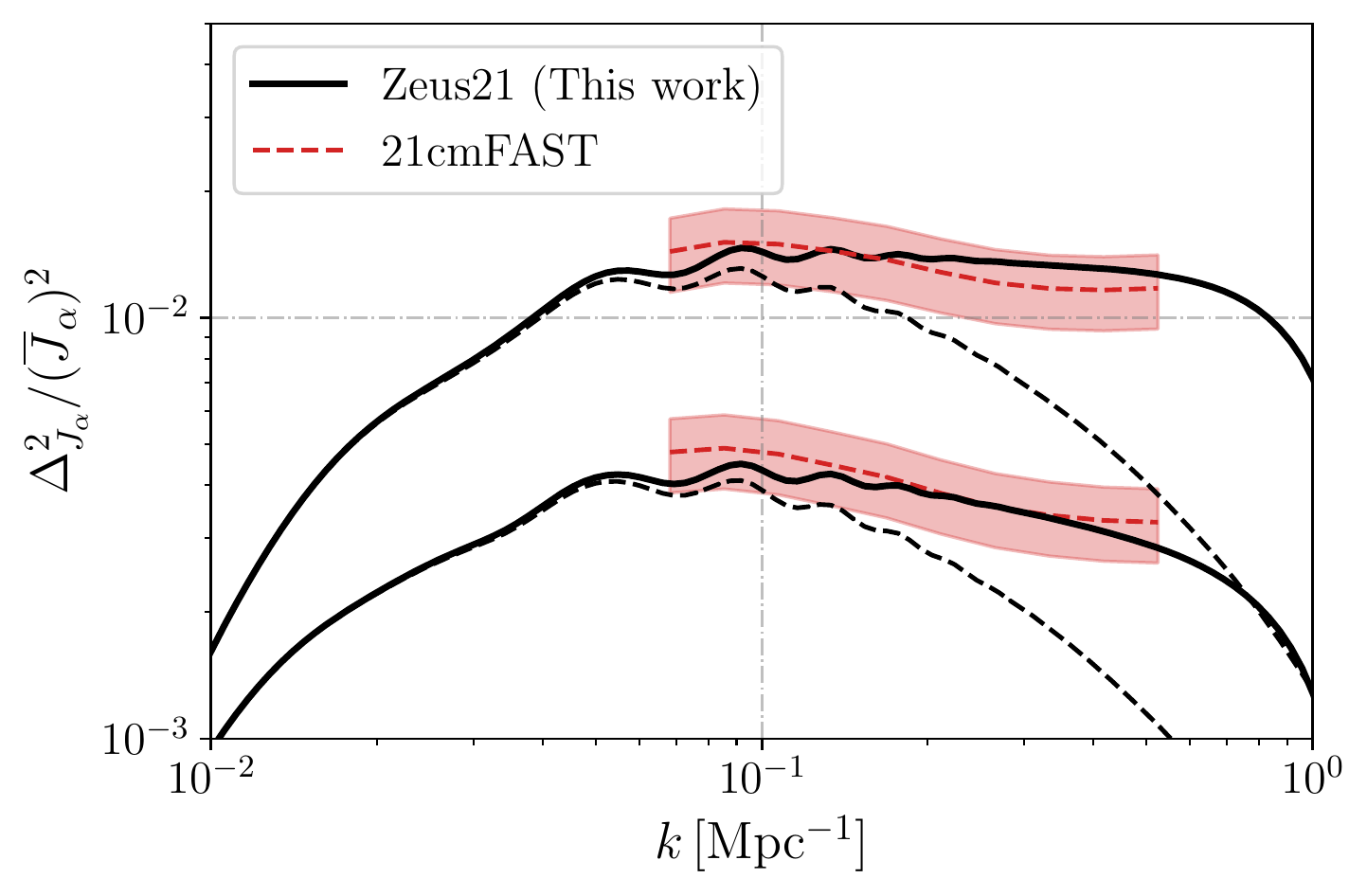}
	\caption{Same as Fig.~\ref{fig:PTk_vs_21cmfast} but for the fluctuations of $J_\alpha$. Here the results from \cmfast\ simulations do not appear converged between simulation runs, but the \codename\ result agrees with the highest-resolution curves towards high $k$.
	}
	\label{fig:PJa_vs_21cmfast}
\end{figure}

We show in Fig.~\ref{fig:PTk_vs_21cmfast} the power spectrum of the IGM temperature fluctuations ($\delta T_k/\overline{T}_k$) due to X-rays, both for our \codename\ run (with the modifications outlined above), and for the average of three \cmfast\ boxes (with 1000, 500, and 300 Mpc in side, all with 100$^3$ cells), to which we assign an error given by their root mean square difference (or 20\%, whichever is larger) to test $k$ convergence.
The two calculations agree very well at both redshifts.
Interestingly, a linear-only calculation (following BLPF) significantly underpredicts the power for $k\gtrsim 0.1$ Mpc$^{-1}$ in Fig.~\ref{fig:PTk_vs_21cmfast}.
This shows the need for a nonlinear approach as ours, and further validates its output against simulated $T_k$ boxes.
At $k\gtrsim 0.8$ Mpc$^{-1}$, however, even the nonlinear calculation deviates from the \cmfast\ result. That is partly due to our minimum radius $R_{\rm min}$ at the cell size, as well as to possible aliasing on the \cmfast\ boxes (as the power spectrum spikes up at high $k$ in a resolution-dependent manner, which is more obvious in Fig.~\ref{fig:PJa_vs_21cmfast}).

We show a similar analysis for the $J_\alpha$ fluctuations in Fig.~\ref{fig:PJa_vs_21cmfast}.
There, the \cmfast\ boxes are less converged, as larger boxes tend to miss some of the high-$k$ power, whereas smaller boxes lack the low-$k$ support (the photons that produce the WF effect can travel a distance comparable to $H^{-1}(z)$ before entering a resonance, see Fig.~\ref{fig:coeffs2_R}).
This translates into a wider \cmfast\ band.
Nevertheless, we see the same trends as for $T_k$, with good agreement between the semi-numeric simulations of \cmfast\ and our analytic power from \codename.
As before, the addition of nonlinearities to \codename\ is critical for inferring the correct power spectrum in the relevant scales of $k\sim 0.1-1$ Mpc$^{-1}$.

These tests show that our approach is able to capture the relevant physics at cosmic dawn, including nonlinearities.
We now move to compare the 21-cm signal between both approaches.

\subsection{The 21-cm signal}

\begin{figure}
	\centering
	\includegraphics[width=0.48\textwidth]{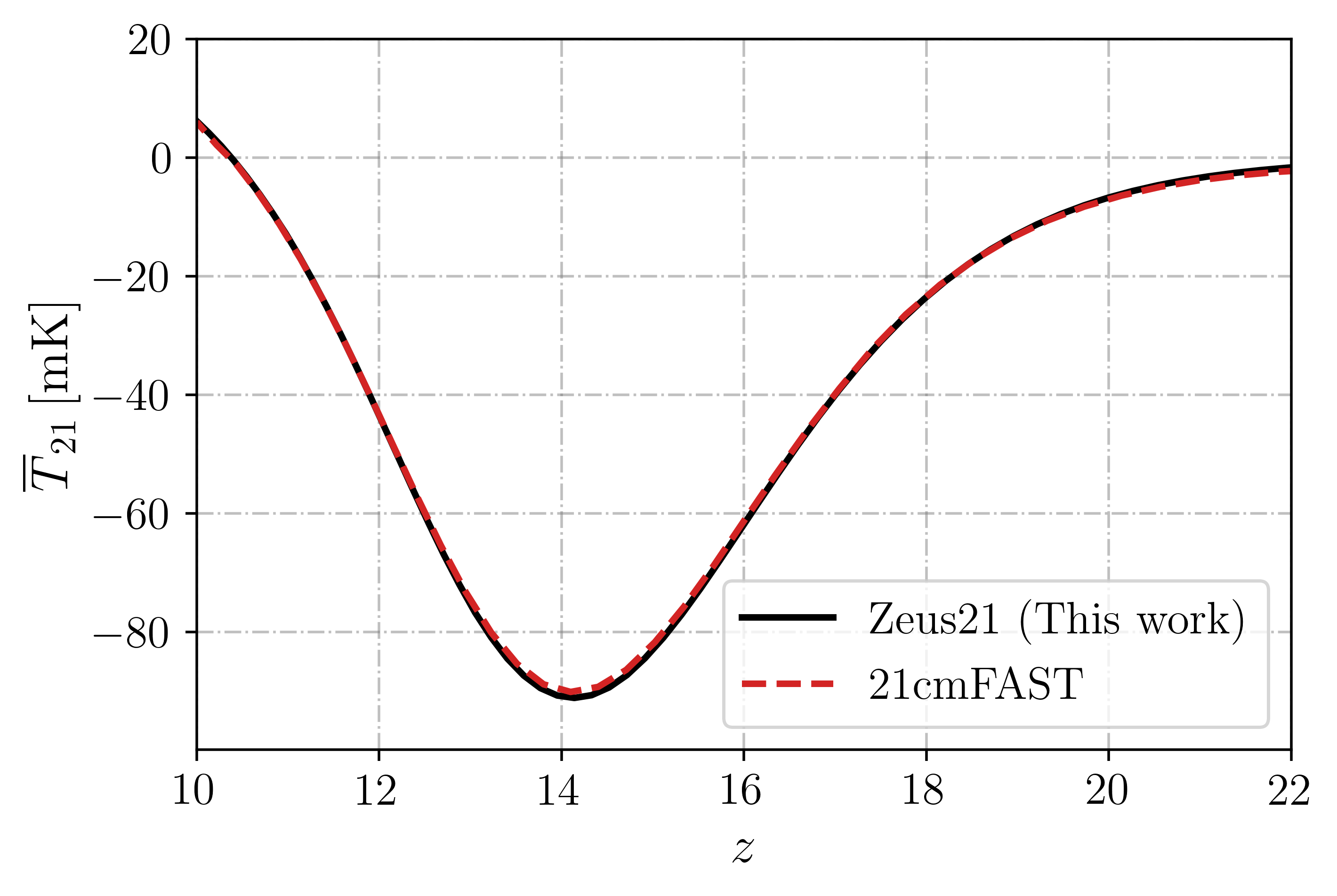}
	\caption{Global signal from our {\codename} (black) versus \cmfast\ (red dashed), given the same input parameters.
	}
	\label{fig:GS_vs_21cmFAST}
\end{figure}

We show our predicted 21-cm global signal in Fig.~\ref{fig:GS_vs_21cmFAST}, along with the \cmfast\ result.
These runs have been performed on a 120$^3$ box, with $180$ Mpc on the side (to yield a standard 1.5 Mpc resolution, which we have tested to be converged at the scales of interest, see Appendix~\ref{app:other21cmfast}).
Both cases show the same broad features, namely a cosmic-dawn trough peaking at $z\sim 14$ (as expected with atomic-cooling haloes only, cf.~Fig.~\ref{fig:T21_P21_z}), with Lyman-$\alpha$ taking the 21-cm signal into absorption at $z\sim 20$, and the gas being fully heated by $z\sim 10$.
The two global signals agree remarkably well across the entire cosmic-dawn.
We remind the reader we ignore reionization (setting $f_{\rm esc}=0$), and thus also photo-ionization feedback (fixing $M_{\rm turn} = 10^{7.5} \Msun$).

\begin{figure}
	\centering
	\includegraphics[width=0.46\textwidth]{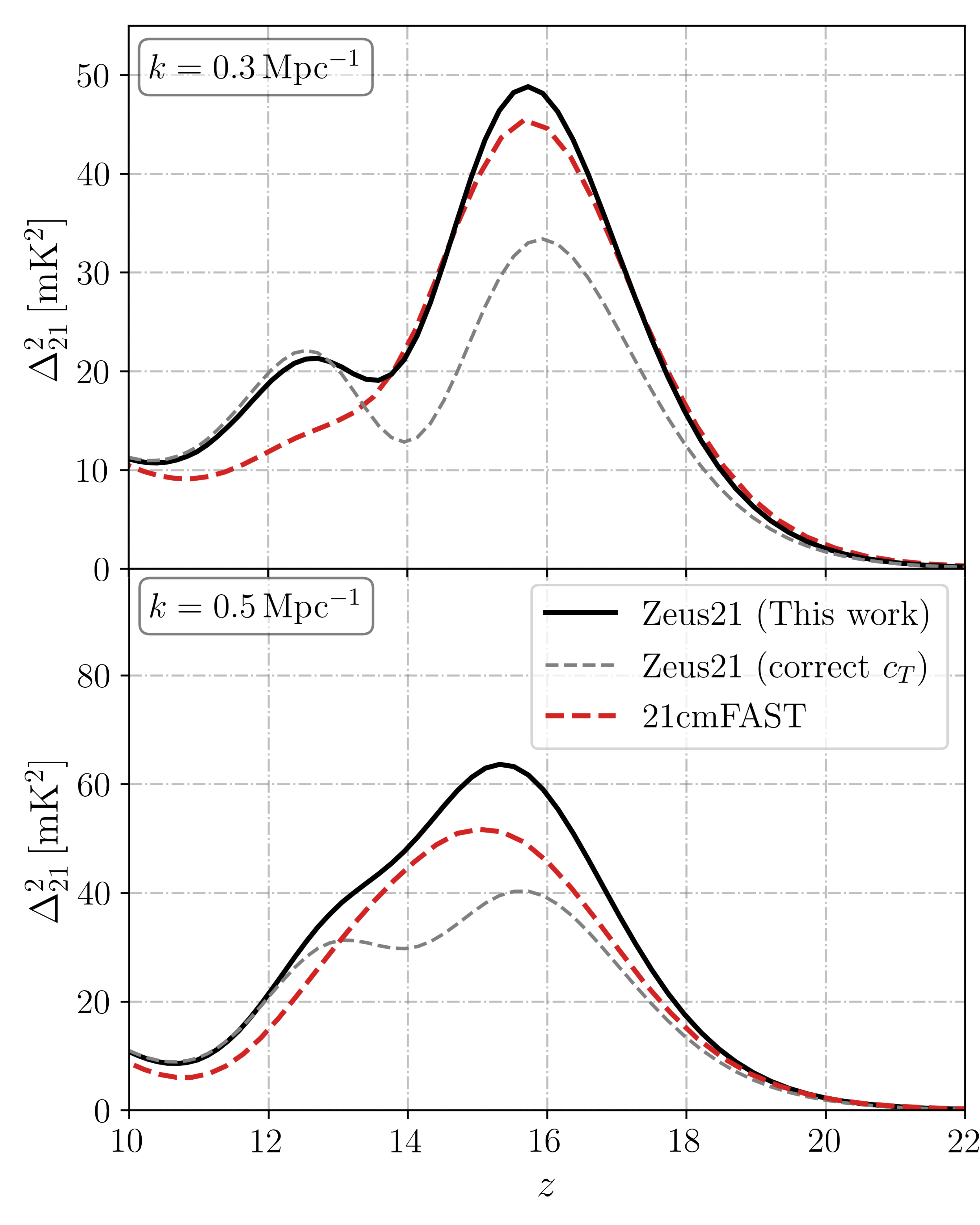}
	\caption{Evolution of the the 21-cm power spectrum across cosmic dawn at two wavenumbers, $k=0.3\,\rm Mpc^{-1}$ ({\bf top}) and $k=0.5\,\rm Mpc^{-1}$  ({\bf bottom}).
	Gray thin lines represent the result from assuming the full correct evolution of the adiabatic index $c_T$ (see Fig.~\ref{fig:cT_adiabatic}), rather than the \cmfast\ assumption of $T_k$ homogeneous at $z=35$, which we emulate in our black lines.
	}
	\label{fig:PS_vs_21cmFAST}
\end{figure}

The true test of \codename\ is in the 21-cm fluctuations, though.
We show  the 21-cm power spectrum at two wavenumbers, $k=0.3$ and 0.5 Mpc$^{-1}$, in Fig.~\ref{fig:PS_vs_21cmFAST}.
The output of the two codes agrees within $\sim 10\%$ through most of cosmic history, though they can deviate from each other at particular $z$ slices.
For instance, the $k=0.3\,\Mpcinv$ result from \codename\ shows a small bump at $z\sim 12$ that is negligible in \cmfast, and our $k=0.5\,\Mpcinv$ power is a bit larger. This may be due to the $k$ binning in simulations, or to some remaining assumption in \cmfast\ that we have not ported to \codename, including the treatment of adiabatic fluctuations (as our approach to match the behavior of $c_T$ in \cmfast\  is only approximate, and linear) or our simplified Lyman-$\alpha$ SED.
We have also plotted in Fig.~\ref{fig:PS_vs_21cmFAST} our prediction when correctly modeling the adiabatic evolution, as opposed to the \cmfast\ assumption that $T_k$ starts homogeneous at $z=35$. 
A full treatment of adiabatic fluctuations results in a 50\% downward revision of the 21-cm power spectrum during cosmic dawn, which ought to be corrected when doing inference with current and upcoming data\footnote{We suggest a solution for \cmfast, which consists of initializing the $T_k$ box with linear adiabatic fluctuations with our fitted index $c_T$ from Eq.~\eqref{eq:cT_fit}, and then evolving it normally.}.

This agreement between semi-numerical simulations and our fully analytic approach is promising, especially since we have not ``tweaked" any of the astrophysical parameters.
That is, the two calculations have the same set of inputs.
The fact that the fluctuations of each component (X-ray and Lyman-$\alpha$) agree with \cmfast\  (Figs.~\ref{fig:PTk_vs_21cmfast} and ~\ref{fig:PJa_vs_21cmfast}) builds confidence in our treatment of nonlinearities and nonlocalities.
Moreover, the 21-cm signal, both global and in the power spectrum, agrees to a similar $\sim$10\% accuracy.
Agreement beyond 10\% is technically difficult, and would require a dedicated code-comparison project, beyond the scope of this work.
Regardless, the semi-numeric treatment of codes like \cmfast\ is likely accurate only to the 10\% level~\citep{Zahn:2010yw,Park:2018ljd}.

We have stopped our calculations at $z=10$.
Below that the 21-cm signal will begin to saturate ($T_S \gg \Tcmb$), so the 21-cm fluctuations will just trace the LSS (given we are not yet modeling reionization fluctuations in \codename).
We show this in Appendix~\ref{app:other21cmfast}, where we compare against \cmfast\ and find agreement down to $z=5$.
Note that we have focused on our fiducial astrophysical parameters here, but also invite the reader to visit the same Appendix~\ref{app:other21cmfast} to see similar accuracy for other parameter sets.

We note that there may be lingering numerical issues in \cmfast\ near the cell size (e.g.~\citealt{Mao:2011xp,Georgiev:2021yvq}, as the power artificially rises near the resolution limit in Figs.~\ref{fig:PTk_vs_21cmfast} and~\ref{fig:PJa_vs_21cmfast}), as well as missing features on the \codename\ side (as for instance we are only considering linear adiabatic fluctuations, and are ignoring inhomogeneities on the WF correction from~\citealt{Hirata:2005mz}).
Further, \cmfast\ takes some approximations on the cosmology side, including the growth factor.
Nevertheless, our approach in \codename\ can reproduce simulation-based results with a much reduced computational cost ($<5$ s in a laptop), and negligible memory requirements.
This meets the benchmark required for next-generation 21-cm interferometers, making \codename\ an ideal 21-cm analysis tool.

\section{Discussion and Conclusions}
\label{sec:conclusions}

In this paper we have presented an {\it effective} model for the formation of the first structures, and thus for the 21-cm line of neutral hydrogen during cosmic dawn.
The effective nature of our approach consists of an approximation to the star-formation-rate density (SFRD), which we show traces the over/underdensities roughly as an exponential.
As such, it is a lognormal variable, and we can calculate its fluctuations analytically.
The power spectra of  quantities derived from the SFRD, such as the Lyman-$\alpha$ ($J_\alpha$) and X-ray $(J_X)$ fluxes, follows straightforwardly, and so does the 21-cm power spectrum.
We have implemented our lognormal model in the public package \codename, which can predict the 21-cm global signal and power spectrum including nonlinearities and nonlocalities from photon propagation during cosmic dawn.
We have shown remarkable agreement when comparing against \cmfast\ semi-numerical simulations.
Unlike \cmfast, however, a run of \codename\ takes $<5$ s in a single core.
We want to emphasize that \codename\ does not have the same purposes as simulations, like \cmfast.
We do not aim to produce 3D maps of $T_{21}$ so we are currently not computing higher-order correlations (e.g., bispectra), and we rely on our effective lognormal model for the SFRD (which for instance \cmfast\ does not need).
Nevertheless, \codename\ is built to produce a cheap -- but fully nonlinear -- computation of the 21-cm global signal and fluctuations.
\codename\ is fully built in Python and interfaces with cosmological codes like {\tt CLASS}.
We hope this work encourages community development and usage of the public \codename\ for modeling the 21-cm signal.

The most obvious use code for \codename\ is running inference on upcoming 21-cm data.
Usual Metropolis-Hastings MCMCs will become possible, given the speed of \codename~(comparable to \class). This will allow us to analyze upcoming data from 21-cm telescopes such as HERA~\citep[e.g.,][]{HERA:2022wmy} at a much reduced computational cost. 
In addition, we want to highlight here a few other possible applications.

$\bullet$ Beyond standard-model cosmology. Adding new parameters self-consistently in \codename\ comes at a relatively low cost, especially if their effects are already encoded in {\tt CLASS}, as \codename\ reads the matter transfer functions and adiabatic gas temperatures from its output. As a consequence, models of non-cold dark matter, such as ETHOS~\citep{Cyr-Racine:2015ihg}, warm~\citep{Bode:2000gq}, and fuzzy dark-matter~\citep{Marsh:2015xka} can be implemented during cosmic dawn with ease (see e.g.~\citealt{Schneider:2018xba,Munoz:2020mue,Jones:2021mrs} for some previous work in this direction). Likewise, models that heat or cool the gas (such as millicharged particles~\citealt{Munoz:2018pzp}, and annihilating or decaying dark matter~\citealt{Furlanetto:2006wp,Lopez-Honorez:2016sur}) are directly included as long as they are in {\tt CLASS}.

$\bullet$ Testing new astrophysical models. \codename\ is built to be modular, so one can modify each key astrophysical assumption independently. As an example, we have implemented two different models for the halo-galaxy connection at high $z$ (see Sec.~\ref{sec:effective_model}), and adding new ones is straightforward. Likewise about new effects like Lyman-$\alpha$ scattering or heating~\citep{Venumadhav:2018uwn}.

$\bullet$ Find an effective description of the first galaxies. Our lognormal model relies on the effective biases $\gamma_R$. Rather than predict them from a given halo-galaxy connection and cosmology, one can use them directly as free parameters and find them from the data. Our Fig.~\ref{fig:gamma_vs_R} hints at them being roughly $z$-independent, and thus a possibly simplified model for cosmic dawn.

$\bullet$ Constrain the SEDs of the first galaxies. Due to computational constraints, standard 21-cm tools such as \cmfast\ often assume that the X-ray spectra of the first galaxies is a power-law in energy (see, however,~\citealt{Das:2017fys}), which we have followed in this work. 
With \codename, however, we can implement any arbitrary X-ray (or UV) SED for the first sources, and find their effects in the 21-cm signal. While this question has been explored for the 21-cm global signal in \citet{Mirocha:2014faa}, ours is to our knowledge the first public code that can compute the 21-cm fluctuations for arbitrary X-ray SEDs.

$\bullet$ Studies of cross terms. We can easily find the effect of each term on the final 21-cm signal, enabling us to study their correlations. For instance, this has allowed us to find a missing assumption on the standard adiabatic cooling in \cmfast\ (see Fig.~\ref{fig:cT_adiabatic}).

$\bullet$ Predict fluctuations on other tracers. While we have focused on the 21-cm line as a tracer of the SFRD, there are other tracers of this quantity that could benefit from our effective approach.
For instance, line-intensity mapping of other high-$z$ lines is also expected to trace the SFR of each galaxy~\citep{Kovetz:2017agg}.
Likewise for galaxy luminosity functions at high $z$ from space telescopes like HST, JWST, and {\it Roman}, or Earth-based like {\it Subaru}.
Cross correlations to other observables ought to be modeled within our formalism, which we will examine in future work.

$\bullet$ Joint 21-cm and CMB analyses. As \codename\ interfaces with {\tt CLASS} in Python it is straightforward to jointly consider 21-cm data with CMB and other cosmological data-sets, including the large-scale structure and supernovae.

To summarize, we have presented an effective lognormal model for cosmic dawn, which allows us to find the evolution of the 21-cm signal including nonlinearities and nonlocalities fully analytically.
Our model is available as the open-source \codename\ software package.
This is a new tool to model the star-formation rate density and the 21-cm signal during cosmic dawn, and it sets the basis upon which to build an efficient 21-cm pipeline.
As such, \codename\ represents a key step towards unearthing astrophysical and cosmological information from the deluge of cosmic-dawn data ahead of us.

\subsection*{Acknowledgements}

JBM acknowledges partial support by the University of Texas at Austin and by a Clay Fellowship at the Smithsonian Astrophysical Observatory.
We are thankful to Yacine Ali-Ha\"imoud, Cari Cesarotti, Daniel Eisenstein, and Jordan Mirocha for enlightening discussions;
as well as Jordan Flitter, Ely Kovetz, Adrian Liu, Andrei Mesinger, Jordan Mirocha, and the anonymous referee for insightful comments on a previous version of this manuscript.

\subsection*{Data Availability}

Our code is publicly available on GitHub. The data underlying this article will be shared on reasonable request to the author.

{\bf Software}:~numpy~\citep{numpy}, scipy~\citep{scipy}, powerbox~\citep{powerbox_Murray}, mcfit, classy~\citep{Blas:2011rf}, hyrec~\citep{AliHaimoud:2010dx}, 21cmFAST~\citep{Mesinger:2010ne}.

\bibliographystyle{mnras}
\bibliography{21cm_hiz}

\appendix

\section{Normalization of the lognormal SFRD}
\label{app:normalization}

Through this work we have assumed a functional form
\be
{\rm SFRD}(\delta_R) \propto e^{\gamma_R  \delta_R}
\ee
for the SFRD at each $R$.
Since $\delta_R$ is a Gaussian variable with norm zero, this makes the SFRD a log-normal variable.
However, its average will not necessarily be the $\delta_R=0$ result, which raises a potential issue when normalizing the SFRD against simulations.
In the main text we defined the $\tilde \delta_R$ variable in Eq.~\eqref{eq:tilde_deltaR}, so that $\VEV{e^{\gamma_R \tilde \delta_R}} = 1$ and the factor in front of Eq.~\eqref{eq:sfrd_exponential} is simply $\rm \overline{SFRD}(z)$, the SFRD obtained from the spatially averaged Sheth-Tormen HMF (see Eq.~\ref{eq:sfrd}).
	
This would be the end of the story if we worked in Lagrangian space (as that is where the Sheth-Tormen fit we use in Eq.~\ref{eq:sfrd} is calibrated).
However, the SFRD is evaluated in Eulerian space, which gives rise to an extra factor of $(1+\delta)$ (see Eq.~\ref{eq:SFRD_EPS}).
As such, we ought to correct the pre-factor of $\rm \overline{SFRD}$ to provide the true Eulerian-space average.

This correction will have the form
\be
\phi_R = \dfrac{\VEV{(1 + \delta_R) \times \rm SFRD^L(\delta_R)}}{\VEV{\rm SFRD^L(\delta_R)} } = 1 + \VEV{\delta_R\,e^{(\gamma_R-1) \tilde \delta_R}},
\ee
where we have used Eqs.~(\ref{eq:SFRD_EPS}) and (\ref{eq:sfrd_exponential}), from Sec.~\ref{sec:effective_model} and the $^L$ superscript implies Lagrangian space.	For a linear model of the SFRD this correction will vanish.	
Of course our model is nonlinear, but its analytic nature allows us to calculate this correction in a closed form to approximately be
\be
\phi_R \approx 1 + (\gamma_R - 1) \sigma_R^2,
\ee
which grows towards smaller $R$, where the SFRD is more nonlinear against $\delta_R$.
Added over all $R$, this correction is $\sim 5$\% for the Lyman-$\alpha$ and X-ray power spectra in our fiducial case. 
While it is below our 10\% precision goal, we include it in \codename\, and allow the user to turn it on or off with a flag.

\section{Non-linear Expansions}
\label{app:Taylor}

In this appendix we expand the different terms of the 21-cm signal to higher order, and study how well our approximations work in describing both the 21-cm global signal and the power spectrum.

We begin with the expansion of temperature term,
\be
1 - \dfrac{\Tcmb}{T_c} = 1 - \dfrac{\Tcmb}{\overline T_c} + \dfrac{\Tcmb}{\overline T_c} \sum_{n=1}^\infty (-1)^{n+1} \left(\dfrac{\delta T_k}{\overline T_k} \right)^n,
\label{eq:Tk_Taylor_appendix}
\ee
where, as in the main text, we ignore fluctuations on the $T_c-T_k$ relation (from \citealt{Hirata:2005mz}), in which case $\delta T_c/\overline T_c = \delta T_k/\overline T_k$.
In practice, Eq.~\eqref{eq:Tk_Taylor_appendix} may not converge as one sums over larger $n$, as this is an asymptotic series (since there is a pole at $T_k=0$, so the radius of convergence of this series around that point is formally zero in the complex plane).
This is, nevertheless, physically irrelevant, as $T_k$ does not vanish during cosmic evolution, and when expanding around $T_k=T_k^{\rm ad}$ the radius of convergence will be finite (but still we expect the series to break at large $\delta T_k$).
Likewise, we expand the Lyman-$\alpha$ term as
\be
\dfrac{x_\alpha}{1 + x_\alpha} = \dfrac{\overline x_\alpha}{1 + \overline  x_\alpha} + \dfrac{1}{1 + \overline x_\alpha} \sum_{n=1}^\infty (-1)^{n+1} \left(\dfrac{\delta x_\alpha}{1 + \overline x_\alpha} \right)^n.
\ee
This case has two interesting regimes.
For $x_\alpha \ll 1$ the linear term suffices, even for large fluctuations, due to the $1+$ term on the denominator.
Later on, however, we expect $x_\alpha \gtrsim 1$, in which case higher-order terms can matter.
As such, we expect that corrections from non-linearities will grow at lower $z$.
We have worked to $n=1$ in the text, and we will study the corrections from higher $n$ to both terms here.

\begin{figure}
	\centering
	\includegraphics[width=0.46\textwidth]{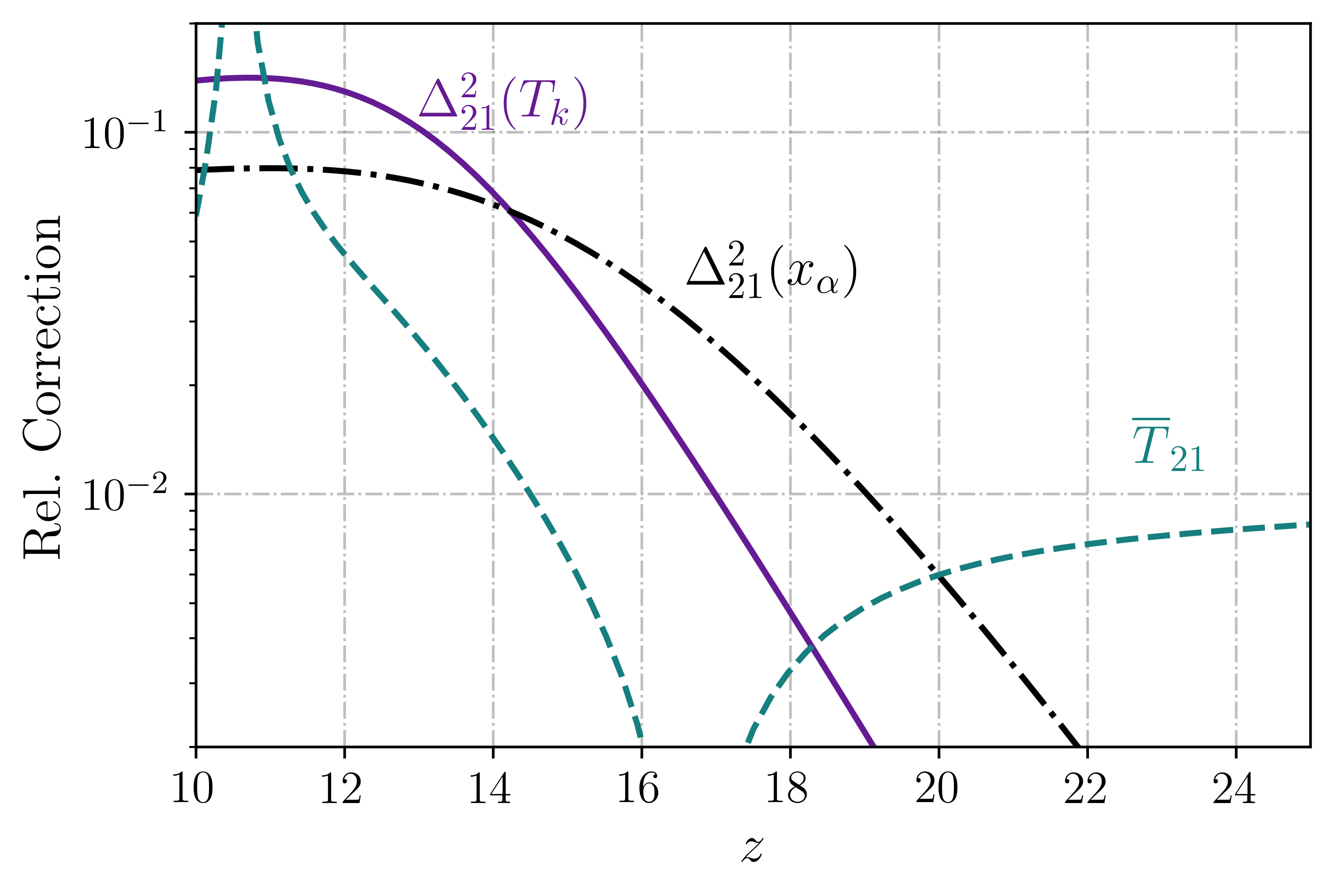}
	\caption{Corrections to the 21-cm power spectrum and the global signal from higher-order terms that we ignore in this work, as a function of $z$.
		We show our estimate of the relative contribution to $\Delta^2_{21}$ at observables scales $k$ from non-linear terms in the kinetic temperature $T_k$ (purple solid), and in the Lyman-$\alpha$ coupling (black dot-dashed).
		We also show a similar estimate for the 21-cm global signal $\overline T_{21}$ (teal dashed, from all contributions).
		It blows up at $z\sim 11$ where the global signal vanishes, but the correction is under 1 mK at all $z$ for this model.
	}
	\label{fig:corrs_quadr}
\end{figure}

\subsection*{Corrections to the Power Spectrum}

Let us estimate the size of the correction to the 21-cm power spectrum from the terms ignored.
For that, we compare the 2nd- and 1st-order  terms in the above equations to each other.
In this estimate we will simply take $\delta q = \sigma(q)$ for both $q=\{\xa,T_k\}$, with $\sigma$ averaged over $R=10$ Mpc in order to represent the typical scales observed in 21-cm experiments.
We will work in real space, and ignore adiabatic fluctuations in $T_k$, both for simplicity and to home in on the impact of our nonlinear SFRD model.
We show the ratios for both $\xa$ and $T_k$ in Fig.~\ref{fig:corrs_quadr}, which are safely below 10\% for $z > 12$, during cosmic dawn.
They rise to roughly 10\% at lower $z\sim 10$, though this error is still comparable to the expected uncertainties in fully nonlinear semi-numerical simulations (e.g., \cmfast\ analyses often include a 20\% ``systematic" uncertainty~\citealt{Park:2018ljd}).
As such, we have ensured that our Taylor expansion of the $T_k$ and $\xa$ terms within $T_{21}$ is accurate enough for determining the 21-cm power spectrum during cosmic dawn.
At $z\lesssim 12$ one may need to model quadratic terms in $T_{21}$ to go below 10\% error.
We note, however, that precisely at those lower $z$ the 21-cm fluctuations from reionization will start to grow, and they are expected to overshadow the $T_k$ and $\xa$ fluctuations in Fig.~\ref{fig:corrs_quadr}, so this is largely an academic issue.

\subsection*{Corrections to the Global Signal}

Let us now calculate the correction $\Delta\overline{T}_{21}$ to the 21-cm global signal.
We expand Eq.~\eqref{eq:T21def} to second order in fluctuations as
\be
\dfrac{\Delta\overline{T}_{21}}{\overline{T}_{21}} = \sum_{i>j} \beta_i \beta_j \VEV{\delta q_i\delta q_j} + \beta_\alpha^{(2)} \sigma^2(x_\alpha)+ \beta_T^{(2)} \sigma^2(T_k),
\ee
where the sum runs over all cross terms (with $q_i=\{\delta,x_\alpha,T_k\}$), and the last two terms in this Equation include the higher-order Taylor expansion (note there is not one for $q=\delta$ in real space).
We define the second-order $\beta_i^{(2)}$ parameters to be
$\beta_T^{(2)} = \beta_T/\overline T_k$ and $\beta_\alpha^{(2)} = \beta_\alpha/(1 + \overline x_\alpha$) in terms of the first-order $\beta_i$ (extracted from Eqs.~\eqref{eq:Taylor_expansions} in the main text, see also BLPF).
We show this correction in Fig.~\ref{fig:corrs_quadr}, which is safely below 10\% during the entire cosmic dawn, barring a spike at $z\sim 11$ where the global signal vanishes.
The absolute correction is below 1 mK for the entire cosmic-dawn range we study.

We thus conclude that corrections from quadratic (and higher) terms in the definition of $T_{21}$ are largely unimportant during cosmic dawn, both for the global signal and the power spectrum.
This is in strike contrast with the corrections from the nonlinearities arising from the SFRD (see e.g.~Figs.~\ref{fig:Deltasq_xa} and~\ref{fig:Deltasq_Tk}).

\begin{figure}
	\centering
	\includegraphics[width=0.46\textwidth]{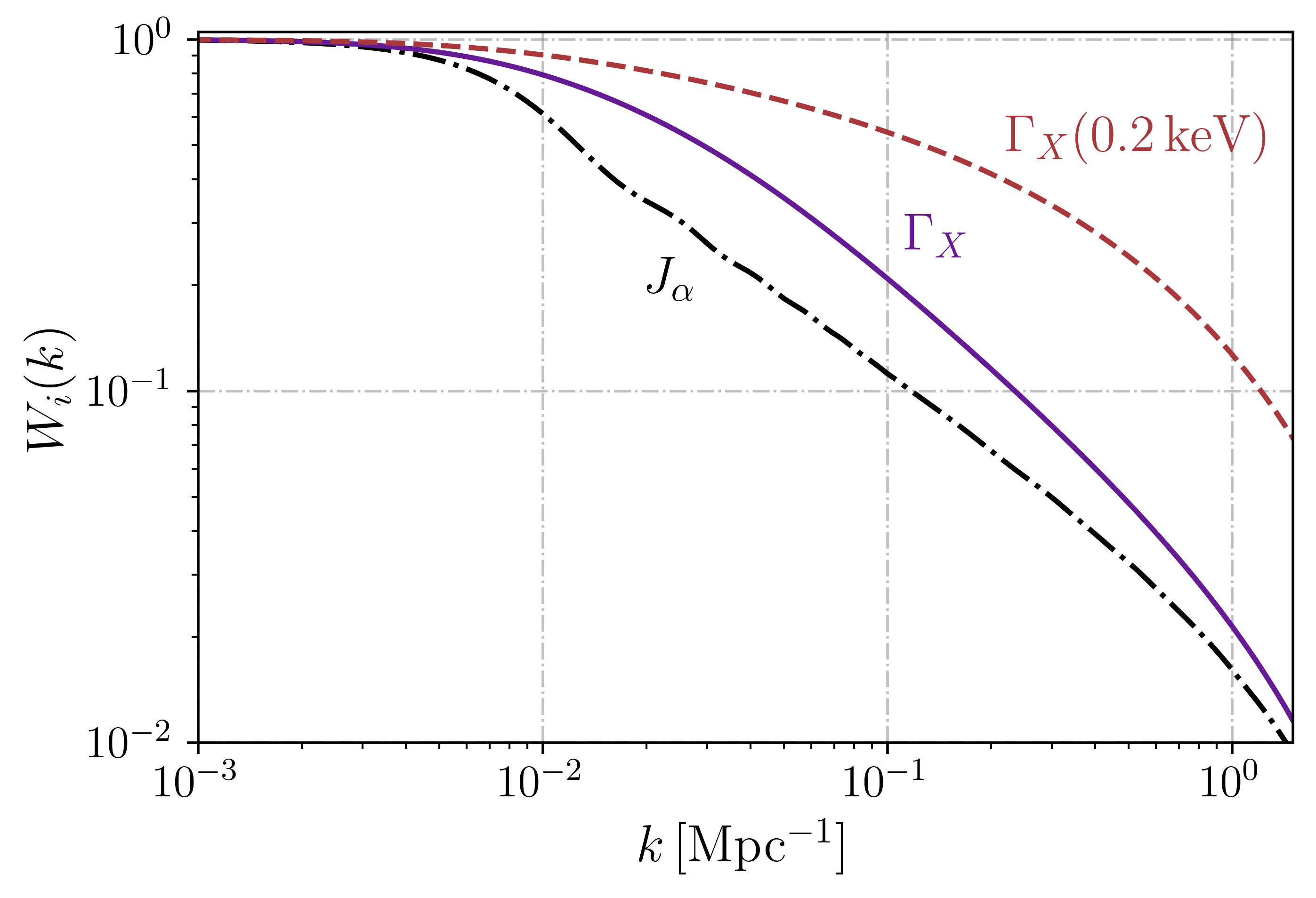}
	\caption{Window function for (linear) fluctuations in the Lyman-$\alpha$ flux (black dot-dashed), and in the X-ray heating term $\Gamma_X$ (purple) for our fiducial case of $E_0=0.5$ keV.
		If we set a lower $E_0=0.2$ keV (red dashed) the fluctuations survive undamped to higher $k$, as lower-energy X-rays are absorbed more locally. All cases are at $z=10$.
	}
	\label{fig:windows_app}
\end{figure}

\section{Impact of the X-ray SED}
\label{app:Xrays}

Here we illustrate the large impact that the X-ray SED has in cosmic dawn, by showcasing a case with a smaller energy cutoff $E_0$ in the SED.

Rather than show the entire 21-cm evolution,  let us focus on the window function in Fourier space in order to isolate the effect of X-ray propagation (as both are easy to calculate with \codename, and the latter gives us direct access to the astrophysics of cosmic dawn).
The linear window functions $W_i$ are defined in terms of the (linear) power spectra as
\be
\Delta^2_i(k) = W_i^2(k) \Delta^2_m(k)
\ee
for any quantity $i$, and as such they can be understood as describing how much power at each $k$ is lost due to effects like photon propagation.

\begin{figure}
	\centering
	\includegraphics[width=0.46\textwidth]{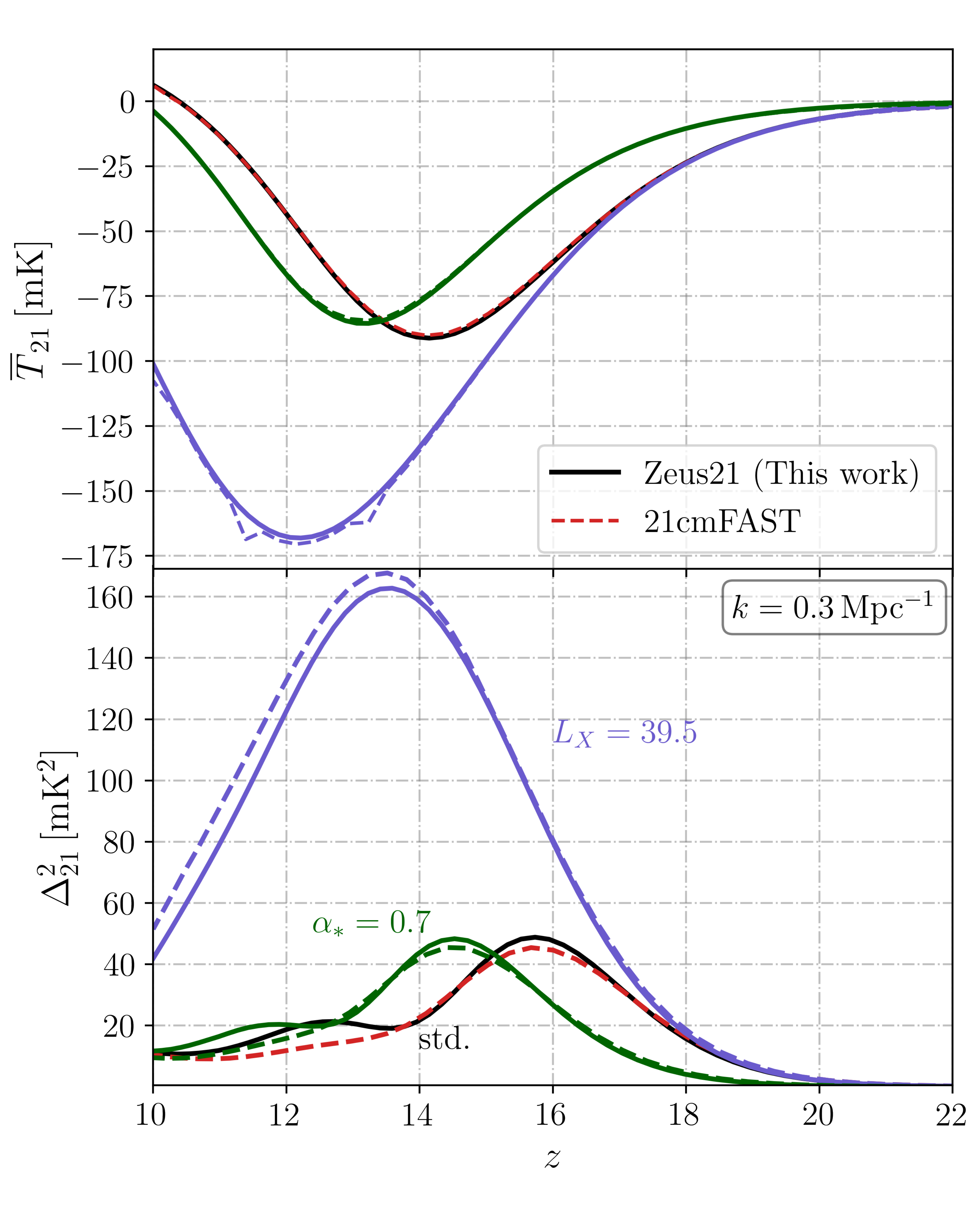}
	\caption{Comparison of the 21-cm global signal ({\bf top}) and power spectrum ({\bf bottom}, at $k=0.3\,\Mpcinv$) between \codename\ (solid) and \cmfast\ (dashed).
		The black/red lines correspond to our fiducial parameter set chosen through the paper, whereas the blue lines have a factor of 10 lower X-ray luminosity per unit SFR $(L_X$, which produces some numerical noise on the \cmfast\ simulations), and the green lines a steeper faint-end of the luminosity function ($\alpha_*$).
		In all cases our effective approach in \codename\ reproduces the \cmfast\ outputs accurately.
	}
	\label{fig:params_21cmfast}
\end{figure}

We show the linear windows $W_i$ in Fig.~\ref{fig:windows_app} for three different cases, all at $z=10$.
First, for our fiducial X-ray case with $E_0=0.5$ keV the window stays near unity until $k\sim 1/\lambda_{\rm mfp}$, where $\lambda_{\rm mfp}$ is the mean-free path of X-ray photons around the threshold ($\lambda_{\rm mfp}\sim 35$ Mpc for this fiducial).
This is similar to the Lyman-$\alpha$ case, which drops at slightly larger scales (lower $k$). In that case the drop is due to the distance that a photon can travel between two Lyman resonances, which is typically $\sim 100$ Mpc comoving, rather than the X-ray cross section.
A smaller cutoff $E_0=0.2$ keV on the X-ray SED makes a large impact on the 21-cm fluctuations, as the last window function we show in Fig.~\ref{fig:windows_app} illustrates.
In that case $\lambda_{\rm mfp}\sim 2$ Mpc, and as thus fluctuations are retained to larger $k$.
These windows are useful to build intuition, though of course they will be modified by our nonlinear SFRD apporach in \codename.

\section{Broader Comparison to \cmfast}
\label{app:other21cmfast}

\begin{figure}
	\centering
	\includegraphics[width=0.46\textwidth]{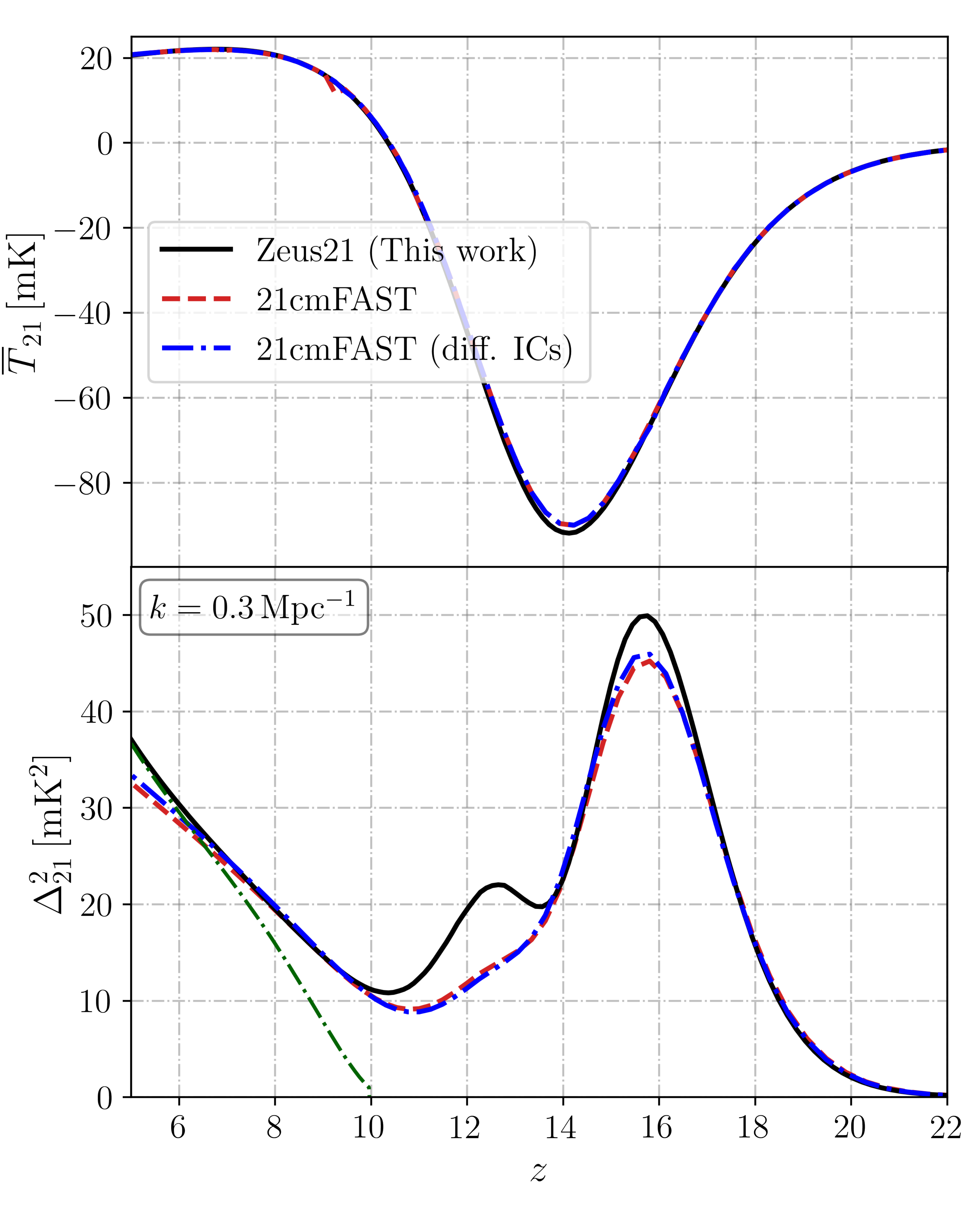}
	\caption{Same as Fig.~\ref{fig:params_21cmfast} but all the way down to $z=5$. We have set our fiducial parameters as in the main text, including $\fesc=0$ (as we are not yet modeling reionization bubbles). 
		Below $z=10$ the spin temperature begins to saturate ($T_S \gg \Tcmb$), so in the absence of reionization the 21-cm signal will just trace densities.
		The green dash-dotted line denotes $\overline T_{21}^2\Delta_m^2$, which indeed matches the 21-cm power at low $z$ under our assumptions.
		Finally, the blue dot-dashed curves correspond to a different set of initial conditions on \cmfast, which shows convergence at the scales and $z$ of interest.
	}
	\label{fig:compare_21cmfast_z5}
\end{figure}

In the main text we compared the output of \codename\ to \cmfast\ under one fiducial parameter set, finding agreement down to $z=10$.
In this appendix we show how that agreement generalizes to other astrophysical situations, to lower  redshifts, and check that our choice of box-size is appropriate.

Rather than perform an entire parameter scan, which is computationally expensive in \cmfast\ and beyond the scope of this work, we will simply vary two of the main input parameters.
We run two simulations, one with a power-law $\alpha_*=0.7$ for the faint end of the halo-galaxy connection (steeper than the fiducial $\alpha_*=0.5$), and one with a different X-ray efficiency $L_{40}=10^{-0.5}$, equivalent to $\log_{10}(L_X/\rm SFR) = 39.5$ (a factor of 10 lower than usual).
We show the output of both codes (global signal and power spectra) for these parameters in Fig.~\ref{fig:params_21cmfast}.
The case with lower $L_{40}$ produces a much deeper 21-cm global signal, and consequently a stronger 21-cm power spectrum during cosmic dawn, which \codename\ captures (the spikes in the \cmfast\ output are likely due to numerical noise).
Likewise, the case with a steeper index $\alpha_*$ delays structure formation by reducing the brightness of small-mass objects, which shows up as a lower-$z$ 21-cm trough.

As for lower $z$, we show in Fig.~\ref{fig:compare_21cmfast_z5} the result of running both codes down to $z=5$, under our assumption of $\fesc=0$ (i.e., no reionization).
This shows the same level of agreement at $z<10$ as we found during cosmic dawn.
This is to be expected, as for $z<10$ our 21-cm signal is fully saturated ($T_S\gg \Tcmb$), in which case the 21-cm fluctuations will simply trace densities.
We show this is true in Fig.~\ref{fig:compare_21cmfast_z5}, as we also show a curve proportional to the matter power spectrum.

Finally, through this work we have compared \codename\ with a \cmfast\ simulation box that is 180 Mpc on a side, with 1.5 Mpc resolution. While this is smaller than recommended in~\citet{Kaur:2020qsa}, it more than suffices to capture the 21-cm signal at the scales that we study.
To show this, we run a \cmfast\ box with different initial conditions and show the results in Fig.~\ref{fig:compare_21cmfast_z5}.
It is clear that any sample variance from the box size is small during cosmic dawn (see \citealt{Giri:2022ijp} for a similar discussion during reionization).

To summarize, in all cases we see remarkable agreement between \codename\ and \cmfast, building confidence in our effective model for the cosmic-dawn 21-cm signal.

\end{document}